\documentclass[twoside,twocolumn,9pt]{article}
\usepackage{extsizes}
\usepackage[super,sort&compress,comma]{natbib} 
\usepackage[version=3]{mhchem}
\usepackage[left=1.5cm, right=1.5cm, top=1.785cm, bottom=2.0cm]{geometry}
\usepackage{balance}
\usepackage{mathptmx}
\usepackage{sectsty}
\usepackage{graphicx} 
\usepackage{lastpage}
\usepackage[format=plain,justification=justified,singlelinecheck=false,font={stretch=1.125,small,sf},labelfont=bf,labelsep=space]{caption}
\usepackage{float}
\usepackage{fancyhdr}
\usepackage{fnpos}
\usepackage[english]{babel}
\addto{\captionsenglish}{%
  
}
\usepackage{array}
\usepackage{droidsans}
\usepackage{charter}
\usepackage[T1]{fontenc}
\usepackage[usenames,dvipsnames]{xcolor}
\usepackage{setspace}
\usepackage[compact]{titlesec}
\usepackage{hyperref}
\usepackage{upgreek}
\usepackage{array}
\usepackage{placeins}
\usepackage{textgreek}
\usepackage{pdfpages} 
\usepackage{pgffor} 

\usepackage{xr}
\makeatletter
\AtBeginDocument{\let\LS@rot\@undefined}

\newcommand*{\addFileDependency}[1]{
\typeout{(#1)}
\@addtofilelist{#1}
\IfFileExists{#1}{}{\typeout{No file #1.}}
}\makeatother

\newcommand*{\myexternaldocument}[1]{%
\externaldocument{#1}%
\addFileDependency{#1.tex}%
\addFileDependency{#1.aux}%
}

\myexternaldocument{supporting_info}

\makeatother

\def\supplementfilename{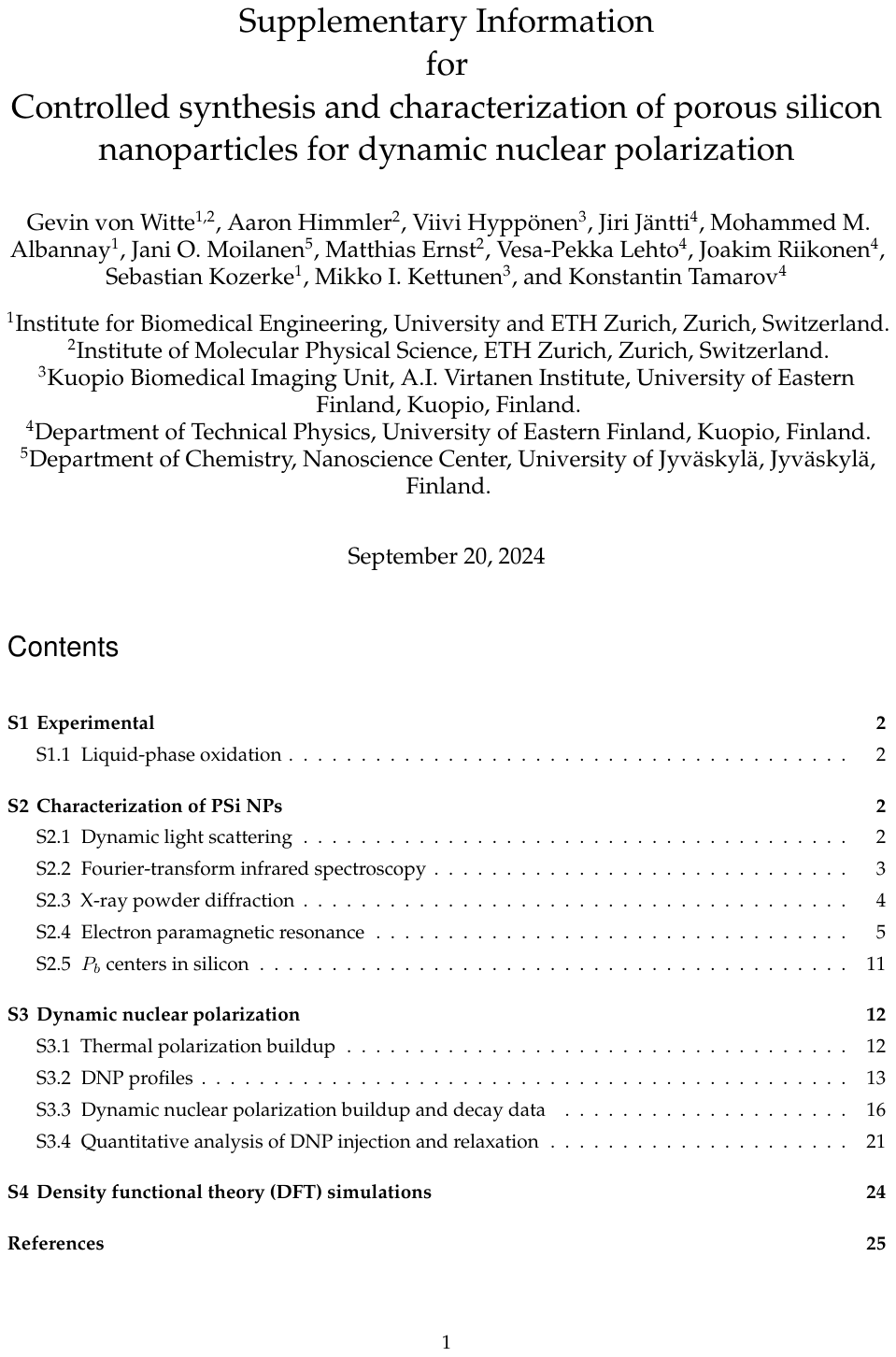}

\pdfximage{\supplementfilename}
\def\numbersupplementpages{\the\pdflastximagepages}

\newif\ifarXiv
\arXivtrue

\definecolor{cream}{RGB}{222,217,201}

\begin{document}

\pagestyle{fancy}
\thispagestyle{plain}
\fancypagestyle{plain}{
\renewcommand{\headrulewidth}{0pt}
}

\makeFNbottom
\makeatletter
\renewcommand\LARGE{\@setfontsize\LARGE{15pt}{17}}
\renewcommand\Large{\@setfontsize\Large{12pt}{14}}
\renewcommand\large{\@setfontsize\large{10pt}{12}}
\renewcommand\footnotesize{\@setfontsize\footnotesize{7pt}{10}}
\makeatother

\renewcommand{\thefootnote}{\fnsymbol{footnote}}
\renewcommand\footnoterule{\vspace*{1pt}%
\color{cream}\hrule width 3.5in height 0.4pt \color{black}\vspace*{5pt}} 
\setcounter{secnumdepth}{5}

\makeatletter 
\renewcommand\@biblabel[1]{#1}            
\renewcommand\@makefntext[1]%
{\noindent\makebox[0pt][r]{\@thefnmark\,}#1}
\makeatother 
\renewcommand{\figurename}{\small{Fig.}~}
\sectionfont{\sffamily\Large}
\subsectionfont{\normalsize}
\subsubsectionfont{\bf}
\setstretch{1.125} 
\setlength{\skip\footins}{0.8cm}
\setlength{\footnotesep}{0.25cm}
\setlength{\jot}{10pt}
\titlespacing*{\section}{0pt}{4pt}{4pt}
\titlespacing*{\subsection}{0pt}{15pt}{1pt}

\fancyfoot{}
\fancyfoot[LO,RE]{\vspace{-7.1pt}\includegraphics[height=9pt]{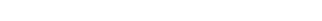}}
\fancyfoot[CO]{\vspace{-7.1pt}\hspace{13.2cm}\includegraphics{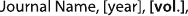}}
\fancyfoot[CE]{\vspace{-7.2pt}\hspace{-14.2cm}\includegraphics{head_foot/RF}}
\fancyfoot[RO]{\footnotesize{\sffamily{1--\pageref{LastPage} ~\textbar  \hspace{2pt}\thepage}}}
\fancyfoot[LE]{\footnotesize{\sffamily{\thepage~\textbar\hspace{3.45cm} 1--\pageref{LastPage}}}}
\fancyhead{}
\renewcommand{\headrulewidth}{0pt} 
\renewcommand{\footrulewidth}{0pt}
\setlength{\arrayrulewidth}{1pt}
\setlength{\columnsep}{6.5mm}
\setlength\bibsep{1pt}

\makeatletter 
\newlength{\figrulesep} 
\setlength{\figrulesep}{0.5\textfloatsep} 

\newcommand{\topfigrule}{\vspace*{-1pt}%
\noindent{\color{cream}\rule[-\figrulesep]{\columnwidth}{1.5pt}} }

\newcommand{\botfigrule}{\vspace*{-2pt}%
\noindent{\color{cream}\rule[\figrulesep]{\columnwidth}{1.5pt}} }

\newcommand{\dblfigrule}{\vspace*{-1pt}%
\noindent{\color{cream}\rule[-\figrulesep]{\textwidth}{1.5pt}} }

\makeatother

\twocolumn[
  \begin{@twocolumnfalse}
{\includegraphics[height=30pt]{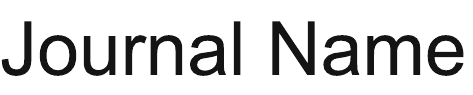}\hfill\raisebox{0pt}[0pt][0pt]{\includegraphics[height=55pt]{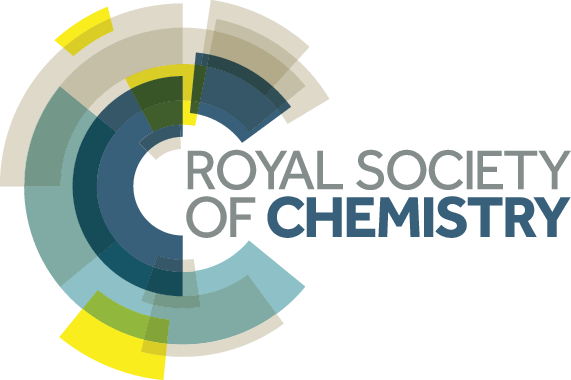}}\\[1ex]
\includegraphics[width=18.5cm]{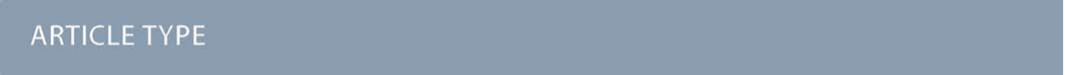}}\par
\vspace{1em}
\sffamily
\begin{tabular}{m{4.5cm} p{13.5cm} }

\includegraphics{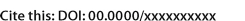} & \noindent\LARGE{\textbf{Controlled synthesis and characterization of porous silicon nanoparticles for dynamic nuclear polarization$^\dag$}} \\
\vspace{0.3cm} & \vspace{0.3cm} \\

 & \noindent\large{Gevin von Witte,\textit{$^{a,b}$} Aaron Himmler,\textit{$^{b}$} Viivi Hyppönen\textit{$^{c}$}, Jiri Jäntti\textit{$^{d}$}, Mohammed M. Albannay\textit{$^{a}$}, Jani O. Moilanen\textit{$^{e}$}, Matthias Ernst\textit{$^{b}$}, Vesa-Pekka Lehto\textit{$^{d}$}, Joakim Riikonen\textit{$^{d}$}, Sebastian Kozerke\textit{$^{a}$}, Mikko I. Kettunen\textit{$^{c\ddag}$} and Konstantin Tamarov$^{\ast}$\textit{$^{d\ddag}$}} \\

\includegraphics{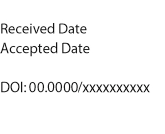} & \noindent\normalsize{
Si nanoparticles (NPs) have been actively developed as a hyperpolarized magnetic resonance imaging (MRI) contrast agent with an imaging window close to one hour.
However, the progress in the development of NPs has been hampered by the incomplete understanding of their structural properties that correspond to efficient hyperpolarization buildup and long polarization decays.
In this work we study dynamic nuclear polarization (DNP) of single crystal porous Si (PSi) NPs with defined doping densities ranging from nominally undoped to highly doped with boron or phosphorus.
To develop such PSi NPs we perform low-load metal-assisted catalytic etching for electronic grade Si powder followed by thermal oxidation to form the dangling bonds in the Si/\ce{SiO2} interface, the $P_b$ centers.
$P_b$ centers are the endogenous source of the unpaired electron spins necessary for DNP.
The controlled fabrication and oxidation procedures allow us to thoroughly investigate the impact of the magnetic field, temperature and doping on the DNP process.
We argue that the buildup and decay rate constants are independent of size of Si crystals between approximately 10 and 60\,nm.
Instead, the rates are limited by the polarization transfer across the nuclear spin diffusion barrier determined by the large hyperfine shift of the central \textsuperscript{29}Si nuclei of the $P_b$ centers.
The size-independent rates are then weakly affected by the doping degree for low and moderately doped Si although slight doping is required to achieve the highest polarization.
Thus, we find the room temperature relaxation of low boron doped PSi NPs reaching $75 \pm 3$\,minutes and nuclear polarization levels exceeding $\sim 6$\% when polarized at 6.7\,T and 1.4\,K.
Our study thus establishes solid grounds for further development of Si NPs as hyperpolarized contrast agents.} \\

\end{tabular}

 \end{@twocolumnfalse} \vspace{0.6cm}

  ]

\renewcommand*\rmdefault{bch}\normalfont\upshape
\rmfamily
\section*{}
\vspace{-1cm}


\footnotetext{\textit{$^{\ast}$E-mail: konstantin.tamarov@uef.fi}}
\footnotetext{\textit{$^{a}$~Institute for Biomedical Engineering, University and ETH Zurich, 
Zurich, Switzerland. }}
\footnotetext{\textit{$^{b}$~Institute of Molecular Physical Science, ETH Zurich, Zurich, Switzerland. }}
\footnotetext{\textit{$^{c}$~Kuopio Biomedical Imaging Unit, A.I. Virtanen Institute, 
University of Eastern Finland, Kuopio, Finland. }}
\footnotetext{\textit{$^{d}$~Department of Technical Physics, University of Eastern Finland, 
Kuopio, Finland. }}
\footnotetext{\textit{$^{e}$~Department of Chemistry, Nanoscience Center, 
University of Jyväskylä, Jyväskylä, Finland. }}

\footnotetext{\dag~Supplementary Information available: FTIR, XRPD, EPR characterization data for all the samples as well as the additional DNP data including DNP sweep spectra, polarization buildup and decay data, rate constants for various experimental conditions. See DOI: 00.0000/00000000.}

\footnotetext{\ddag~Joint senior authorship.}


\section{Introduction}

Magnetic resonance imaging (MRI) is a non-invasive versatile technique that can provide anatomical images with either sub-millimeter spatial\cite{Dumoulin2018} or milliseconds temporal\cite{Zhong2021} resolution.
Applying recent advances in artificial intelligence based image reconstruction and enhancement methods\cite{Rudie2022}, low field MRI has recently reached real world adoption even in the mobile setting\cite{Deoni2022}.
MRI, however, is inherently insensitive at room temperature due to low thermal polarization of nuclei, which complicates the observation of nuclei other than \textsuperscript{1}H.
Detecting low-abundant nuclei, such as \textsuperscript{13}C, \textsuperscript{15}N or \textsuperscript{29}Si, brings additional versatility to MRI allowing to \textit{e.g.}, image tumor metabolism\cite{Day2007}, locally detect pH\cite{Gallagher2008,Jiang2015}, detect Si particles \textit{in-vivo} during prolonged time window\cite{Cassidy2013,Kwiatkowski2017,Seo2018}.
Porous Si nanoparticles (NPs) hold particular promise due to their biocompatibility and numerous treatment modalities\cite{Santos2021}.

To detect Si NPs in an MRI scanner, their \textsuperscript{29}Si nuclei require hyperpolarization \textit{i.e.}, a polarization significantly beyond thermal equilibrium at body temperature.
A mature method to hyperpolarize various nuclei in the solid state is dynamic nuclear polarization (DNP)\cite{Wenckebach2016}.
DNP requires the presence of polarized unpaired electronic spins, whose polarization is subsequently transferred to hyperfine (HF) coupled nuclei by (near-) resonant microwave (MW) irradiation.

In Si, the unbound electrons required for DNP can originate from substitutional donor dopant atoms, such as group V (P, As, Sb, Bi) or group VI (S) atoms, which carry one or more donor electrons.
As each dopant carries extra electron(s), the majority carriers are negatively charged electrons and Si is named \textit{n}-type.
Spins of electrons bound to \textsuperscript{31}P donors have been widely used to polarize \textsuperscript{29}Si nuclear spins and to study polarization dynamics in bulk Si samples with different \textsuperscript{29}Si and \textsuperscript{31}P content\cite{Hayashi2009,Dementyev2011,Jarvinen2015,Jarvinen2020}.
With the variation of \textsuperscript{31}P and \textsuperscript{29}Si content, well resolved solid effect (SE)\cite{Hayashi2009,Jarvinen2015,Jarvinen2020}, differential SE\cite{Hayashi2009} and Overhauser effect (OE)\cite{Dementyev2011,Jarvinen2015,Jarvinen2020} DNP mechanisms of \textsuperscript{29}Si hyperpolarization have been identified.
More sophisticated protocols, such as resonant polarization transfer from polarized \textsuperscript{31}P to \textsuperscript{29}Si nuclei\cite{Dluhy2015}, have been demonstrated.

If Si is doped with group III atoms, in particular boron, each dopant atom binds an electron leaving a hole in the valence band.
The majority carriers are the positively charged holes and the Si is called \textit{p}-type.
Hole states in the valence band from the $p$-orbitals as opposed to the $s$-orbitals of electrons in the conduction band.
The need to satisfy the 3-fold degeneracy of the $p$-orbital results in the splitting of the valence band into heavy and light hole bands\cite{Luttinger1956}.
The degeneracy of these bands combined with the dopant atom-induced local random stresses broadens the electron paramagnetic resonance (EPR) spectrum making it hard to observe in B-doped Si unless uniaxial strain is applied\cite{Feher1960,Henstra1988,Dirksen1989}.
Strained single crystal Si:B has been used to study the integrated solid effect\cite{Henstra1988}.

Another source of electron spins are defect sites found in amorphous Si\cite{Stutzmann1989}, oxidized Si surfaces\cite{Poindexter1984,Stesmans1987,Brower1989} and elemental Si particles\cite{Cassidy2013,Kwiatkowski2017,Kwiatkowski2018}.
Such defect sites are characterized by a broken Si bond with an unpaired electron mostly localized on the central Si atom\cite{Brower1989}.
When the defect is located at the Si/\ce{SiO2} interface, it is called the $P_b$ center\cite{Poindexter1984,Stesmans1987,Brower1989}.
$P_b$ centers and $P_b$-like centers in amorphous Si have been used to hyperpolarize various Si particles and applied them as background-free contrast agents for MRI\cite{Cassidy2013,Whiting2016}.
The long spin-lattice ($T_\mathrm{1n}$) relaxation times of Si particles around $\sim$40\,min at room temperature offered extended imaging time windows compared to about 30\,s in $^{13}$C molecules\cite{Gallagher2008} or 145\,s (15\,min) in nanodiamonds (microdiamonds)\cite{Kwiatkowski2018b}.
In diamonds, the substitutional nitrogen defects in the particle's bulk (often called C or P1 center) are responsible for DNP while surface dangling bonds commonly cause strong relaxation.
The surface dangling bonds thus are detrimental for nanodiamonds with high surface-to-bulk ratio leading to lower polarization levels and faster relaxation compared to microdiamonds\cite{Boele2020}.
This is different from the case in Si with the $P_b$ centers located on the interface to the naturally forming surface oxide which allows the hyperpolarization of 50\,nm particles with identical enhancements compared to \textmu m-sized particles\cite{Kwiatkowski2018}. 

Despite the demonstrated high nuclear polarization and long nuclear $T_\mathrm{1n}$ relaxation times at room temperature\cite{Atkins2013,Kwiatkowski2018} in bulk Si particles, the proposed slow spin diffusion fails to explain the similar $T_\mathrm{1n}$ in micro- and nanoparticles. 
The diversity of fabrication methods further complicates the identification of the structural properties, their comparison between different particles and influence on $T_\mathrm{1n}$.
In this study, we apply a top-down fabrication approach\cite{Tamarov2020a,Tamarov2020b} to produce porous silicon nanoparticles (PSi NPs, sometimes denoted as nanobeads) with a high surface area from doping controlled, single crystal Si wafers.
The role of the high surface area is twofold.
First, it enables the controlled formation of a relatively large number of endogenous $P_b$ centers to drive the DNP process.
To the best of our knowledge, previous attempts to hyperpolarize PSi NPs required the use of external radicals for DNP to be efficient\cite{Seo2018}, complicating possible MRI applications of those NPs.
Second, the large surface area combines good biocompatibility with a well understood diverse chemistry for (targeted) nanocarrier capabilities\cite{Santos2021} making the developed PSi NPs suitable both for imaging and drug delivery\cite{Santos2021}.
Herein, we prove that endogenous $P_b$ centers in PSi NPs are capable of providing DNP enhancements similar to state-of-the-art bulk particles\cite{Kwiatkowski2018}.
Furthermore, we demonstrate that PSi nanoparticles from slightly doped Si wafers can achieve room temperature hyperpolarization decay times ($\tau_\mathrm{dec}$) exceeding one hour and \textsuperscript{29}Si polarization levels around 6\,\%.

\section{Methods}
\label{sec:Methods}

\subsection{Silicon}

\begin{table*}
\small
  \caption{Summary of Si grade abbreviations used to fabricate PSi NPs.}
  \label{tbl:silicon}
  \begin{tabular*}{\textwidth}{p{0.12\textwidth} *{4}{>{\centering\arraybackslash}p{0.18\textwidth}}}
    \hline
    Abbreviation & Resistivity ($\Upomega$$\cdot$cm) & Dopant & Doping density (cm$^{-3}$)\textsuperscript{\emph{a}} & Average dopant distance (nm)\textsuperscript{\emph{b}} \\
    \hline
    P++\textsuperscript{\emph{c}} & 0.0186 & Boron & $4 \cdot 10^{18}$ & 3.49 \\
    P+\textsuperscript{\emph{c}} & 0.105 & Boron & $3 \cdot 10^{17}$ & 8.27 \\
    P\textsuperscript{\emph{c}} & 25 & Boron & $5 \cdot 10^{14}$ & 69.8 \\
    UW\textsuperscript{\emph{c}} & $>$ 5000 & Boron & $< 10^{12}$ &  $>$ 554 \\
    N\textsuperscript{\emph{c}} & 19.7 & Phosphorus & $2.3 \cdot 10^{14}$ & 90.4 \\
    N+\textsuperscript{\emph{c}} & 1.15 & Phosphorus & $3 \cdot 10^{15}$ & 38.4 \\
    N++\textsuperscript{\emph{c}} & 0.0144 & Phosphorus & $3 \cdot 10^{18}$ & 3.84 \\
    MC10\textsuperscript{\emph{d}} & \multicolumn{4}{p{0.8\textwidth}}{Metallurgical grade powder, polycrystalline, 99.997\% purity. Impurities: Al, Fe, Ca, Ti} \\
    \hline
    \multicolumn{5}{p{\textwidth}}{\textsuperscript{\emph{a}}  Dopant densities were calculated using Caughey-Thomas expression\cite{Klaassen1992} for electron and hole mobilities. Effective Bohr radii are 1.3 (3.8) and 2.1\,nm for heavy (light) holes and electrons in B doped and P doped Si, respectively. The effective Bohr radius of the P electron assumes the pancake-like wavefunction ansatz proposed by Kohn and Luttinger\cite{Koiller2002}.
  \textsuperscript{\emph{b}} Average distance between the dopant atoms was calculated from their density using the random probability distribution in three dimensions\cite{Chandrasekhar1943};
  \textsuperscript{\emph{c}} Powder from single crystal (100) wafers, Okmetic;
  \textsuperscript{\emph{d}} Elkem Silicon Products.}
  \end{tabular*}
\end{table*}

Previous studies on the DNP of Si NPs relied on either commercially available\cite{Atkins2013,Kwiatkowski2017,Kwiatkowski2018,Aptekar2009} or on in-house bottom-up fabrication approaches\cite{Atkins2013,Kim2021,Seo2018,Hu2018}.
In contrast, we selected single crystal Si wafers as the starting material to precisely control crystallinity and doping level (Table~\ref{tbl:silicon}).
Specifically, we used electronics grade single crystal (100) silicon wafers of different doping (Okmetic Oy, Finland).
The samples were denoted according to the doping type and doping density.
Doping type was indicated by P (positive) and N (negative) letters for boron and phosphorus doping, respectively.
The doping density ranged from $4\cdot10^{18}$\,cm$^{-3}$ for P++ and N++ porous Si (PSi) NPs down to less than $10^{12}$\,cm$^{-3}$ for the nominally undoped wafer (UW).
The doping density was below the insulator-to-metal transition for all Si wafers considered.
In addition to wafers, we prepared PSi NPs from a relatively cheaper commercially available polycrystalline (1--10)\,\textmu m Silgrain Supreme MC10 SB powder (Elkem Silicon Products, Norway) with known concentration of impurities (MC10 sample):
The purity of the powder was 99.997\,\% determined by the supplier, where the main impurities were Fe (14\,ppm), Al (6\,ppm), Ca (3\,ppm), Ti (1\,ppm), B ($< 1$\,ppm), and P ($< 1$\,ppm).

Dopant type of the Si wafers was verified by hot point probe method\cite{Sailor2012}. Specific resistivity was calculated using a MATLAB (The MathWorks, Inc., USA) script using wafer thickness and resistivity measured with a four-point probe (Jandel Engineering Ltd, UK) connected to a Cropico DO5000 microhmmeter (Seaward Electronics Ltd, UK)\cite{Sailor2012}.
The dopant concentrations were estimated by comparing the measured specific resistivities with the ones calculated using Caughey-Thomas expression\cite{Klaassen1992} from electron and hole mobilities at 300\,K assuming full ionization of dopant atoms.
The average distances between dopant atoms were calculated from the doping densities using the probability density function to find the atom at a distance $r$\cite{Chandrasekhar1943}.
Assuming the uniform random distribution of the dopant atoms, the average distance is $\langle r \rangle \approx 0.554\cdot N_c^{-1/3} $, where $N_c$ is the density of atoms.
The equivalent Bohr radii for acceptors were estimated using the expression derived from the hydrogen atom-like model of donor (acceptor): $a_{A} = \epsilon_r m_0 / m_{eff} \cdot a_0$, where $a_0$ is the Bohr radius of hydrogen atom, $\epsilon_r$ is the relative dielectric permittivity of Si, $m_0$ is the electron mass, and $m_{eff}$ is the effective mass of a hole.
For the donors, a more precise value of the electron's effective Bohr radius is given by the geometric mean $a_{D} = a_\parallel^{1/3}a_\perp^{2/3} \approx 2.087$\,nm, where $a_\parallel \approx 1.44$\,nm and $a_\perp \approx 2.51$\,nm are the two radii of the pancake-like wavefunction ansatz for the electron ground state proposed by Kohn and Luttinger\cite{Koiller2002}.
The information about Si types and abbreviations used in the text are summarized in Table~\ref{tbl:silicon}.

\subsection{Porous Si powders}

(10--25)\,\textmu m powders were prepared by ball-milling Si wafers using Fritsch Pulverisette 7 Premium Line (Fritsch GmbH, Germany).
Obtained powders were washed in 3\% wt. aqueous \ce{H2O2} by sonicating them for 1\,h in an ultrasound bath\cite{Tamarov2020b}.
Such washing removes possible surface contaminations and ensures reproducibility.
The powders were then dried and used to produce porous Si by low-load metal-assisted catalytic etching (LL-MACE) as reported before\cite{Tamarov2020a,Tamarov2020b}.
The protocol was scaled up to perform etching of 2\,g Si powder batches.
Briefly, 2\,g of Si powder was first dispersed in 30\,ml of acetic acid (Ph. Eur., VWR Chemicals) inside of a 50\,ml PTFE dish by 5\,min sonication.
Then, 30\,ml of hydrogen fluoride  (HF, 30--40\,\%, Merck) was added, and the dish was placed in a water bath on a heat plate under stirring.
Next, Au NPs were nucleated on Si powder surfaces using a syringe pump injection of 8.334\,ml ($=50$\,\textmu mol) of 0.006\,M Au$^{3+}$ ion solution, which was prepared by dissolving \ce{HAuCl4\cdot \text{3} H2O} (99.99\%, Alfa Aesar, Thermo Fisher GmbH) in water.
Injection rate was 333.3\,\textmu l$\cdot$min$^{-1}$; after it was completed, Si powder suspension was stirred for 5\,min more to complete the nucleation of Au$^{3+}$ NPs. 
The temperature of the water bath was kept at about 39\,\textcelsius \ to retain the temperature of the suspension in the range of (51--53)\,\textcelsius \ during etching.
The etching was performed by injecting \ce{H2O2}/\ce{H2O} solution using the syringe pump at a rate of 133.3\,\textmu l$\cdot$min$^{-1}$ (injection time equals to 90\,min).
The \ce{H2O2} volume (35\,wt.\%, Acros Organics, Thermo Fisher GmbH) in the solution was selected to match the \ce{H2O2\text{/}Si} molar ratio of 1.03.
The open end of the plastic tube going from the syringe was immersed into the suspension with Si particles.

After the etching finished, porous Si particles were washed in Büchner-style funnel on a 55\,mm diameter Grade 2 Whatman\textsuperscript{\textregistered} filter.
After the etching solution was removed, porous Si particles were washed three times on the filter using deionized water.
Next, about 10\,ml of n-pentane ($\geq$\,99\%, VWR Chemicals) was poured on the filter with porous powder and it was allowed to dry for a few min under the fume hood.
N-pentane reduced the surface tension of water inside the pores and prevented the collapse of porous structure during the final drying which was completed overnight in an oven at 65\,\textcelsius.
Obtained microscale PSi powders were then stored in glass vials.

\subsection{Surface oxidation and preparation of nanoparticles}

After LL-MACE surfaces of PSi powders were hydrogen terminated (Figs.~\ref{fig:supp-ftir-p-type} and~\ref{fig:supp-ftir-n-type}, Suppl. Inf.).
Localized electronic defects ($P_b$ centers) formed at the \ce{Si}/\ce{SiO2} interface during thermal oxidation of PSi particles.
This approach gives the highest number of $P_b$ defects among other methods\cite{Riikonen2012}.
Thermal oxidation was done in NaberTherm R50/500/12 tube furnace (Nabertherm GmbH) at 345\,\textcelsius \ in air\cite{Riikonen2012}.

Thermally oxidized PSi powders were then milled into NPs using a dedicated system\cite{Nissinen2016b}.
About 400\,mg of a PSi powder was placed into a 4\,ml glass vial which was subsequently filled with 1\,mm \ce{ZrO2} milling balls.
The milling was then performed in 5\,min cycles at 900\,rpm to avoid overheating and leaks. After each cycle, pressure was released from the vials.
Typically, 10 cycles were enough to obtain PSi NPs with most of the particles below 200\,nm in hydrodynamic diameter (Fig.~\ref{fig:characterization}b).

In addition to thermal oxidation, two-step liquid-phase oxidation (oxidation in \ce{H2O2}/ \ce{NH4OH} solution followed by oxidation in \ce{H2O2}/HCl solution)\cite{Riikonen2012}, or one-step (only \ce{H2O2}/HCl solution) was performed for thermally oxidized PSi NPs (\textit{i.e.}, after milling of thermally oxidized PSi powders, details in Suppl. Inf.).
Liquid-phase oxidation removed the remaining hydrogen in $-\text{O}_{\text{y}}\text{SiH}_{\text{x}}$ groups (Figures~\ref{fig:supp-ftir-p-type} and~\ref{fig:supp-ftir-n-type}, Suppl. Inf.) as well as induced additional backbond oxidation. Liquid-phase oxidation was tested because it would be needed in future surface modification with PEG molecules to prolong the systemic circulation time and enabling the use of the PSi NPs e.g. as MR imaging agents\cite{Nissinen2016}.

\subsection{Au removal}
The absence of Au NPs influence on the DNP was confirmed with the N sample.
Au NPs were dissolved using the KI/\ce{I2} gold etchant for the porous Si powder after LL-MACE.
Gold etchant solution was prepared by dissolving 6.08\,g of KI and 1.51\,g of \ce{I2} in 30\,ml of 5\,M HCl. Use of HCl as solvent demonstrated better Au dissolution compared to water.
To dissolve Au NPs, about 250\,mg of N powder after LL-MACE was dispersed in 3\,ml of ethanol to wet the hydrophobic surfaces.
Then, 15\,ml of gold etchant was slowly added to the Si powder suspension. Particles were then stirred for 2\,h at 75\,\textcelsius.
Au amount before and after the dissolution was measured using a home build portable XRF setup\cite{Tiihonen2022} and calculated using the calibration standards prepared with Au deposition step of LL-MACE.
Finally, particles were washed 3 times with water in a Büchner-style funnel, wetted with n-pentane and dried in an oven as above.
Then the powder was milled to NPs and denoted as N-Au.

\subsection{Characterization}

Morphology of PSi NPs was examined by transmission electron microscopy (JEOL JEM-2100F, JEOL Ltd.).
A 2.5\,\textmu l drop of suspension diluted in ethanol to a concentration of 20\,\textmu g$\cdot$ml$^{-1}$ was dried on 400 mesh Cu holey carbon grid (Agar Scientific Ltd.) and the grid was examined in the instrument.
Hydrodynamic sizes of PSi NP were measured using dynamic light scattering (Zetasizer Nano ZS, Malvern Panalytical Ltd) after redispersion in water at 0.1\,mg$\cdot$ml$^{-1}$ concentration. 

Specific surface area, specific pore volume and pore size distributions of PSi powders after LL-MACE were characterized by N$_2$ sorption (Tristar II, Micromeritics Instrument Corp.).
Specific surface areas were calculated from the linear part of adsorption isotherm using Brunauer-Emmett-Teller theory.
Specific pore volumes were obtained from the total adsorbed amount at relative pressure of 0.97.
Pore size distributions were calculated from desorption isotherm using Barrett-Joyner-Halenda model.

Pore sizes and sizes of catalytic Au NPs were further measured with X-ray powder diffraction (XRD, D8 Discover, Bruker Corp.).
Powders were placed on a zero-background holder and scanned in (25--60)$^\circ$ two-theta angle range with step size of $0.0057^\circ$ and time per step of 0.205\,s.
Crystalline sizes of two Si phases and one Au phase were then calculated with Rietvield refinement method using TOPAS\textsuperscript{\textregistered} 4.6 software (Section~\ref{sec:supp-xrpd}, Suppl. Inf.).
The sizes calculated from the Si peak broadenings corresponded to the two pore sizes according to the Babinet's principle in single crystals\cite{Born2019}.

Surface chemical species and $P_b$ centers formed by oxidation were studied with Fourier-transform infrared spectroscopy (FTIR, Thermo Nicolet iS50, ThermoFisher Scientific Corp.) and electron paramagnetic resonance spectroscopy (EPR, Magnettech MiniScope MS5000, Bruker Corp.).
In FTIR, KBr tablets with dried PSi NPs were measured in transmission mode (Suppl. Inf.).
For EPR measurements, the first 7\,mm of an EPR tube were filled with dried PSi NP powder.
The tube was placed in the spectrometer at the same height for each measurement with the following parameters: (1) $B_0=336$\,mT, $B_0^{\text{scan}} = 15.5$\,mT, $B_0^{\text{modulation}}=0.2$\,mT, $t_{\text{scan}}=60$\,s, MW attenuation 25\,dB and gain 10 for the full spectra; (2) $B_0=336$\,mT, $B_0^{\text{scan}} = 35$\,mT, $B_0^{\text{modulation}}=0.7$\,mT, $t_{\text{scan}}=60$\,s averaged 3 times, MW attenuation 15\,dB and gain 500 to resolve hyperfine peaks.
To calculate the concentration of $P_b$ centers and the $g$-factor, a standard 2,2,6,6-tetramethylpiperidinyloxyl (TEMPO) radical (99\%, Sigma-Aldrich) sample with known number of paramagnetic centers and $g$-factors was used.
EPR spectra were fitted using EasySpin 5.2.35 by simulating solid-state continuous-wave powder spectra using a combination of anisotropic $P_b^\mathrm{(111)}$ and isotropic $P_b^\mathrm{iso}$ centers.
$g$-factor strain, hyperfine coupling and Voigtian line broadening were included (Section~\ref{sec:supp-EPR}, Suppl. Inf.)\cite{Konstantinova2018}.

\subsection{Dynamic nuclear polarization}

Hyperpolarization of PSi NPs was studied using three different polarizer designs: SpinAligner (Polarize ApS) operating at 3.35\,T or 6.7\,T and a base temperature of 1.4\,K as well as with two home-built setups with 3.34\, and 7\,T \cite{batel_cross-polarization_2014,jahnig_spin-thermodynamic_2019,Himmler2022} with both operating at a base temperature of 3.4\,K. 
About 100\,mg of dried PSi NP powder was packed into a polymer sample container for measurements with the SpinAligner compared to (50--60)\,mg for the home-built set-ups. 
Microwave radiation was delivered through a waveguide elbow to directly irradiate the sample.
The microwave irradiation\cite{Kwiatkowski2017,Kwiatkowski2018} was frequency modulated in all polarizers.
Magnetic field strength, temperature, microwave power $W$, frequency modulation bandwidth $\Delta \nu_\mathrm{MW}$ and frequency of modulation $\nu_\mathrm{MW}$ are summarized in Tbl.~\ref{tbl:DNP-cond}.

\begin{table}[h]
\small
  \caption{Summary of the DNP conditions.}
  \begin{tabular*}{0.48\textwidth}{*{5}{>{\centering\arraybackslash}p{0.075\textwidth}}}
    \hline
    $B_0$, T & $T$, K & $W$\textsuperscript{\emph{a}}, mW & $\Delta\nu_\mathrm{FM}$\textsuperscript{\emph{b}}, MHz & $\nu_\mathrm{FM}$\textsuperscript{\emph{c}}, kHz \\
    \hline
    3.34 & 3.4 & 200 & $\sim$150 & 1 \\
    3.35 & 1.4 & 80 & 100 & 1 \\
    6.7 & 1.4 & 30 & 200 & 3 \\
    7 & 3.4 & 200\textsuperscript{\emph{d}} & 300 & 10 \\
    \hline
    \multicolumn{5}{p{0.48\textwidth}}{\textsuperscript{\emph{a}} Microwave power; \textsuperscript{\emph{b}} Frequency modulation bandwidth; \textsuperscript{\emph{c}} Frequency of the modulation; \textsuperscript{\emph{d}} Silver-plating the waveguide approximately doubled the MW power reaching the sample\cite{Himmler2022} for the nominal 200\,mW output of the source}
  \end{tabular*}
  \label{tbl:DNP-cond}
\end{table}

To monitor the \textsuperscript{29}Si signal, a flip angle of $\sim 2.8^\circ$ was used in the SpinAligner with varied time intervals between the measurements.
Flip angles of $\sim 1.5^\circ$ each 20\,min at 3.34\,T and $\sim 6.9^\circ$ every 6 to 10\, min at 7\,T were used.
Obtained data was analyzed using MATLAB scripts, where either the real part of the time-domain free induction decay (FID) was fitted with an exponential ansatz or the real part after fast Fourier transform (FFT) in the frequency-domain was fitted with pseudo-Voigt functions.
Polarization enhancements and absolute polarization values were calculated from the thermal equilibrium signal taken in the hyperpolarization conditions after 72\,h of polarization with microwave irradiation switched off for the 6.7\,T (1.4\,K) measurements (Sec.~\ref{sec:supp-thermal}, Suppl. Inf.).
For the 3.34 and 7\,T (3.4\,K) measurements, the thermal equilibrium signal at 300\,K of a fully \textsuperscript{29}Si isotope labeled sample (Isoflex, Russia) was measured and adjusted for temperature upon calculation of enhancements and absolute polarization in the DNP experiments.
Both the polarization buildup data and the relaxation data was corrected for the perturbations by the monitoring RF pulses\cite{Witte2023} (except for the 3.34\,T due to the small flip angle used and difficulties in measuring such small flip angles with high relative accuracy).

\subsection{Correlation analysis}

\begin{figure*}
\centering
  \includegraphics[width=\textwidth]{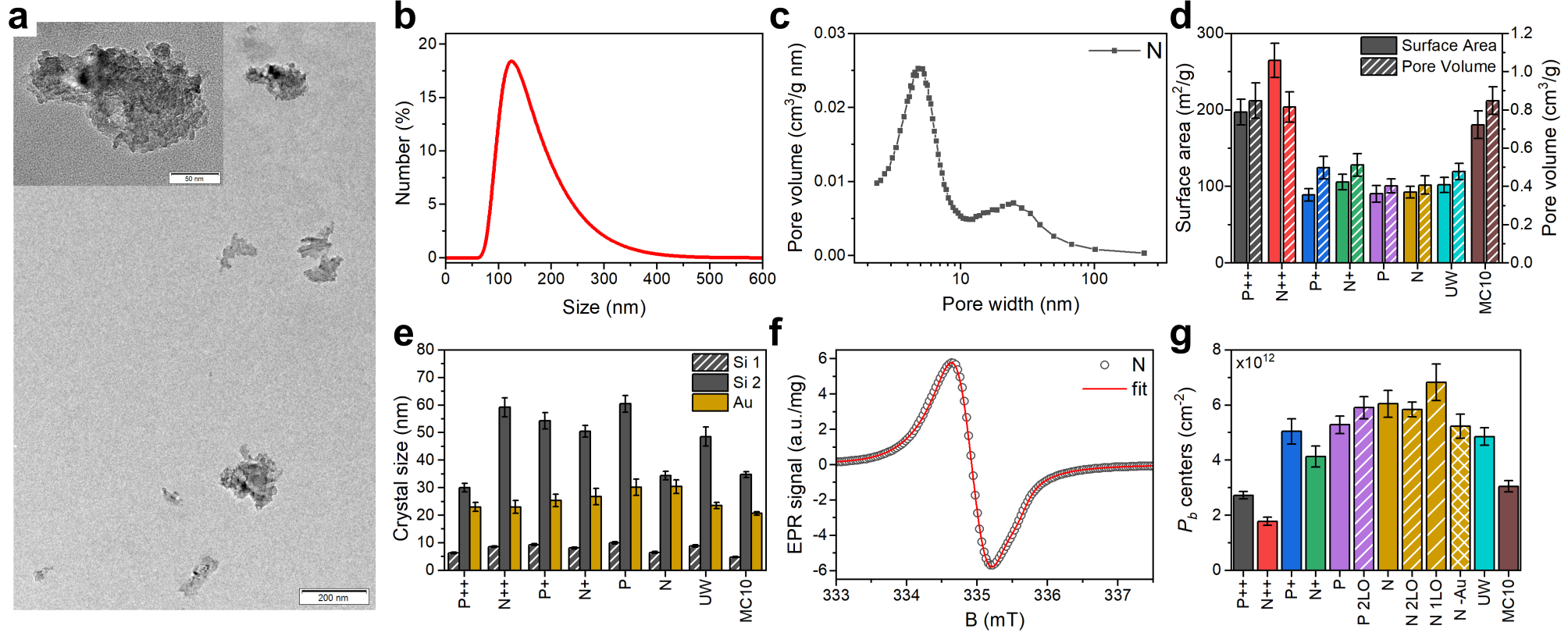}
  \caption{Characterization of PSi NPs. (a) Typical transmission electron microscopy image of PSi NPs dried out of suspension; the inset shows the high magnification view. (b) Hydrodynamic size distribution of the N PSi NPs in water suspension. (c) Pore size distribution of N Si powder after LL-MACE. (d) Specific surface areas and pore volumes obtained from N$_2$ sorption measurements of Si powders after LL-MACE. (e) Crystalline sizes of pore walls in PSi NPs and sizes of Au NPs calculated from X-ray powder diffraction spectra. (f) Electron paramagnetic resonance spectrum of N PSi NPs. The experimental data (black circles) was fitted (red lines) using trigonal $P_b^\text{(111)}$ and isotropic $P_b^\text{iso}$ defects (details see text and Sec.~\ref{sec:supp-EPR}, Suppl. Inf.) (g) $P_b$ defect density of PSi NPs formed by thermal oxidation (no label), thermal and two-step liquid-phase oxidation (2LO label), thermal and one-step liquid-phase oxidation (1LO label), and oxidation induced by Au dissolving solution (-Au label).}
  \label{fig:characterization}
\end{figure*}

Correlation analysis was performed using the Matlab corrcoef() function.
The data supplied to the function consisted of sample characterization data (Fig.~\ref{fig:characterization}), EPR data obtained from fitting (Tbl.~\ref{tbl:supp-epr} and \ref{tbl:supp-easyspin}, Suppl. Inf.), DNP data obtained from fitting and the rate-equation model (Fig.~\ref{fig:exp-comparison} and \ref{fig:supp-rates-6p7T}, Suppl. Inf.).

\section{Results}

The applied fabrication procedure (Experimental section) results in irregular shaped PSi NPs with average particle sizes of $150\pm 65\,\text{nm}$ (Fig. \ref{fig:characterization}a,b).
Additional milling and centrifugal selection could further reduce particle sizes if required for a specific (biological or medical) application (Sec.~\ref{sec:supp-DLS}, Suppl. Inf.).
The porous structure with two distinct pore sizes was formed during the Au-catalyzed LL-MACE (Fig. \ref{fig:characterization}c).
Etch track pores ($>10$\,nm) were produced by Au NPs boring into Si, while tortuous pores ($<10$\,nm) were formed by hole escape from space-charge layers to distant Si surfaces\cite{Tamarov2020a,Tamarov2020b}.
This porosity resulted in a high surface area and a high number of surface $P_b$ centers after oxidation (Fig. \ref{fig:characterization}d,g).
X-ray powder diffraction (Fig.~\ref{fig:characterization}e and Sec.~\ref{fig:supp-xrpd-topas}, Suppl. Inf.) showed Si peaks with distinct superimposed peak profiles.
Typically, the peak broadening of small crystals is dependent on the crystallite size but the porous nature of the PSi NPs complicates the picture.
Since the Si particles are single crystals before etching (except for the MC10 sample) and preserve the crystallinity during the etching, the pores also give a contribution to the peak broadening according to the Babinet’s principlecite\cite{Born2019}.
Therefore, three distinct contributions to the peak broadening would be expected for the PSi NPs caused by the crystallite size, population of wide etch track pores and the population of narrow tortuous pores.
It was not possible to reliably fit the data with three peak profiles but instead fit with two profiles was done.
The narrower peak profiles ((30--60)\,nm bars) were attributed to the etch track pores and the broadening from the small crystal size, while the wider ones ((5--10)\,nm bars) were due to tortuous pores penetrating the large crystals (Fig.~\ref{fig:characterization}e)\cite{Tamarov2020a,Tamarov2020b}.
The wide and narrow XRD peaks were on the order of the corresponding pore sizes measured by \ce{N2} sorption and depicted in Fig.~\ref{fig:characterization}c. 

The subsequent thermal oxidation (Methods and Sec.~\ref{sec:supp-FTIR}, Suppl. Inf.) stabilized the H-terminated surface of freshly etched samples simultaneously making them hydrophilic.
The created core-shell structure of PSi NPs thus consisted of the crystalline cores of pore walls (bulk) with a thin oxide shell (surface).
The lattice constant mismatch between the Si and \ce{SiO2} led to the formation of paramagnetic centers in the Si/\ce{SiO2} interface.

Electron paramagnetic resonance (EPR) spectra (Fig.~\ref{fig:characterization}f, Discussion of $P_b$ centers below and Sec.~\ref{sec:supp-EPR}, Suppl. Inf.) showed the presence of two typical paramagnetic centers found on oxidized (porous) Si surfaces:
(i) trigonal $P_b^\text{(111)}$ centers with axial symmetry similar to defects found in oxidized planar (111) and porous Si surfaces ($g_\parallel = 2.00185\,g_\perp = 2.0081$)\cite{Brower1989,VonBardeleben1996,Stesmans1998} and (ii) isotropic $P_b^\text{iso}$ defects commonly observed in oxidized porous Si ($g=2.0055$)\cite{Poindexter1983,Rong1993a,VonBardeleben1993a,VonBardeleben1993b,Laiho1994,VonBardeleben1996}.
EasySpin\cite{Stoll2006} was used to simulate the experimental EPR spectra to obtain the relative weights of the $P_b^\text{(111)}$ and $P_b^\text{iso}$ centers in our samples (Section~\ref{sec:supp-EPR}, Suppl. Inf.).
The simulations gave typical weights of $(10\textrm{--}20)$\,\% for the $P_b^\text{(111)}$ and $(80\textrm{--}90)$\,\% for the $P_b^\text{iso}$ defects.
It was expected that $P_b^\text{iso}$ is the dominant defect center due to the random nature of pore formation in LL-MACE and thermal oxidation in air.
Hyperfine (HF) interaction with the central \textsuperscript{29}Si was also observed (Sec.~\ref{sec:supp-EPR}, Suppl. Inf.) and measured to be in the range of $A = (325\textrm{--}431)$\,MHz, which coincided well with $A_\parallel = 210$\,MHz and $A_\perp = 417$\,MHz for the planar $P_b^\text{(111)}$ center\cite{Stesmans1998}.
The number of all types of $P_b$ centers per unit area and per mass varied between $(1.8\textrm{--}6.8)\cdot 10^{12}$\,cm$^{-2}$ (Fig.~\ref{fig:characterization}g) and $(4.4\textrm{--}6.3)\cdot 10^{15}$\,mg$^{-1}$, respectively (Sec.~\ref{sec:supp-EPR}, Suppl. Inf.).
These values corresponded to the fraction of total $P_b$ centers per silicon interface atoms of $f\equiv [P_b]/N_a = (0.23\textrm{--}0.87)$\,\% (where $N_a=7.83\cdot 10^{14}$\,cm$^{-2}$ is the density of lattice sites in the (111) plane).
The average distance between the $P_b$ centers was calculated from the concentration per unit area using the nearest neighbors distribution\cite{Chandrasekhar1943} derived for the 2D case.
The average distances varied between 1.9\,nm (N 1LO PSi NPs) and 3.7\,nm (N++ PSi NPs).
Correspondingly, the maximum dipolar interaction between electron spins of $P_b$ centers ranges from 1.0 to 7.4\,MHz if a uniform surface distribution of $P_b$ centers is assumed.

\begin{figure*}
\centering
  \includegraphics[width=\textwidth]{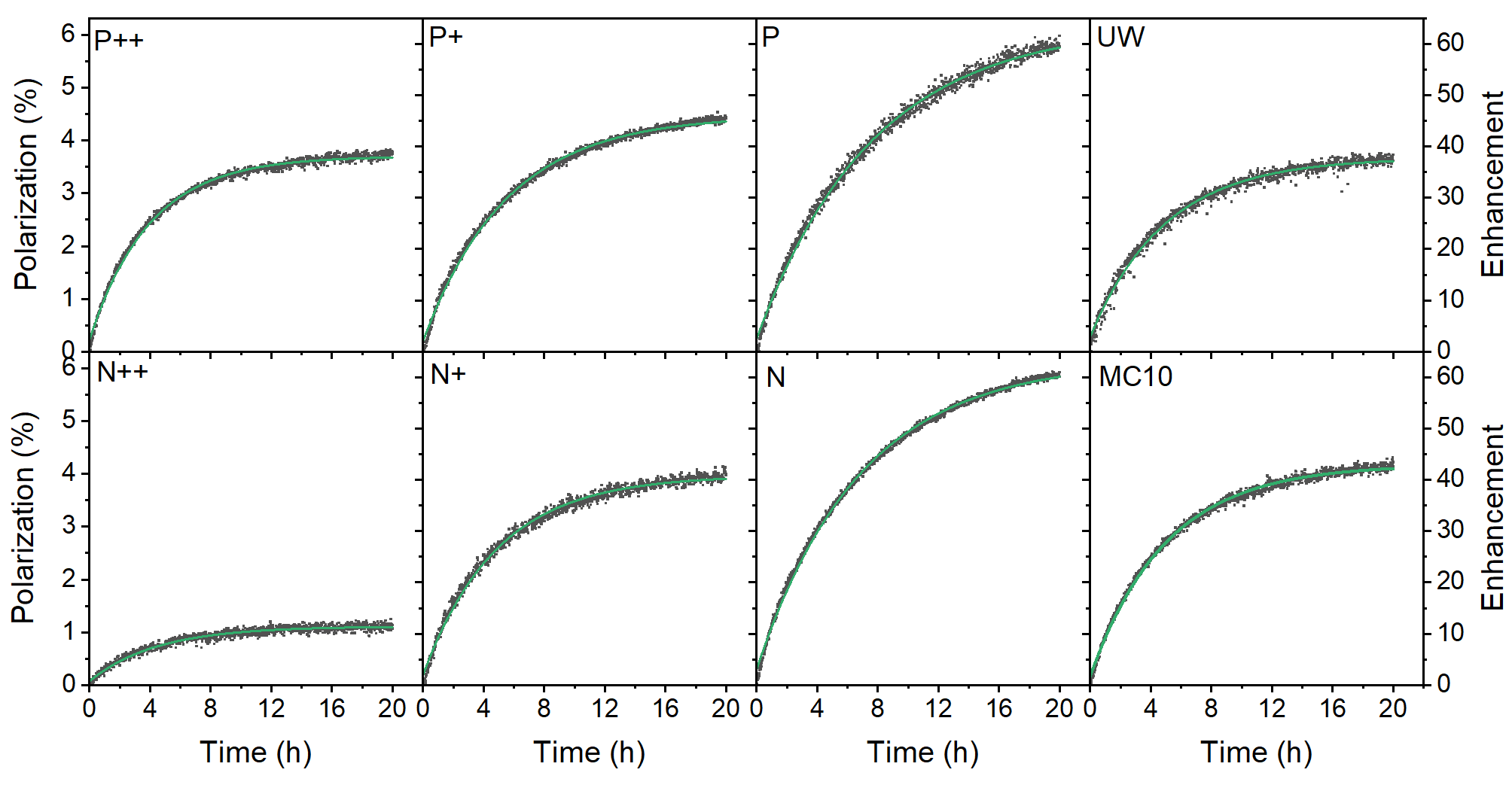}
  \caption{Dynamic nuclear polarization of thermally oxidized PSi NPs with different dopants after correcting for perturbation by the RF pulses\cite{Witte2023} (dark squares) and single exponential fit (green lines) at 6.7\,T and 1.4\,K. The microwave frequency was set to 187.82\,GHz with a 150\,MHz bandwidth, 3\,kHz modulation and 30\,mW microwave output power. The enhancement is relative to the thermal polarization of \textsuperscript{29}Si nuclear at the polarization buildup conditions. For a characterization of the various samples, see Table~\ref{tbl:silicon}.}
  \label{fig:bup-grid-6p7T}
\end{figure*}

We performed DNP NMR hyperpolarization and relaxation studies at four different conditions (3.34\,T and 7\,T at 3.4\,K, 3.35\,T and 6.7\,T at 1.4\,K, Tbl.~\ref{tbl:DNP-cond}) with only selected samples evaluated at all the experimental conditions.
The measured DNP profiles followed the symmetry of the EPR spectrum with the positive and negative DNP lobes located at a similar distance to the central zero crossing of the DNP enhancement (Sec.~\ref{sec:supp-DNP}, Suppl. Inf.).
The zero crossing of the DNP enhancement coincided with the center of the EPR line in agreement with previous works with endogenous defects in Si\cite{Atkins2013,Kwiatkowski2018}. 

\textsuperscript{29}Si polarization buildup data at 6.7\,T (1.4\,K) for the thermally oxidized PSi NPs with various dopants are depicted in Fig.~\ref{fig:bup-grid-6p7T}.
The data was corrected for the perturbations by the monitoring RF pulses\cite{Witte2023}.
We confirmed that the algorithm correctly recovered the genuine buildup dynamics from high sampling rate data in Fig.~\ref{fig:bup-grid-6p7T} using a low sampling rate of 30\,min for the P sample (Fig.~\ref{fig:supp-one-compartment}, Suppl. Inf.).
The one-compartment model underlying the RF correction assumes a mono-exponential buildup and decay dynamics \cite{Witte2023} as observed in all our samples and experimental conditions (Sec.~\ref{sec:supp-DNP}, Suppl. Inf.).

\begin{figure}[!htb]
\centering
  \includegraphics[width=0.8\linewidth]{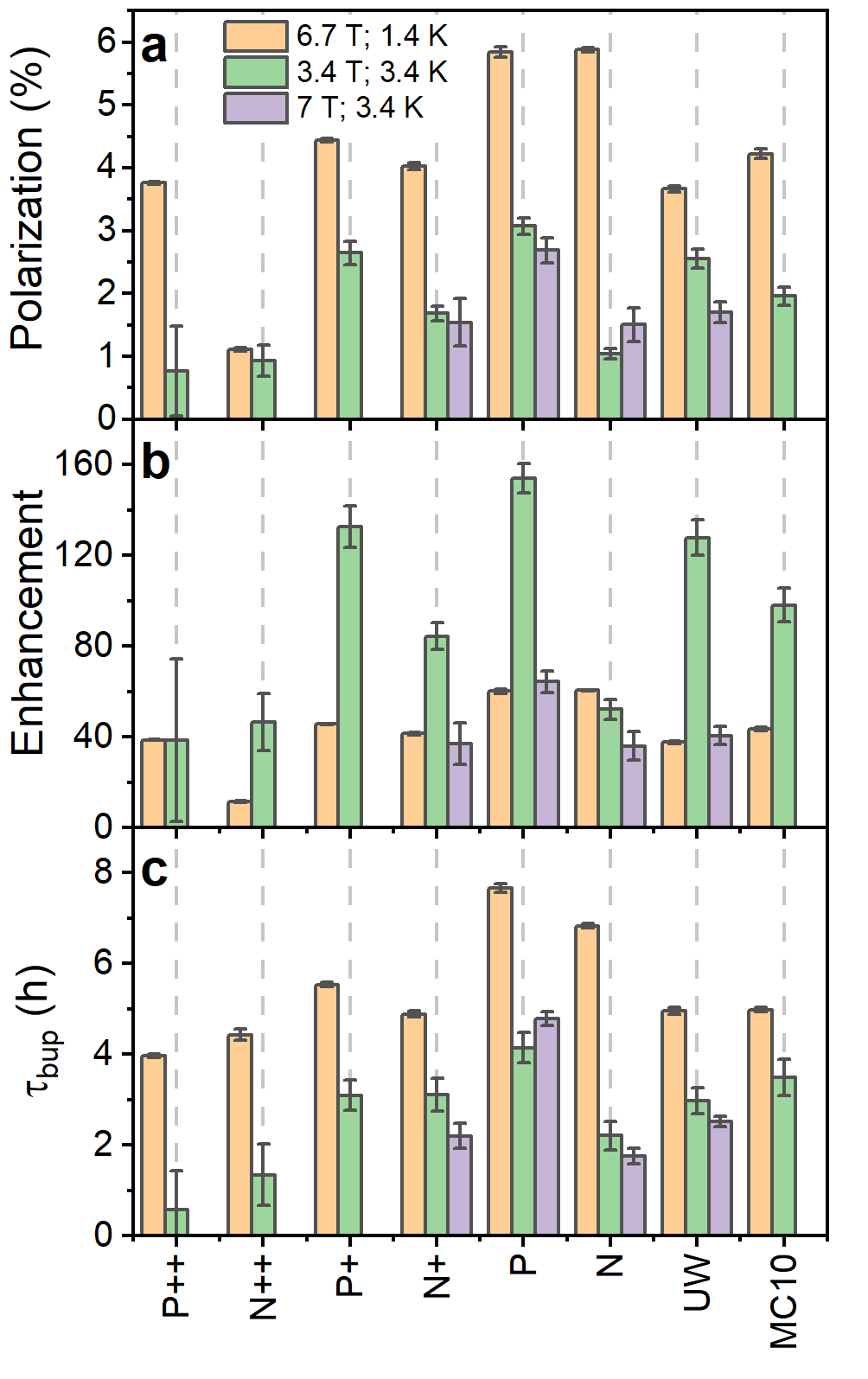}
  \caption{Comparison of \textsuperscript{29}Si nuclear polarization (a), the enhancement over the thermal signal (b) and polarization buildup time (c) for PSi NPs at 6.7\,T (1.4\,K) (orange bars) as well as 3.34\,T (3.4\,K) (green bars) and 7\,T (3.4\,K) (violet bars).
  Temperature decrease or increase of magnetic field strength increase the thermal nuclear polarization used to calculate the enhancement from the nuclear polarization. The polarization, enhancement and buildup time are corrected for perturbations by the RF pulses\cite{Witte2023}. MW frequency modulation was employed in all the experiments.}
  \label{fig:exp-comparison}
\end{figure}

The polarization buildup (at 6.7\,T and 1.4\,K) depended on the doping degree.
The lowest polarization was found for the highly doped P++ and N++ samples but with significant difference between them despite the similar doping level of the starting Si powder (Tbl.~\ref{tbl:silicon}).
With the decrease of doping density, the gained polarization levels tended to equalize between different doping types (P+ and N+ PSi NPs).
Interestingly, the nominally undoped UW PSi NPs did not show the highest absolute \textsuperscript{29}Si polarization; the highest polarization levels were obtained for lightly doped P and N samples.
Moreover, the relatively impure polycrystalline MC10 PSi NPs showed slightly better DNP polarization and similar buildup times than moderately doped P+ and N+ samples.
Such polycrystalline grades could thus be a cheaper alternative to electronics grade sample with sufficiently good DNP properties.

The DNP characteristics changed significantly at 3.34\,T and 3.4\,K (Fig.~\ref{fig:exp-comparison} and \ref{fig:supp-bup-3p4T}, Suppl. Inf.).
The polarization buildup times (Fig.~\ref{fig:exp-comparison}c) for all the samples almost halved compared to 6.7\,T (1.4\,K).
The observed enhancements  (Fig.~\ref{fig:exp-comparison}b) were significantly higher at 3.34\,T especially for the low B doped PSi NPs compared to the 6.7\,T data.
The $n$-type samples demonstrated only moderate enhancement increases with the N sample showing even lower enhancement than at 6.7\,T.
Despite the higher enhancements at 3.34\,T, the estimated absolute \textsuperscript{29}Si polarization was still higher at 6.7\,T (1.4\,K) compared to 3.34\,T (3.4\,K) (Figs.~\ref{fig:bup-grid-6p7T}, Fig.~\ref{fig:exp-comparison}a and \ref{fig:supp-bup-3p4T}, Suppl. Inf.) due to the higher thermal nuclear polarization.

In order to clarify the influence of the experimental conditions on DNP, we performed selected measurements at 7\,T (3.4\,K) to discriminate between field and temperature dependent changes (Fig.~\ref{fig:exp-comparison} and \ref{fig:supp-bup-decay-7T}, Suppl. Inf.).
The decreased polarization for the N PSi NPs clearly followed the same trend as at 3.34\,T while the absolute enhancement values and buildup times for P and UW samples were close to the 6.7\,T data.
The similarities for P and UW samples were even more striking provided the MW power was 30\,mW at 6.7\,T compared to 200\,mW at 7\,T.
We then verified at 7\,T (3.4\,K) that 200\,mW and 20\,mW provided similar enhancements at 7\,T making the comparison between 6.7\,T and 7\,T possible despite the large difference in MW power (Fig.~\ref{fig:supp-power-7T}, Suppl. Inf.).
We, therefore, conclude that temperature plays the crucial role in DNP performance of $n$-type PSi NPs, while it has less influence on the $p$-type samples.
The temperature dependence for $p$-type samples was further investigated at 3.35\,T (1.4\,K) (Fig.~\ref{fig:supp-bup-dec-3p35T}, Suppl. Inf.).
We found a significant decrease of enhancement levels compared to the other conditions with minor differences between P and P++ PSi NPs.

\begin{figure}[h]
\centering
  \includegraphics[width=0.8\linewidth]{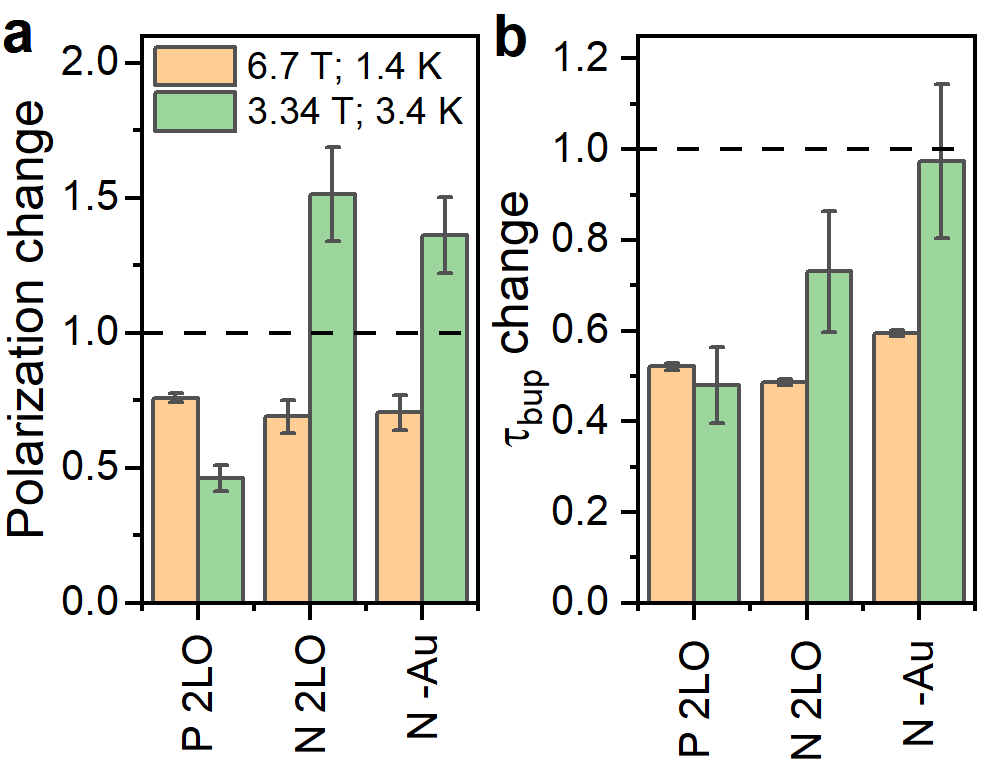}
  \caption{Relative change of the \textsuperscript{29}Si steady-state polarization (enhancement) (a) and polarization build up time (b) due to oxidations for P and N samples. The 2LO oxidation indicates the two-step liquid oxidation (Sec.~\ref{sec:supp-LO}, Suppl. Inf.) performed after the thermal oxidation. For the N-Au, oxidation  emerged during the Au removal after LL-MACE (Experimental section). The dashed line indicates no change  \textit{i.e.}, the same measured value compared to thermally oxidized samples. The absolute values are in Fig.~\ref{fig:supp-enh-bup-oxy}, Suppl. Inf.}
  \label{fig:enh-bup-oxy}
\end{figure}

In addition to the thermal oxidation used to create $P_b$ centers on differently doped PSi NPs, we applied liquid oxidation\cite{Riikonen2012} to the P and N PSi NPs.
Liquid oxidation reduced the number of surface hydrogen in -Si$_\text{y}$H$_\text{x}$-Si-H and -O$_3$SiH surface groups (Section~\ref{sec:supp-FTIR}, Suppl. Inf.), which is an important step towards an improved surface coating for biomedical applications\cite{Nissinen2016}.
We note that liquid oxidation affected the $p$- and $n$-type Si samples differently (Section~\ref{sec:supp-FTIR}, Suppl. Inf.).
The same is true for the measurements with different DNP conditions (Fig.~\ref{fig:enh-bup-oxy} and Figs.~\ref{fig:supp-bup-3p4T}, \ref{fig:supp-bup-oxy-6p7T}, \ref{fig:supp-enh-bup-oxy}, Suppl. Inf.):
For the P sample, enhancement dropped significantly at 6.7\,T (1.4\,K) and 3.34\,T (3.4\,K).
Contrary to the P sample, liquid oxidation of the N sample increased the enhancement about 1.4 times at 3.34\,T (3.4\,K), while at 6.7\,T (1.4\,K) the enhancement decreased.
The polarization build up times were affected in a more consistent manner (Fig.~\ref{fig:enh-bup-oxy}b):
For all the samples and liquid oxidations, the buildup times shortened to $(0.5\textrm{--}0.7)$ times the buildup time of the thermally oxidized N or P PSi NPs.
Future studies might explore the influence of oxidation, doping, and DNP conditions on the DNP via $P_b$ centers further.

Finally, we verified that the presence of Au NPs left in PSi NPs after LL-MACE had little impact on DNP performance.
For verification, we applied an iodine-based Au etchant to the N PSi NPs directly after LL-MACE (no thermal oxidation).
The Au dissolution resulted in decrease of Au content from 0.37\% for N PSi NPs to 0.02\% for N-Au PSi NPs as measured by XRF. 
Since the Au etchant is a strong oxidative solution, the dissolution process also oxidized the PSi NP surfaces which are hydrogen terminated and hydrophobic after LL-MACE.
For N PSi NPs, the etchant-induced oxidation had similar effects as liquid oxidation  (Fig.~\ref{fig:enh-bup-oxy}).

After collecting the DNP data for the various samples at 3.34\,T and 6.7\,T, we selected the P, UW and N samples for room temperature relaxation measurements (Fig.~\ref{fig:supp-decay-room-t}, Suppl. Inf.).
For this, the samples were hyperpolarized at 3.34\,T (3.4\,K) for around 20\,h and subsequently transferred (dry, tightly packed sample container) to the nearby temperature-controlled (300\,K) 7\,T setup. 
At room temperature, the differences between the decay times $\tau_\mathrm{dec}$ of the selected samples diminished compared to liquid helium temperatures (Table~\ref{tbl:supp-decay-t}, Suppl. Inf.).
Nevertheless, a smaller $\tau_\mathrm{dec}$ for the N sample compared to the P and UW samples was observed.
The hyperpolarized decay times at room temperatures of the P and UW samples were around 70\,min.

\begin{table}[h]
\small
  \caption{Relaxation time of the selected PSi NPs at 7\,T and room temperature after DNP at 3.34\,T and 3.4\,K.}
  \begin{tabular*}{0.48\textwidth}{*{2}{>{\centering\arraybackslash}p{0.22\textwidth}}}
    \hline
    Abbreviation & $\tau_\mathrm{dec}$, min \\
    \hline
    P & $75\pm 3$ \\
    UW & $67\pm 8$ \\
    N & $52\pm 5$ \\
    \hline
  \end{tabular*}
  \label{tbl:decay-room-t}
\end{table}

\section{Discussion}

\begin{figure}[h]
    \centering
    \includegraphics[width=0.8\linewidth]{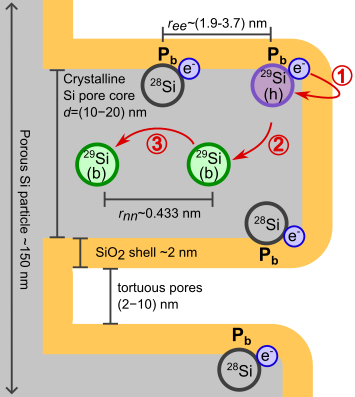}
    \caption{Sketch of the PSi NPs with $\sim 150$\,nm particle size and a large number of torturous pores (not to scale).
    $P_b$ centers form at the interface between the surface oxide shell and the crystalline pore wall cores.
    Average electron-electron ($r_\mathrm{ee}$) and (\textsuperscript{29}Si) nuclear-nuclear ($r_\mathrm{nn}$) distances for 4.7\,\% natural abundance \textsuperscript{29}Si are indicated.
    The hyperpolarization pathway is indicated in red. 
    The polarization is transferred from the electron to the hypershifted \textsuperscript{29}Si (h) nucleus of a $P_b$ center (step 1) and from there to a nearby bulk (b), NMR visible \textsuperscript{29}Si spin (step 2).
    Within the crystalline pore wall core (step 3), the nuclear hyperpolarization is spread via nuclear spin diffusion.
    Only 4.7\,\% of $P_b$ centers have \textsuperscript{29}Si nucleus and, therefore, directly participate in DNP.
    }
    \label{fig:PSi_sketch}
\end{figure}

The following discussion is organised along Fig.~\ref{fig:PSi_sketch}, which sketches the relevant length scales and the proposed polarization pathway in the PSi NPs.
The $P_b$ centers at the interface between the surface oxide shell and the crystalline silicon core provide the unbound electrons required for DNP.
Thus, understanding DNP in PSi NPs requires a basic understanding of $P_b$ centers, which will be provided in Sec.~\ref{sec:discussion_Pb}.

To achieve a hyperpolarized nuclear state, the high thermal electron polarization is transferred via hyperfine (HF) coupling to \textsuperscript{29}Si nuclei of a $P_b$ center located on the interface between the bulk pore walls and oxide shell  (step 1 of hyperpolarization buildup sketched in Fig.~\ref{fig:PSi_sketch}).
The HF coupling shifts the resonance frequency of the $P_b$ nuclear spins rendering it difficult to observe these spins with NMR (hypershifted spins\cite{Pang2024}).
The hypershifted spins have a a resonance frequency (energy) discrepancy to the bulk nuclear spins in the pore walls (visible by NMR).
The frequency discrepancy suppresses the nuclear spin diffusion between the hypershifted and bulk spins (step 2 in Fig.~\ref{fig:PSi_sketch}) making the step to be the time limiting as further argued below in Sec.~\ref{sec:discussion_bup}.
Between the bulk \textsuperscript{29}Si in the pore wall cores, nuclear spin diffusion (nSD) spreads the nuclear hyperpolarization throughout the crystalline pore wall cores (step 3 in Fig.~\ref{fig:PSi_sketch}).
The discussion of the different steps is then extended to the hyperpolarization decay in Sec.~\ref{sec:discussion_decay}.
Finally, the effects of different samples and experimental conditions are discussed in sec.~\ref{sec:discussion_samples}

\subsection{$P_b$ centers} \label{sec:discussion_Pb}

In Sec.~\ref{sec:supp-Pb}, Suppl. Inf., we concisely review existing literature on the interfacial $P_b$ centers in Si/\ce{SiO2}.
Based on this review, the measured EPR spectra are fitted with two types of $P_b$ centers: (i) a $P_b^\text{(111)}$ center with trigonal symmetry and (ii) $P_b^\mathrm{iso}$ with spatially isotropic $g$-factor.
If a $P_b$ center has a \textsuperscript{29}Si nucleus at its central site, it possesses a Fermi-contact (isotropic) hyperfine coupling of several hundred MHz ($A_\parallel = 210$\,MHz; $A_\perp = 417$\,MHz) to the unbound electron owing to the localized $P_b$ electron's wavefunction. 
The next nearest neighboring \textsuperscript{29}Si atoms have hyperfine (HF) coupling of a few tens of MHz.
Si nuclei further away from the $P_b$ center are assumed to be mostly coupled via dipolar HF interactions to the electron and to other Si nuclei.
In naturally abundant Si with 4.7\,\% \textsuperscript{29}Si, only a fraction of $P_b$ centers has a HF coupling exceeding tens of MHz and only 4.7\,\% will have HF couplings exceeding hundreds of MHz.

The HF couplings of hundreds of MHz split the EPR line into three lines: a strong central EPR line of $P_b$ centers with non-magnetic Si nuclei and weak HF doublet of the 4.7\,\% of $P_b$ centers with a \textsuperscript{29}Si nucleus at its central site.
In our experiments, the doublet outer lines are shifted by roughly $\pm A_\mathrm{ave}/2 = \pm (A_\perp + A_\parallel)/4 \approx \pm 162$\,MHz with respect to the central electron line in a good agreement with the literature (Sec.~\ref{sec:supp-Pb}).
Each of the three lines is anisotropically broadened due to $g$-factor strain in the irregular Si/\ce{SiO2} interface, which leads to a full electron line consisting of three connected EPR lines ($m_I = {-1/2, 0, 1/2}$).
Taken together, the anisotropic line broadening provided by the HF interaction and the $g$-factor strain (together $>300$\,MHz) is larger than the nuclear Larmor frequency $\omega_\mathrm{0n}$ (between 28 and 60\,MHz).

From the fitted $P_b$ signal (Fig.~\ref{fig:characterization}d, Tab.~\ref{tbl:supp-epr}, Suppl. Inf.) and the measured surface area (Fig.~\ref{fig:characterization}d), the estimated average distance between the $P_b$ centers assuming their uniform distribution is $r_\mathrm{ee} = (1.9\textrm{--}3.7)$\,nm.
This distance gives the estimated electron dipolar coupling $D_\mathrm{ee}$ on the order of $D_\mathrm{ee}=(1.0\textrm{--}7.4)$\,MHz.
The electron dipolar coupling is about $(1-10)$ times lower than the homogeneous line broadening (Tbl.~\ref{tbl:supp-easyspin}, Suppl. Inf.) which might indicate a non-uniform distribution and with that larger electronic couplings.
The estimated electron dipolar coupling values $D_\mathrm{ee}$ are strong enough to induce mutual electron spin flip-flops within the EPR line\cite{Kundu2019}.



The summarized EPR data satisfies the three main conditions for the triple spin (2 electron spins and 1 nuclear spin) family of DNP mechanisms.
First, the dipolar interaction is strong enough within the EPR line to result in electron-electron flip-flops.
Second, the EPR line is broader than the nuclear Larmor frequency at all the experimental conditions.
Third, part of the electron spins in $P_b$ centers are HF coupled to \textsuperscript{29}Si nuclei.
Following the ongoing theoretical efforts to understand triple spin flips in DNP\cite{Kundu2019,Wenckebach2019a,Wenckebach2019b,Wenckebach2021,Wenckebach2021a,Wenckebach2021b,Wenckebach2023}, we refrain ourselves from going into the specific variants, such as cross effect or thermal mixing DNP.
We highlight that $D_\mathrm{ee}$ values in our samples support cross effect DNP according to recent quantum mechanical simulations\cite{Kundu2019}.
Finally, we also note the results from previous study of nominally undoped Si microparticles, in which the decay of nuclear hyperpolarization was explained through triple spin flips\cite{Lee2011}, emphasizing the importance of triple spin flips in the Si/\ce{SiO2} interface.

\subsection{Rate limiting step for the polarization buildup} \label{sec:discussion_bup}

To achieve the polarization levels up to a few percent, the polarization needs to penetrate from the surface nuclei into the pore wall cores of the PSi NPs for which we invoke the concept of nuclear spin diffusion (nSD)\cite{Bloembergen1949,Wenckebach2016}.
The dipolar interaction between nuclei induces nuclear spin flip-flops --- a zero-quantum (ZQ) process with no net change of the total magnetic quantum number.
This ZQ process causes an effective spatial transport of magnetization that can be described by a diffusion equation if a nuclear polarization gradient is present in the sample.

The nSD constant in Si was previously approximated with $D_\mathrm{diff} \approx a^2 / (c T_\mathrm{2n})$, where $c = 30$\cite{Hayashi2009} or $c = 50$\cite{Dementyev2008}, $a$ is the average distance between \textsuperscript{29}Si nuclei in a cubic lattice and $T_\mathrm{2n}$ is a measure for the inhomogeneous SQ line width in the spectra. 
In the approximation of $D_\mathrm{diff}$ in Ref.~\cite{Dementyev2008}, it was implicitly assumed that the experimentally measured single quantum (SQ) Hahn echo decay characterized by $T_\mathrm{2n}^\prime \approx 5.6$\,ms\cite{dementyev_anomalies_2003} characterizes also the width of the ZQ line, which is relevant for nSD.
In Ref.~\cite{Hayashi2009}, the decay constant of the FID ($T_\mathrm{2n}^*$) was assumed to characterize the width of the ZQ line \cite{Hayashi2009}.
We note that all of these decay-time constants are not relaxation times in the strict sense of stochastic processes that lead to decoherence.
Nonetheless, both approaches lead to similar nSD coefficients of $D_\mathrm{diff}\approx 0.5 - 1.7$\,nm$^2$s$^{-1}$.
Therefore, for the polarization to diffuse from the surface into the pore wall's cores $r_\mathrm{wall}/2$, a time scale of only $\sim 8$\,s or $\sim 140$\,s ($\tau_\mathrm{diff} = \langle (r_\mathrm{wall}/2)^2 \rangle / 6D_\mathrm{diff})$ is required for the tortuous or etch track pores, respectively (Fig.~\ref{fig:characterization}c,e). 
These time scales are orders of magnitude shorter than the liquid helium build-up times of hours ( Fig.~\ref{fig:bup-grid-6p7T} and Sec.~\ref{sec:supp-DNP}, Suppl. Inf.)) or the room temperature decay times of around one hour (Tab.~\ref{tbl:decay-room-t} and Fig.~\ref{fig:supp-decay-room-t}, Suppl. Inf.).
Hence, we conclude that the nuclear spin diffusion (step 3 of the hyperpolarization buildup in Fig.~\ref{fig:PSi_sketch}) is not limiting the hyperpolarization process.

\begin{figure}[!htb]
\centering
    \includegraphics[width=0.8\linewidth]{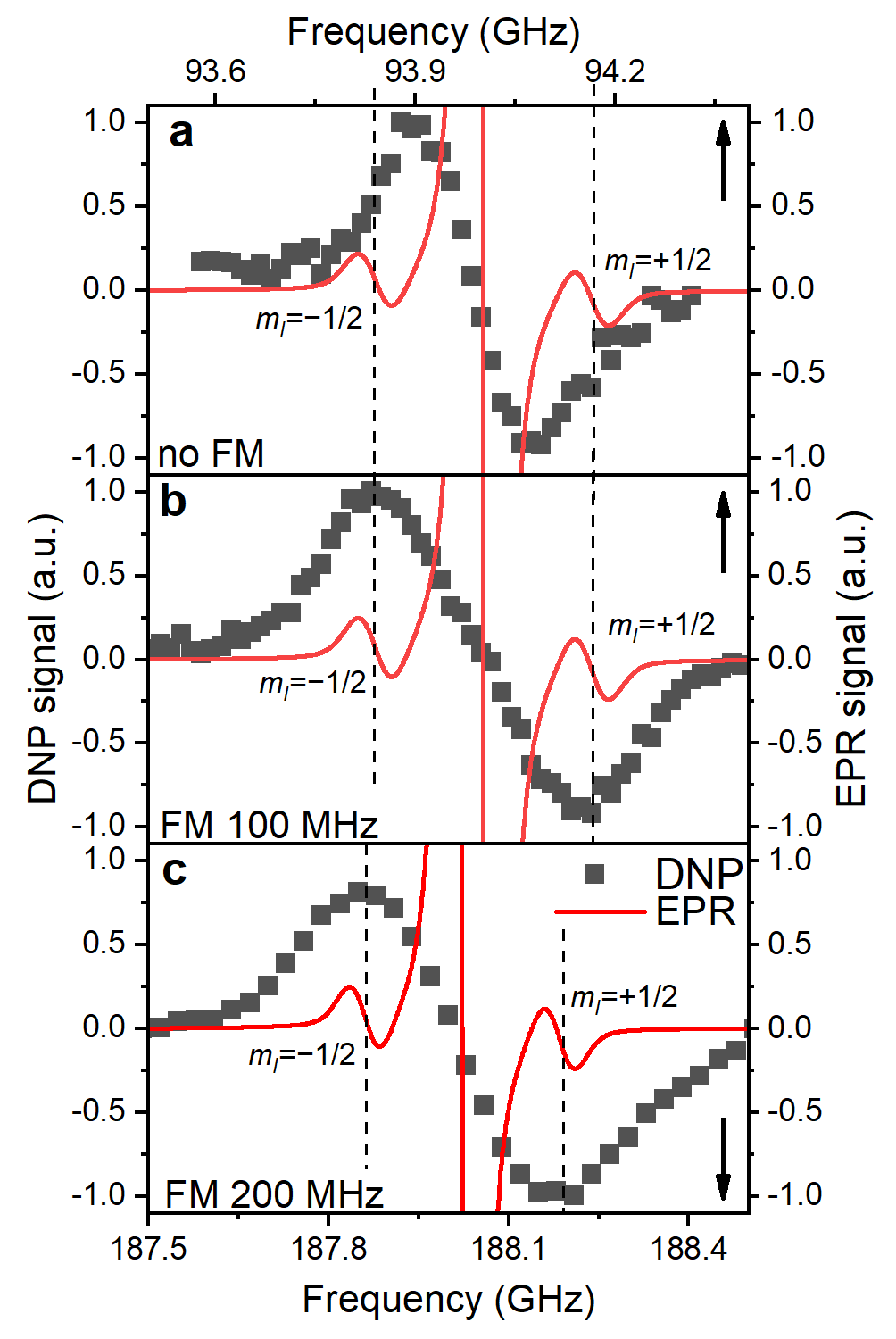}
    \caption{Overlay of the simulated EPR and experimental DNP spectra for the P sample at (a) 3.35\,T, 1.4\,K without MW modulation (frequency modulation - FM), (b) 3.35\,T, 1.4\,K with 100\,MHz MW modulation and (c) 6.7\,T, 1.4\,K with 200\,MHz MW modulation.
    The DNP and EPR spectra are normalized separately in each panel.
    EPR spectrum is the frequency-swept spectrum simulated using the model obtained from the experimental data fitting (Fig.~\ref{fig:characterization}f, Section~\ref{sec:supp-EPR}, Suppl. Inf.).
    EPR spectrum consists of the central \textsuperscript{28}Si manifold (clipped) and two hyperfine-split manifolds for $P_b$ centers with \textsuperscript{29}Si nuclei ($m\mathrm{I} = \pm 1/2$, dashed lines).
    The upwards and downwards arrows indicate the X axis for each graph.}
    \label{fig:EPR-DNP}
\end{figure}

The EPR spectrum extrapolated to the DNP field strength of 3.34\,T or 6.7\,T consists of three lines (Fig.~\ref{fig:EPR-DNP}): the central line for $P_b$ at $^{28}$Si nuclei is surrounded by the two HF-split lines for $m_I = \pm 1/2$ whose shape is the same as of the central line.
The DNP profiles show two DNP peaks of positive and negative enhancements with nearly equal amplitude and width (Fig.~\ref{fig:EPR-DNP} and  \ref{fig:supp-sweep-type-6p7T}, Suppl. Inf.).
If MW modulation is applied, the extrema of the DNP enhancement in our samples coincide with the frequencies of the HF-split $m_I = \pm 1/2$ doublet in the EPR spectrum (Fig.~\ref{fig:EPR-DNP}b,c).
Switching off MW modulation (Fig.~\ref{fig:EPR-DNP}a) narrows the DNP profile while retaining its featureless shape with its width far exceeding the nuclear Larmor frequency ($\omega_\mathrm{0n}(3.34\,\mathrm{T})\approx28$\,MHz).

In DNP, the strength of the HF interaction between electron and nuclear spins determines the polarization transfer rate constant, which is proportional to the square of the HF coupling matrix element. 
Owing to the large HF constant between the $P_b$ electron and the central Si atom ($A_\parallel = 210$\,MHz; $A_\perp = 417$\,MHz), the DNP of these nuclei should be efficient and fast (step 1 of the hyperpolarization buildup in Fig.~\ref{fig:PSi_sketch}).
Already for nearest neighbors ($A_\mathrm{2n} \approx 42$\,MHz\cite{Brower1989}) \textsuperscript{29}Si nuclei the roughly ten times lower HF coupling would lead to an approximately hundred-fold lower polarization transfer rate compared to the central \textsuperscript{29}Si, which outweighs the higher number of nearest neighbor lattice sites (between 1.5 and 3 depending on the location of a $P_b$ center).
MW modulation further improves the DNP likely through recruiting more electrons and shifts the positive and negative enhancements apart when applied\cite{Hovav2014,Kwiatkowski2017,Kwiatkowski2018} as observed in Fig.~\ref{fig:EPR-DNP}a,b.
Interestingly, we found the optimal MW modulation bandwidth to be 100\,MHz and 200\,MHz at 3.35\,T and 6.7\,T, respectively.
These bandwidths make the maximum positive and negative DNP enhancements to coincide with the $m_I = \pm 1/2$ EPR manifolds.
We interpret this coincidence as indication for the electron-nuclear polarization transfer pathway occurring preferentially through $P_b$ centers with \textsuperscript{29}Si central nuclei and not through the backbonded nearest neighbor \textsuperscript{29}Si.
The increased transfer efficiency to the central \textsuperscript{29}Si can be understood by the up to ten times larger HF coupling compared to other possible locations of \textsuperscript{29}Si and amplified by the polarization transfer rate scaling approximately with the HF coupling squared.

Taken together, both the polarization of the central $P_b$ \textsuperscript{29}Si and the nuclear spin diffusion throughout pore walls (steps 1 and 3 of the hyperpolarization buildup in Fig.~\ref{fig:PSi_sketch}) are relatively fast processes compared to the measured buildup and decay times at all the DNP conditions (Fig.~\ref{fig:bup-grid-6p7T} and Sec.~\ref{sec:supp-DNP}, Suppl. Inf.).
In order to explain the long polarization buildup and decays we shall recall that there are \textsuperscript{29}Si nuclei with remarkably strong HF interaction --- the central and backbonded $P_b$ nuclei.
Between these strongly hypershifted \textsuperscript{29}Si spins and the bulk spins exists a large shift in frequency/energy, which is further enhanced due to the sparsity of \textsuperscript{29}Si in the naturally abundant PSi NPs.
Such frequency shifts suppress nuclear flip-flop transitions and create a so called spin diffusion barrier\cite{Bloembergen1949,Wenckebach2016}.
The transfer from the hypershifted nuclear spins to the bulk \textsuperscript{29}Si is, therefore, restrained, making it the rate limiting step in the DNP buildup.

For the nuclear polarization to diffuse across the spin diffusion barrier, the electrons need to modify the effective nuclear-nuclear spin interactions\cite{Horvitz1971,Buishvili1975,Sabirov1979,Atsarkin1980,Wittmann2018,Stern2021,Chessari2023,von_witte_two-electron_2024,redrouthu_overcoming_2024}.
Specifically, a coherent electron-nuclear four-spin flip-flop process \cite{von_witte_two-electron_2024,redrouthu_overcoming_2024} can be consistent with low temperatures employed in our experiments. 
The electron-nuclear four-spin flip-flops are very similar to triple spin flips involving an electronic flip-flop and nuclear spin flip but the nuclear spin flip is replaced by a nuclear dipolar flip-flop\cite{von_witte_two-electron_2024}.
The transition matrix element of the electron-nuclear four-spin flip-flops is proportional to the electronic and nuclear dipolar couplings.
Thus, a higher nuclear isotope abundance increases the rate of electron-nuclear four-spin flip-flops by increasing the nuclear dipolar couplings (due to the decrease of the average internuclear distances).
A higher electron-nuclear four-spin flip-flop rate leads to a faster spin transport from the hypershifted to bulk spins (step 2 in Fig.~\ref{fig:PSi_sketch}).

Another possible explanation for the long buildup time invokes the polarization transfer towards weakly HF-coupled spins.
In this case, distant \textsuperscript{29}Si nuclei are polarized directly by the dipolar coupling to a $P_b$ electron spin.
The direct polarization transfer hypothesis, however, possess a few flaws.
First, this process has a low probability since the HF coupling between a $P_b$ electron and a distant \textsuperscript{29}Si nucleus rapidly vanishes with the distance between them.
For example, for a nuclei located at a distance of three lattice constants, the HF interaction is $\sim 3.5$\,kHz, yielding low rates of direct polarization transfer.
Such a low polarization transfer rate might be too slow for the observed buildup times and enhancements.
Second, even a frequency difference of $\sim3.5$\,kHz is large compared to natural abundance SQ NMR line width of around 100\,Hz\cite{Hayashi2009}.
Assuming that the SQ line is a good approximation for the the ZQ line mediating nSD, the spectral density of energy conserving ZQ flip-flops vanishes when the ZQ line width is much smaller than the energy difference between the nuclei\cite{ernst_chapter_1998} as in this case by the given HF couplings.
Hence, nSD would still be suppressed by the HF couplings and would require an electron spin to alter the spin diffusion locally similar to the polarization transfer occurring at the strongly HF coupled nuclei\cite{Horvitz1971,Buishvili1975,Sabirov1979,Atsarkin1980,Wittmann2018,Stern2021,Chessari2023,von_witte_two-electron_2024,redrouthu_overcoming_2024}.
Third, the direct polarization transfer fails to explain the $m_I=\pm 1/2$ DNP enhancements and zero DNP for the central EPR peak (Fig.~\ref{fig:EPR-DNP}).

We highlight that 95.3\,\% of the $P_b$ centers have magnetically inactive $^{28}$Si or $^{30}$Si central nuclei.
Therefore, in addition to the low probability of transfer from the central \textsuperscript{29}Si to the distanced bulk, only 4.7\,\% of $P_b$ contribute to the hyperpolarization buildup if the polarization transfer flows through the central nuclei.
According to the sample characterization data, a typical PSi NP of 150\,nm size, 55\,\% porosity (0.55\,cm$^3$g$^{-1}$ pore volume) and 100\,m$^2$g$^{-1}$ surface area contains on average $2.3\cdot 10^6$ \textsuperscript{29}Si nuclei and $1.3\cdot 10^4$ $P_b$ centers (Fig.~\ref{fig:characterization}b-g). 
Therefore, a straightforward but incorrect calculation yields the number of \textsuperscript{29}Si to be polarized by one $P_b$ center equal to $\sim$180 --- a common value for partially deuterated water glycerol mixtures (DNP juice)\cite{Prisco2021}. 
However, the number of \textsuperscript{29}Si nuclei that are central to the $P_b$ electrons is only 4.7\% of the total number of nuclei. 
This leads to around 3800 nuclei to be polarized per DNP-active $P_b$ center, a much lower value than in typical DNP samples.

\subsection{Rate limiting step for the polarization decay} \label{sec:discussion_decay}

The polarization decay in Si particles with endogenous electronic centers has been commonly considered to be limited by nSD from bulk \textsuperscript{29}Si to the \textsuperscript{29}Si in the core-shell interface\cite{Aptekar2009,Lee2011,Cassidy2013}.
The arguments of low electron polarization at room temperature and orders of magnitude lower $T_\mathrm{1e}$ and $T_\mathrm{2e}$ than at DNP conditions further supported the hypothesis of nSD limiting relaxation.
Although these arguments seem to be a reasonable for \textmu m-sized Si particles, they are hardly applicable to our case of PSi NPs or to other types of Si NPs\cite{Atkins2013,Kwiatkowski2018,Kim2021,kwiatkowski_vivo_2024} with crystalline cores on the order of 20\,nm.
If nSD is the rate limiting step for the relaxation in our samples, the polarization decay time $\tau_\mathrm{dec}$ at room temperature should be tens of seconds at the slowest, according to the estimated $D_\text{diff}\approx 0.5 - 1.7$\,nm$^2$s$^{-1}$.
Unless this estimation is orders of magnitude incorrect, which is unlikely, nSD fails to explain the room temperature relaxation times in nanoscale Si.

\begin{figure}
    \centering
    \includegraphics[width=0.8\linewidth]{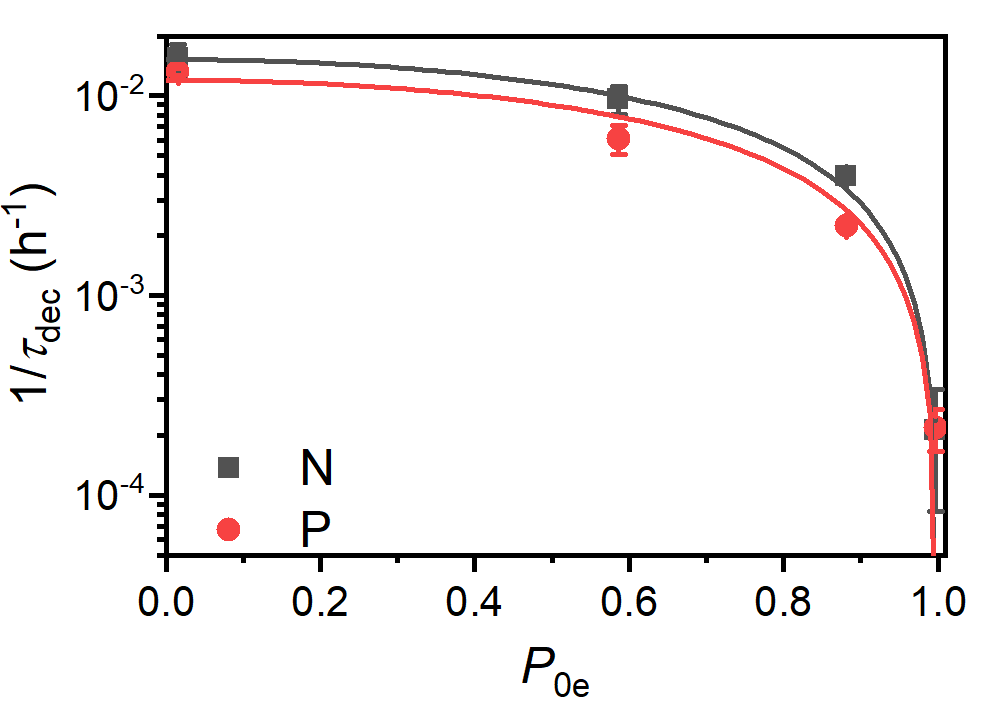}
    \caption{The dependence of the hyperpolarization decay rate $\tau_\mathrm{dec}^{-1}$ on the thermal electron polarization $P_\mathrm{0e}$ for the P and N PSi NPs. 
    The shown experimental data was measured at (ordered with increasing $P_\mathrm{0e}$) 300\,K, 7\,T; 3.4\,K, 3.4\,T; 3.4\,K, 7\,T and 1.4\,K, 6.7\,T.
    For the fits, we assumed $\tau_\mathrm{dec}^{-1} \propto 1-P_\mathrm{0e}^2$ (see main text for discussion).
    }
    \label{fig:relaxationElectronPolarization}
\end{figure}

Fig.~\ref{fig:relaxationElectronPolarization} depicts the dependence of nuclear hyperpolarization decay rates $\tau_\mathrm{dec}^{-1}$ on the thermal electron polarization $P_\mathrm{0e}$ at the given experimental conditions.
The $\tau_\mathrm{dec}^{-1}$ follows a $1-P_\mathrm{0e}^2$ scaling indicating that the nuclear relaxation appears to be governed by paramagnetic effects\cite{AbragamGoldman,Wenckebach2016}.

We note that naturally abundant 160\,nm\cite{Kim2021} and 50\,nm\cite{Kwiatkowski2017,Kwiatkowski2018} non-porous Si NPs as well as the porous particles in this work all give similar room temperature relaxation times of around 50\,min. 
Such a similar relaxation time across particles and the dependence on the \textsuperscript{29}Si abundance might indicate that even at room temperature the nuclear relaxation is governed by the same process as at the DNP conditions, \textit{i.e.}, by the electron modified nSD across the spin diffusion barrier with fast relaxation of strongly hyperfine coupled \textsuperscript{29}Si spins. 
The observed $1-P_\mathrm{0e}^2$ scaling (cf. Fig.~\ref{fig:relaxationElectronPolarization}) would in this case not describe the paramagnetic relaxation itself but the actual spin transport via electron-nuclear four-spin flip-flops\cite{von_witte_two-electron_2024}.
Electron-nuclear four-spin flip-flops have a similar mechanism as triple spin flips causing indirect paramagnetic relaxation\cite{Wenckebach2016} because both mechanisms involve an electron dipolar flip-flop, which provides energy for a nuclear excitation \textit{e.g.}, a nuclear spin flip for triple spin flips and a nuclear flip-flop for electron-nuclear four-spin flip-flops.
Therefore, both triple spin and four-spin flip-flops can be considered to have the same scaling with electron polarization, which for paramagnetic relaxation follows $1-P_\mathrm{0e}^2$ dependence\cite{Wenckebach2016}.

Alternatively, the averaging of the HF couplings in combination with slow paramagnetic relaxation could explain the long room temperature relaxation times.
Both interpretations would be in good agreement with isotope enrichment experiments\cite{Kim2021} which found a 3-fold decrease in $\tau_\mathrm{dec}$ (from 48 to 17 minutes) upon increasing the \textsuperscript{29}Si abundance from 4.7\,\% to 15\,\%.
The higher isotope abundance results in larger nuclear dipolar couplings and reduced frequency differences for nuclei close to the electron which would increase the transport of polarization across the spin diffusion barrier.
Additionally, more $P_b$ centers with a central \textsuperscript{29}Si results in a larger fraction of $P_b$ centers with large (averaged) HF couplings which increases nuclear relaxation.

A major open question concerning the proposed rate limiting step of the room temperature decay is the lack of knowledge of the electron relaxation times with respect to the nuclear Larmor frequency $\omega_\mathrm{0n}$.
If the relaxation rate is much smaller than $\omega_\mathrm{0n}$, the HF couplings are not averaged and the situation is similar to low temperature DNP conditions.
For much faster relaxation rates than $\omega_\mathrm{0n}$, the thermal electron polarization of 1.6\,\% at room temperature and 7\,T leads to (pseudo-)contact shifts due to the partially averaged HF couplings \cite{bloembergen_shift_1950,bertini_magnetic_2002,parigi_magnetic_2019,pell_paramagnetic_2019}.
Even in this averaged case the (pseudo-)contact shift is expected to be large compared to the nuclear dipolar couplings mediating the nSD.

\subsection{Samples and experimental conditions} \label{sec:discussion_samples}

Above we discussed that DNP likely originates from triple spin flips requiring coupling between the involved electrons with a fast polarization transfer to the central \textsuperscript{29}Si nucleus of the $P_b$ center due to its large hyperfine coupling of hundreds of MHz.
This is followed by a slow transfer from this central, strongly hypershifted nucleus to the bulk nuclei, followed by fast spin diffusion in the bulk.
For the decay, the inverse process happens with paramagnetic relaxation instead of triple spin flip DNP.
Similar to the buildup, the decay time appears governed by the hyperpolarization transfer from the bulk to the strongly hypershifted \textsuperscript{29}Si nuclei.
Since both the buildup and the decay involve a single rate limiting step, a mono-exponential hyperpolarization dynamics\cite{Witte2023} is expected and found experimentally (Fig.~\ref{fig:bup-grid-6p7T}, Sec.~\ref{sec:supp-DNP}, Suppl. Inf.).

For a mono-exponential hyperpolarization dynamics, the buildup time, steady-state polarization and thermal electron polarization can be used to define hyperpolarization injection and decay rate constants during buildup\cite{Witte2023}, which we briefly summarize below and in more detail in Sec.~\ref{sec:supp-rates}, Suppl. Inf.
With these rate constants, it is possible to quantify the nuclear relaxation during buildup and compare it to the rate with which hyperpolarization is created.

In the rate-equation model, the buildup time $\tau_\mathrm{bup}$ depends on two competing processes: the nuclear polarization injection rate constant, $k_\mathrm{W}$, and the nuclear relaxation rate constant $k_\mathrm{R}^\mathrm{bup}$. 
Together with the thermal electron polarization $P_\mathrm{0e}$, we can describe the buildup as\cite{Witte2023}:
\begin{subequations} \label{eq:rateEq}
\begin{align}
    \frac{\text{d}P_\mathrm{n}(t)}{\text{d}t} &= (P_\mathrm{0e}-P_\mathrm{n}) k_\mathrm{W} - k_\mathrm{R}^\mathrm{bup} P_\mathrm{n} , \\
    \tau_\mathrm{bup}^{-1} &= k_\mathrm{W} + k_\mathrm{R}^\mathrm{bup} \overset{P_\mathrm{1n}\ll P_\mathrm{0e}}{\approx} k_\mathrm{R}^\mathrm{bup} ,\\
    P_\mathrm{1n} &=  P_\mathrm{0e}\frac{k_\mathrm{W}}{k_\mathrm{W}+k_\mathrm{R}^\mathrm{bup}}  \overset{P_\mathrm{1n}\ll P_\mathrm{0e}}{\approx} P_\mathrm{0e}\frac{k_\mathrm{W}}{k_\mathrm{R}^\mathrm{bup}} ~,
\end{align}
\end{subequations}
where $P_\mathrm{1n}$ is the steady-state nuclear polarization reached by the end of the DNP process.
For the decay, the decay rate constant $k_\mathrm{R}^\mathrm{dec} = \tau_\mathrm{dec}^{-1}$ provides sufficient description since MW radiation is off.

\begin{figure}
    \centering
    \includegraphics[width=0.8\linewidth]{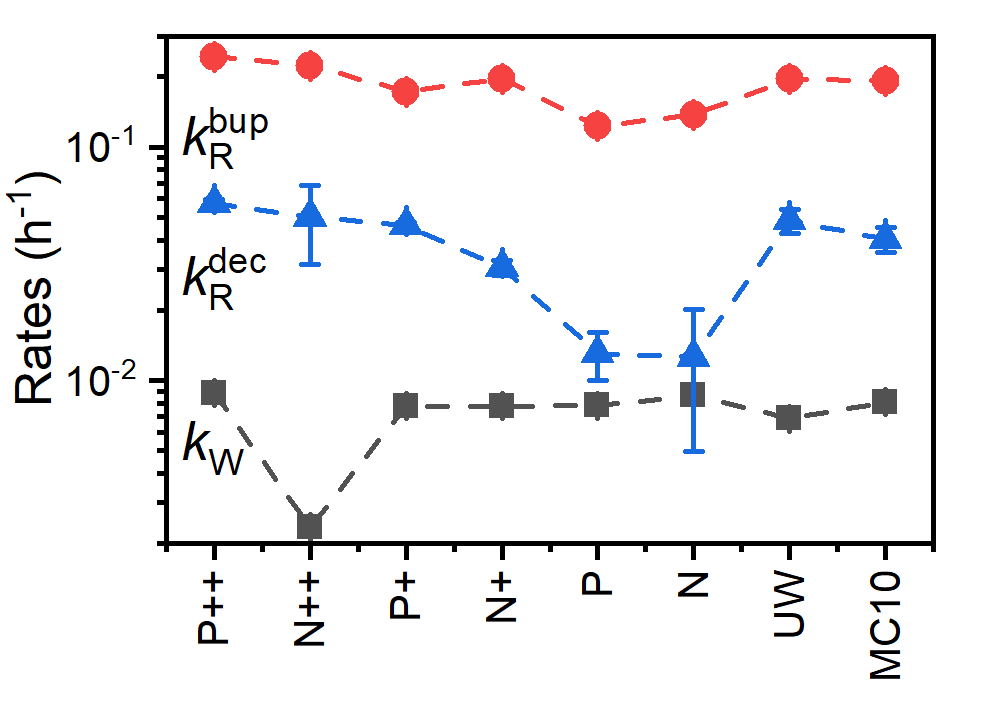}
    \caption{Polarization buildup (dark squares and red circles) and decay (blue triangles) rates (cf. Eqs.~\ref{eq:rateEq}) for the PSi NPs with different doping and oxidation.
    The data was acquired at 6.7\,T, 1.4\,K.
    Lines are a guide for the eye.
    }
    \label{fig:rates6p7T}
\end{figure}

Fig.~\ref{fig:rates6p7T} compares the DNP injection and relaxation rates during build-up and decay at 6.7\,T, 1.4\,K.
The rates at other conditions and a more detailed discussion of these are given in Sec.~\ref{sec:supp-rates}, Suppl. Inf. with the main results summarized below.
The relaxation rates during the buildup are an order of magnitude larger than the injection rates and, hence, govern the buildup time.
The imbalance between between buildup and relaxation rates results in moderate steady-state polarization (and enhancements) compared to \textsuperscript{1}H or \textsuperscript{13}C enhancements under similar conditions (cf. Ref.~\cite{Witte2024} and references therein for state-of-the-art enhancements).
DNP injection appears rather uniform across the samples while the relaxation rates rates show a variation between samples.
Thus, differences between samples mostly originate from different relaxation properties.
Furthermore, the DNP injection shows a less pronounced dependence on the experimental conditions than the relaxation rates (cf. Fig.~\ref{fig:supp-rates-6p7T}, Suppl. Inf. and discussion thereof).
However, suppressing the relaxation with lower temperatures (1.4\,K instead of 3.4\,K) shows at best only a modest improvement because relaxation enhancement by MW irradiation\cite{Witte2024} becomes more pronounced as evident by the much higher relaxation rates during buildup ($k_\mathrm{R}^\mathrm{bup}$) compared to decay ($k_\mathrm{R}^\mathrm{dec}$) as shown in Fig.~\ref{fig:rates6p7T}.
Reduction of this relaxation enhancement \textit{e.g.}, by higher fields, lower temperatures or shortening of electronic relaxation times\cite{Witte2024}, offers the possibility of higher enhancements and polarization levels.

By performing the correlation analysis, we further connect the structural properties of PSi NPs, $P_b$ centers and hyperpolarization.
The nuclear hyperpolarization at 6.7\,T, 1.4\,K (Fig.~\ref{fig:bup-grid-6p7T} is nearly independent from the number of $P_b$ centers (Fig.~\ref{fig:characterization}g) expressed by a correlation coefficient of 0.04 across all samples but correlates (0.44) with the interface density of the $P_b$ centers (number of $P_b$ defects divided by the specific surface area, Fig.~\ref{fig:characterization}d,g).
The density of $P_b$ centers correlates strongly negatively across all the samples with the specific surface area (coefficient equals to $-0.9$) because the increase of surface area does not lead to the proportional increase of the number of $P_b$ centers (Fig.~\ref{fig:characterization}d,g).
Such an independence of the number of $P_b$ centers suggests existence of a limit for their formation at least for the oxidation methods applied here.
A more detailed discussion of the correlation analysis can be found in Sec.~\ref{sec:supp-rates}, Suppl. Inf.


Before concluding, we want to highlight specific aspects of the investigated PSi NPs.
First, the ability to create nuclear hyperpolarization in PSi NP appears exceptionally robust due to the core-shell nature of the particles with the paramagnetic centers protected from the environment.
Specifically, the $P_b$ centers form at the interface between the crystalline pore wall cores and the oxide shell. 
Hence, the $P_b$ centers as well as the nuclear hyperpolarization are largely shielded from everything outside each particle \textit{e.g.}, different particle coating or solution media\cite{kwiatkowski_vivo_2024}.
This is exemplified by the inertness to the presence of the catalytic Au NPs in PSi NPs: removing the Au NPs used as an etching catalyst shows no clear effect on the hyperpolarization process and the observed changes are in line with other additional oxidation steps (cf. Fig.~\ref{fig:enh-bup-oxy}).
This suggests that the pores in PSi NPs with their large surface area available for coating could be used for loading with additional substances to add further diagnostic or therapy capabilities.

Second, the nuclear hyperpolarization in the bulk seems to be inert with respect to a wide range of bulk defects and their densities: both boron and phosphorous doping with densities up to $\sim 10^{16}$\,cm$^{-3}$ show similar high polarization levels (Figs.~\ref{fig:bup-grid-6p7T} and \ref{fig:exp-comparison}).
Furthermore, the light doping of $\sim 5\cdot 10^{14}$\,cm$^{-3}$ is superior compared to the most pure UW sample.
The increase of doping level to the order of $10^{18}$\,cm$^{-3}$ becomes detrimental for achievable hyperpolarization levels due to the onset of the wavefunction overlap of the dopants.
At room temperature, high densities of thermally excited mobile charge carriers from shallow dopants strongly increase the nuclear relaxation\cite{Lee2011}.
The least pure MC10 sample possesses various different dopants, with energy levels often deep in the bandgap of Si and thus with narrow defect wave functions, which makes DNP performance of MC10 sample very similar to the best electronic wafer grade samples.
Such a stable polarization process is important if other, especially bottom-up manufacturing techniques should be employed as these offer a reduced control over the bulk purity compared to the top-down approach of the current work.

\section*{Conclusion}

We employed low-load metal assisted catalytic etching (LL-MACE)\cite{Tamarov2020a,Tamarov2020b} to fabricate a variety of porous Si NPs from electronic grade single crystal Si wafers.
This top-down fabrication approach allowed us to vary dopant type and density while achieving nearly identical surface properties and crystallinity in all the NPs.
A separate oxidation step led to the formation of electronic $P_b$ centers with similar structure and surface density for all the types of PSi NPs.
This resulted in the successful and similar DNP injection in all samples with the polarization differences mostly ascribed to relaxation.
The robustness of the hyperpolarization process to different shallow dopant concentrations and metallurgical grade Si samples containing deep dopants enables and justifies the use of a wide range of manufacturing methods with eventually poor control over the bulk composition, e.g. bottom-up synthesis methods.
The highest steady-state polarization levels were achieved with lightly ($\sim 10^{14}$\,cm\textsuperscript{-3}) phosphorous or boron-doped samples.
Measurements at 7\,T (3.4\,K) and 3.35\,T (1.4 and 3.4\,K) gave lower polarization levels than the 6\,\% achieved at 6.7\,T, 1.4\,K.
Room temperature decay times of the studied samples exceeded one hour --- the longest hyperpolarization decay time obtained so far in Si NPs to our knowledge although slightly longer saturation-recovery $T_1$ times $(102\pm10)$\,min have been reported for 21\,nm (comparable to pore walls in the present work) particles\cite{Thiessen2019}. 

The gained insights about $P_b$ centers enabled us to shed light on the polarization transfer from the electron spins to \textsuperscript{29}Si.
Owing to the core-shell nature of the PSi NPs with the $P_b$ centers at the interface between the core and shell, nuclear spin diffusion is required to transport the hyperpolarization across the sample.
The central \textsuperscript{29}Si nuclei of the $P_b$ centers with hyperfine couplings around 300\,MHz are predominantly hyperpolarized by DNP.
The large difference in frequency compared to the bulk \textsuperscript{29}Si spins seems to cause a slow transport of polarization from the central $P_b$ \textsuperscript{29}Si nuclei towards the bulk, which causes the long buildup and relaxation times in the presence of fast bulk spin diffusion.
Isotope labelling may improve the transport across the spin diffusion barrier and, therefore, improve the NMR signal through the increased enhancement and number of magnetically active spins.
The disadvantages of isotope labelling are high sample cost and shortened room temperature decay times. 
The described hyperpolarization process could be translated to create other slowly relaxing NPs with sizes down to possibly 10\,nm.
For this, three properties of (Si) NPs appear essential: (i) a low bulk relaxation, (ii) DNP on the outer surfaces of the particles and (iii) a slow transport from the bulk to the surface \textit{e.g.}, a strongly localized wavefunction of a surface paramagnetic center with large hyperfine coupling compared to the nuclear line width.

\section*{Author contributions}

Conceptualization: GvW, ME, JR, VPL, SK, MIK, KT.
Data curation: GvW, VH, MIK, KT.
Methodology: GvW, AH, MMA, ME, MIK, KT.
Formal analysis: GvW, KT.
Investigation: GvW, AH, VH, JJ, JOM, SK, MIK, KT.
Funding acquisition: ME, JR, VPL, SK, MIK, KT.
Resources: JOM, ME, SK, VPL, MIK.
Writing – original draft: KT, GvW.
Writing – review and editing: GvW, ME, JR, SK, MIK, KT.
All authors approved the draft.

\section*{Conflicts of interest}

There are no conflicts to declare.

\section*{Data availability}

All the data and analysis scripts for this article are available at ETH Library at \href{https://doi.org/10.3929/ethz-b-000679152}{https://doi.org/10.3929/ethz-b-000679152}.

\section*{Acknowledgements}

The work was supported by Research Council of Finland (grant  nos. 314551, 331371, 322006, and Flagship of Advanced Mathematics for Sensing Imaging and Modelling grant 358944), ETH Zürich and the Schweizerischer Nationalfonds zur Förderung der Wissenschaftlichen Forschung (grant
nos. 200020\_188988 and 200020\_219375), Finnish Cultural Foundation (North Savo regional fund) and Saastamoinen Foundation. The microscopy studies were performed using the facilities of SIB Labs, Laboratory of Microscopy at the University of Eastern Finland.
Part of the work was carried out with the support of Kuopio Biomedical Imaging Unit, University of Eastern Finland, Kuopio, Finland (part of Biocenter Kuopio, Finnish Biomedical Imaging Node, and EuroBioImaging).
Silicon was provided by Elkem Silicon Products (Elkem ASA, Norway) and Okmetic (Okmetik Oy, Finland).



\balance


\bibliography{rsc} 
\bibliographystyle{rsc} 

\clearpage

\ifarXiv
    

\section*{Supplementary material}

# S1 Experimental

## S1.1 Liquid-phase oxidation

Two-step and one-step oxidation were performed for PSi NPs (*i.e.*, after milling of thermally oxidized PSi powders) according to the procedure described in ref [1]. In the first step of oxidation, about 100 mg of PSi NPs stored in ethanol suspension were first redispersed in deionized water by repeating two times the following sequence: centrifugation of PSi NP suspension, supernatant removal, redispersion in water in an ultrasound bath (Elmasonic S10). The final redispersion used only 10 ml of water. Next, 10 ml of $NH_4OH$ (7 wt.%, VWR Chemicals) solution was slowly added under stirring followed by slow pouring of 2 ml $H_2O_2$ (35 wt. %, Acros Organics, Thermo Fisher GmbH). The suspension was then sonicated for 1 min and placed on heating plate and the oxidation reaction proceed for 15 min at 90 °C under stirring. The reaction was then slowed down by diluting the suspension with about 30 ml of water, and the PSi NPs were washed with water by repeating centrifugation-redispersion cycle three times. Again, the final redispersion used only 10 ml of water. In the second step of oxidation, 10 ml of 2 M HCl is added to the NPs under stirring, into which subsequently 2 ml of $H_2O_2$ (35 wt. %) is poured. The reaction is then carried out at 90 °C for 15 min. Finally, washing is performed as in the first step with the final replacement of water with ethanol for NP storage. One-step oxidation employed only the second step of the two-step liquid-phase oxidation.

# S2 Characterization of PSi NPs

## S2.1 Dynamic light scattering

Dynamic light scattering (DLS, Zetasizer Nano ZS, Malvern Panalytical) is a simple and fast method to monitor the integral distribution of hydrodynamic sizes as opposed to the direct (and subjective) observation by transmission electron microscopy. Hydrodynamic size is an important parameter to provides the behavior of NPs in biologically relevant media. Herein, DLS was used to follow the hydrodynamic size distributions of PSi NPs during and after the milling of NPs described in Section 2.3 of the main text. Figure S1 shows the size distributions after 1 hour of milling demonstrating the similar hydrodynamic sizes of all the types of PSi NPs used for hyperpolarization.

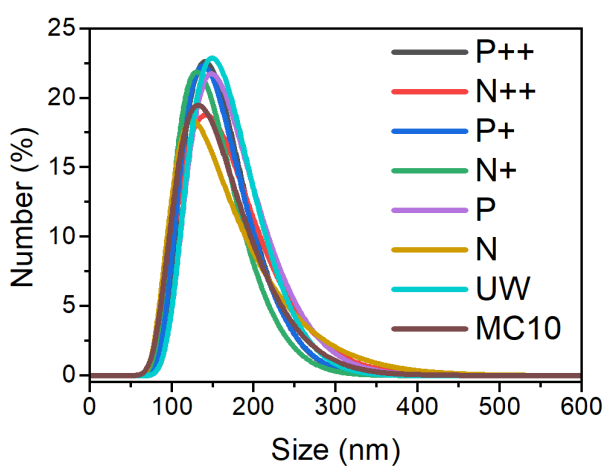

Figure S1: Hydrodynamic size distribution of all the samples after 1 hour milling (see Section 2.3 of the main text).

The top-down approach allows to flexibly alter size distributions by additional milling and centrifugation cycles as required by a specific application. Particularly, for biomedical applications the sizes below 100 nm are preferable[2]. In Figure S2, size distributions of P PSi NPs after centrifugation at 2500 rcf for 20 min show that most of the NPs have sizes below 100 nm. Note that we do not expect any change in the DNP properties of the NPs since the Si crystalline sizes are determined by the pore walls which are

significantly smaller than the hydrodynamic sizes. The use of suspension with relatively large sizes (Figure 1b) made it easier to collect hundreds of milligrams of PSi NPs for the DNP measurements.

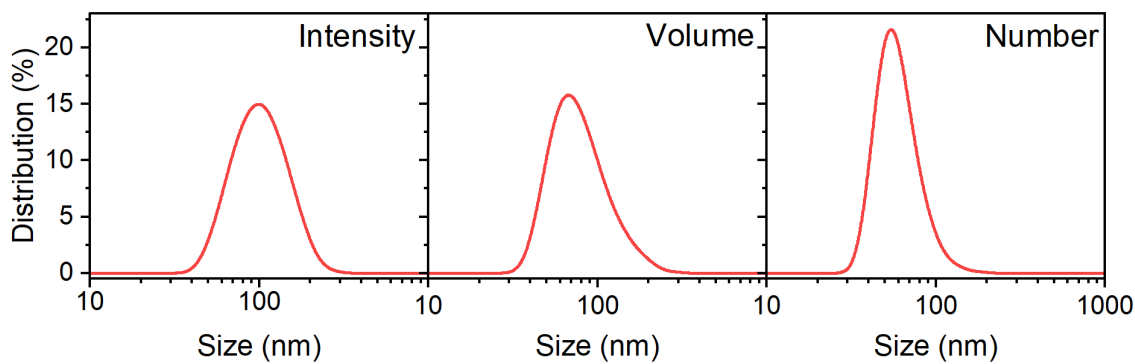

Figure S2: Hydrodynamic size distribution of P PSi NPs after centrifugation with 2500 rcf for 20 min. Intensity is the raw measured signal $\sim r^6$, where $r$ is particle hydrodynamic diameter. Volume ($\sim r^3$) and number ($\sim r^0$) size distributions are calculated from the intensity distribution using the Zetasizer Nano ZS software.

### S2.2 Fourier-transform infrared spectroscopy

Surface oxidation of PSi NPs was characterized using Fourier-transform infrared spectroscopy (FTIR, Thermo Nicolet iS50) of KBr tablets. The tablets were prepared by grinding 200 mg of dried KBr powder together with (1–2) mg of dried PSi NPs in a mortar until the powder color became homogeneous. Manual hydraulic press converted the powder into a uniformly colored KBr tablet. FTIR measurements were then performed in transmission mode.

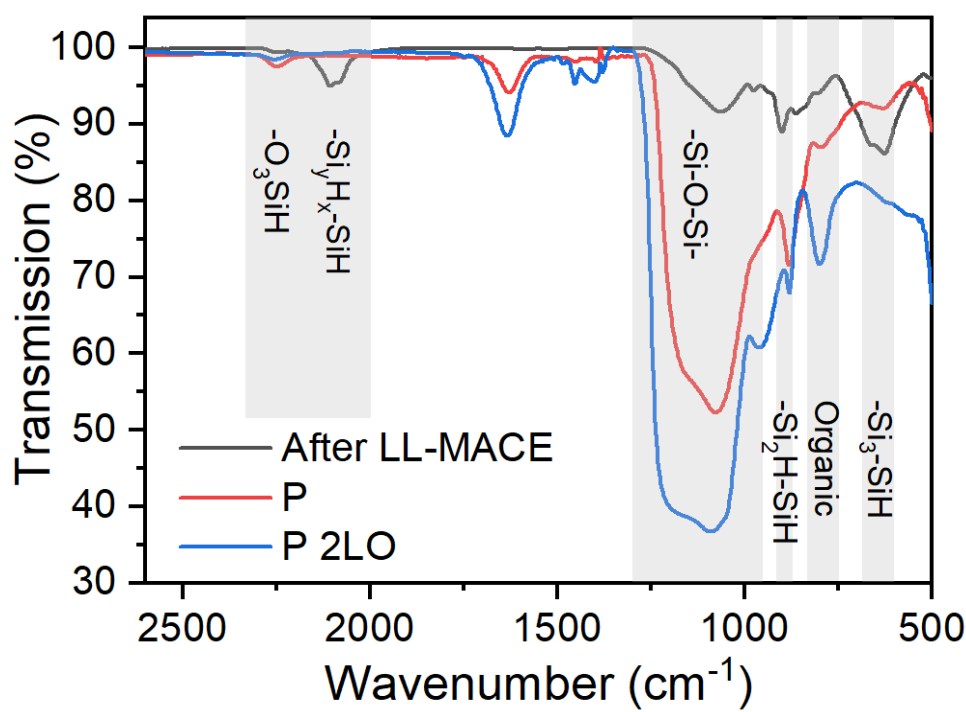

Figure S3: Transmission FTIR spectra of hydrogen-terminated P PSi powder after LL-MACE (dark line), thermally oxidized P PSi NPs (red line), and P PSi NPs after liquid and thermal oxidation (blue line). Grey shaded squares and labels assign FTIR peaks.

3

Figures S3 and S4 depict the difference between the hydrogen-terminated sample after LL-MACE, milled PSi NPs prepared from thermally oxidized PSi powders, and the liquid-phased oxidized PSi NPs. Hydrogen-terminated sample shows strong IR absorption peaks at $(615–625)\,\mathrm{cm}^{-1}$, $948\,\mathrm{cm}^{-1}$, and $(2050–2160)\,\mathrm{cm}^{-1}$ that correspond to various silicon hydride species on PSi surfaces[1]. The wide peak at $(1000–1250)\,\mathrm{cm}^{-1}$ and the peak at $2248\,\mathrm{cm}^{-1}$ demonstrate the native oxidation process during overnight drying in an oven at $65\,°\mathrm{C}$. Thermal oxidation with subsequent milling to NPs created the strong Si−O−Si oxide peak with almost complete disappearance of -Si$_y$H$_x$-SiH hydride species. However, an appearance of hydrogen bound to backbone oxidized Si was observed (−O$_3$SiH species) indicating that not all the hydrogen was removed from PSi surfaces. The presence of hydrogen can impede surface functionalization based on reaction with silanes (for example, PEG-silanes[3] or amine-silanes[4]). Therefore, additional liquid-phase oxidation was applied to PSi NPs to further reduce hydride species on NP surfaces, and the influence of oxidation on hyperpolarization was studied. Both two-step and one-step liquid-phase oxidations efficiently reduced the number of -Si$_y$H$_x$-SiH although decreasing the gain in $^{29}$Si hyperpolarization (see the main text).

Au removal involved highly oxidative iodine solution applied to the N PSi NPs after LL-MACE. The solution effectively oxidized Si surfaces as it can be seen in Fig. S4. The Si oxidative action of the Au etchant was found to be similar to other oxidation types and resulted in nearly full removal of the surface Si-H groups followed by Si backbond oxidation[1].

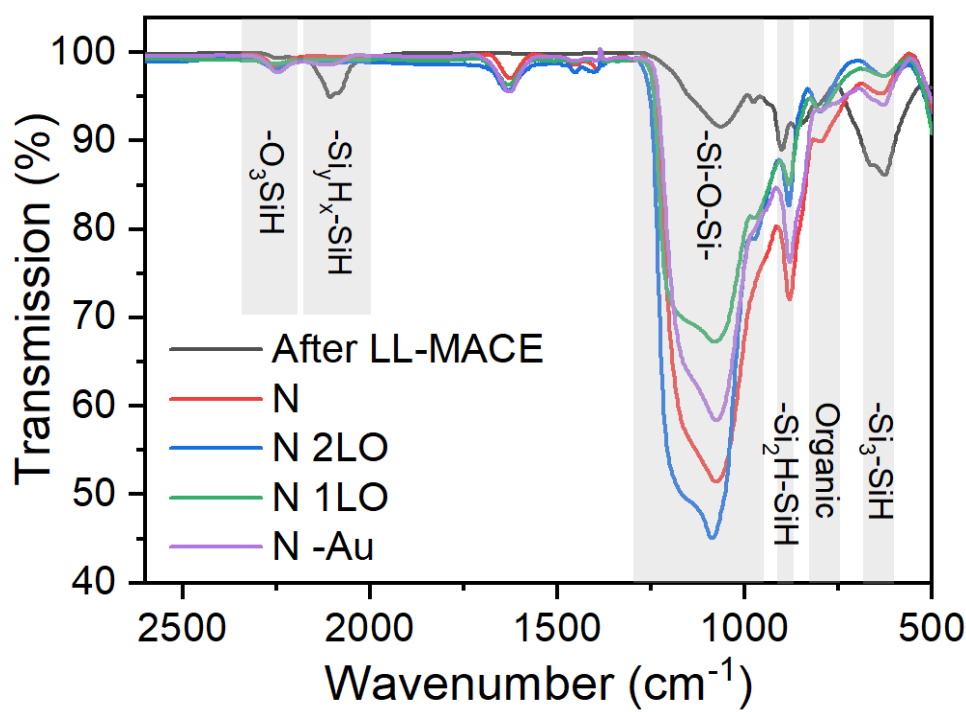

Figure S4: Transmission FTIR spectra of hydrogen-terminated N PSi powder after LL-MACE (dark line), thermally oxidized N PSi NPs (red line), N PSi NPs after two-step liquid and thermal oxidation (blue line), N PSi NPs after one-step liquid and thermal oxidation (green line), and N PSi NPs with Au removed (magenta line). Grey shaded squares and labels assign FTIR peaks.

### S2.3 X-ray powder diffraction

Crystalline sizes of PSi particles after LL-MACE were calculated using Retvield refinement in TOPAS® 4.6 software. Typically, three phases were needed to correctly fit a spectrum: two Si phases and one Au phase (Fig. S5). The Si phases corresponded to the pore walls between etch track pores produced by Au NP movement, and to the pore walls between pores produced by remote etching[5, 6].

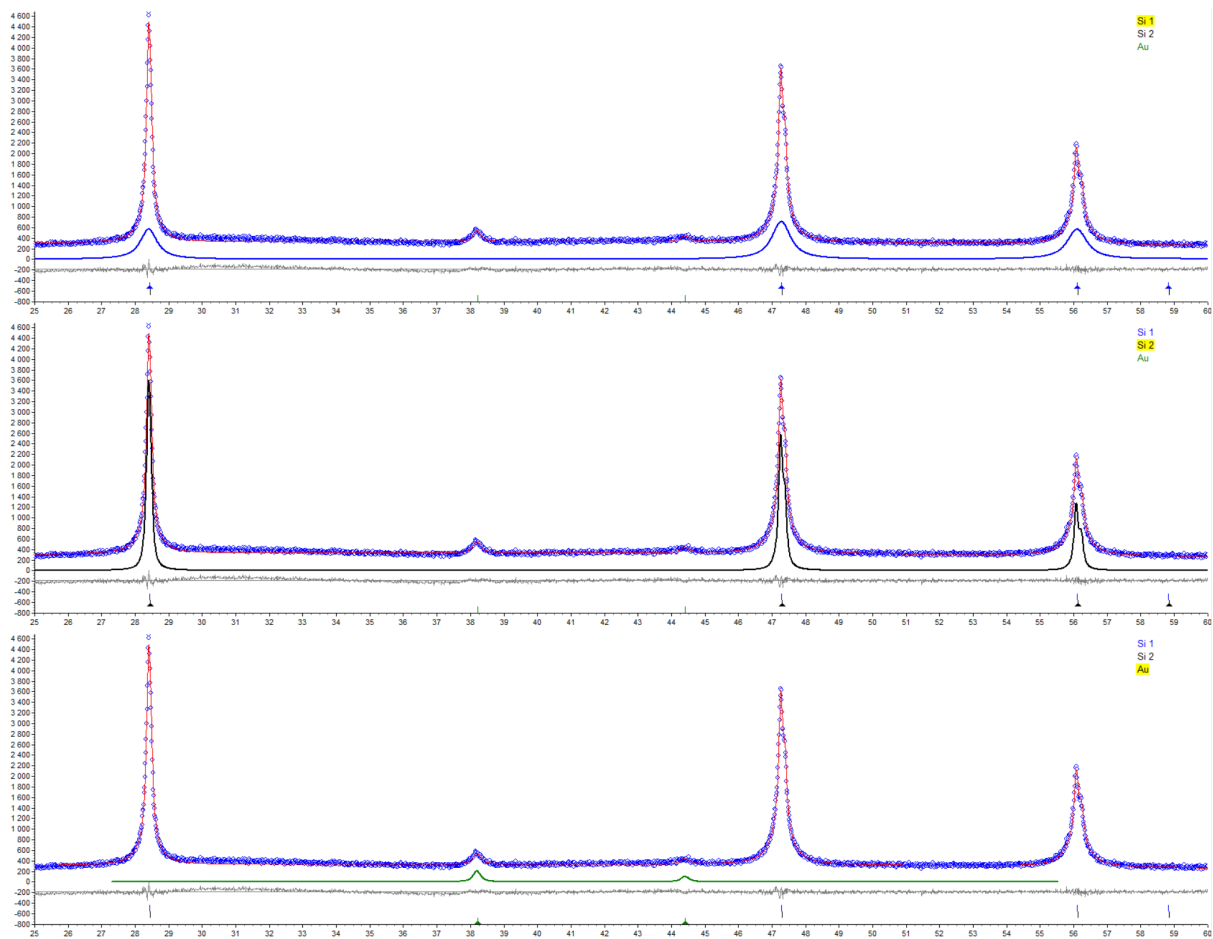

Figure S5: Fitting of XRPD spectrum of P sample with two Si phases and one Au phase. All the samples were processed in the same way.

## S2.4 Electron paramagnetic resonance

Electron paramagnetic resonance (EPR) studies were performed using X-band Magnettech MiniScope MS5000 spectrometer operating at room temperature. The same volume of PSi NPs powder was placed in an EPR tube and the tube was placed at the same height for each measurement. The mass and surface number of paramagnetic centers was calculated by double integration of spectra with subsequent comparison with a TEMPO sample with a known number of radicals. The mass and surface amount of the paramagnetic centers is summarized in Table S1. The concentration of the centers per unit of mass was comparable for all the samples and ranged from $(4.4\pm0.4)\cdot10^{15}$ mg$^{-1}$ for N+ to $(6.3\pm0.6)\cdot10^{15}$ mg$^{-1}$ for N 1LO. The surface density for P++ and N++ samples was about 3 times smaller than for other samples due to their higher surface area (Fig. 1d) while the number of centers per unit mass remained roughly the same.

EPR spectra of thermally oxidized PSi NPs of different Si types and different additional liquid oxidations are presented in Figs. S6 and S7, respectively. All spectra represent a powder average of paramagnetic dangling bond $P_b$ centers that are in turn randomly oriented and located at different Si crystalline planes in Si/SiO$_2$ interface[7]. Hyperfine satellite peaks at $\Delta B = (5.8$–$7.7)$ mT unambiguously demonstrated the presence of $^{29}$Si nuclei at the central $P_b$ position (Fig. S8). The hyperfine constants $A = (325$–$431)$ MHz coincide well with $A_\parallel = 210$ MHz and $A_\perp = 417$ MHz for the (111) $P_b$ center[8]. The superhyperfine interaction typically observable for planar (111) $P_b$ center at about $0.8$ mT or $45$ MHz[9] could not be resolved due to high peak broadening in our samples but was assumed to be present.

Following the discussion of $P_b$ centers in the main text, the measured EPR spectra were fitted with EasySpin 5.2.35[10] using a combination of $P_b^{iso}$ and (111) $P_b$, since these centers are assumed to be the dominant ones in thermally oxidized porous Si[11]. The inclusion of the (111) $P_b$ center in the fitting was the most obvious for the standard thermally oxidized porous Si sample prepared by the

Table S1: Summary of the experimental EPR data. The number of $P_b$ centers was calculated from the known paramagnetic center concentration of TEMPO radical and additionally confirmed using thermally oxidized electrochemically etched PSi sample[1].

| Sample | Calculated from TEMPO sample | | Experimental data | |
|---|---|---|---|---|
| | $P_b$ centers, $\cdot 10^{15}$ mg$^{-1}$ | $P_b$ centers, $\cdot 10^{12}$ cm$^{-2}$ | EPR lwpp[a], mT | EPR FWHM[b], mT |
| PSi | 4.6 ± 0.4 | 1.9 ± 0.1 | 1.10 | 1.13 |
| P++ | 5.4 ± 0.3 | 2.7 ± 0.1 | 0.76 | 1.03 |
| P+ | 4.5 ± 0.4 | 5.0 ± 0.5 | 0.55 | 0.92 |
| P | 4.8 ± 0.3 | 5.3 ± 0.3 | 0.57 | 0.93 |
| P 2LO | 5.3 ± 0.3 | 5.9 ± 0.4 | 0.55 | 0.90 |
| UW | 5.2 ± 0.3 | 4.9 ± 0.3 | 0.55 | 0.90 |
| N | 5.6 ± 0.4 | 6.1 ± 0.5 | 0.55 | 0.90 |
| N 2LO | 5.4 ± 0.2 | 5.8 ± 0.3 | 0.55 | 0.92 |
| N 1LO | 6.3 ± 0.6 | 6.8 ± 0.7 | 0.55 | 0.90 |
| N -Au | 4.9 ± 0.4 | 5.2 ± 0.4 | 0.53 | 0.87 |
| N+ | 4.4 ± 0.4 | 4.1 ± 0.4 | 0.55 | 0.94 |
| N++ | 4.7 ± 0.4 | 1.8 ± 0.1 | 0.60 | 0.95 |
| MC10 | 5.5 ± 0.3 | 3.1 ± 0.2 | 0.60 | 0.96 |

[a] peak-to-peak linewidth [b] full width at half maximum

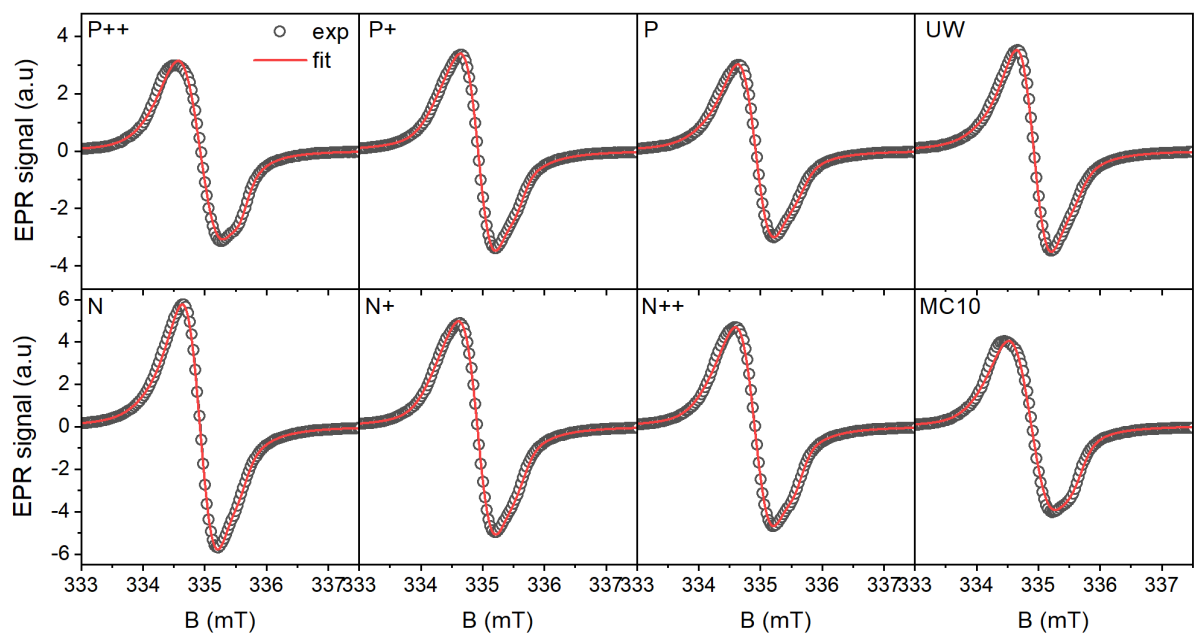

Figure S6: Electron paramagnetic resonance spectra of thermally oxidized PSi NPs of different Si types. The experimental data (dark lines) was fitted (red lines) as discussed in the text.

conventional electrochemical anodization of (100) P++ Si wafer (Fig. S9)[12] with subsequent thermal oxidation. During the electrochemical anodization, the etched pores are formed normal to the (100) surface, which results in more pronounced signal from (111) $P_b$ centers compared to much less ordered pores in the LL-MACE samples. Nevertheless, even in the LL-MACE samples the anisotropy of the EPR spectrum at about 336 mT is due to the presence of (111) $P_b$ centers (Fig. S6 and S7).

The resulting fitting parameters for the anodized PSi sample give reasonable values. The weights for the $P_b$ and $P_b^{iso}$ components are 0.37 and 0.63, respectively, which show the presence of relatively high fraction of well-defined (111) $P_b$ centers. As it is expected, the $g$-factor strain for $\mathbf{B} \perp [111]$ is much higher than for $\mathbf{B} \parallel [111]$ with the strain values close to the ones measured for planar (111) $P_b$

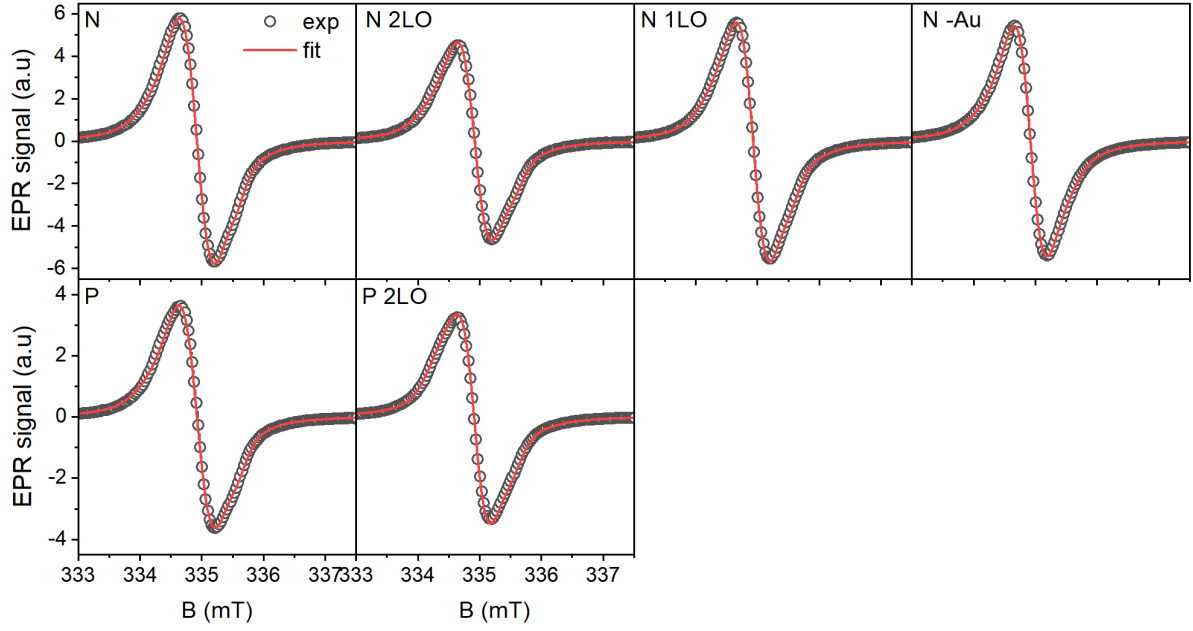

Figure S7: Electron paramagnetic resonance spectra of thermally oxidized P and N PSi NPs, after additional two-step (2LO) or one-step (1LO) liquid oxidation, or PSi NPs prepared by milling N PSi powder oxidized by gold etchant. The experimental data (dark lines) was fitted (red lines) as discussed in the text.

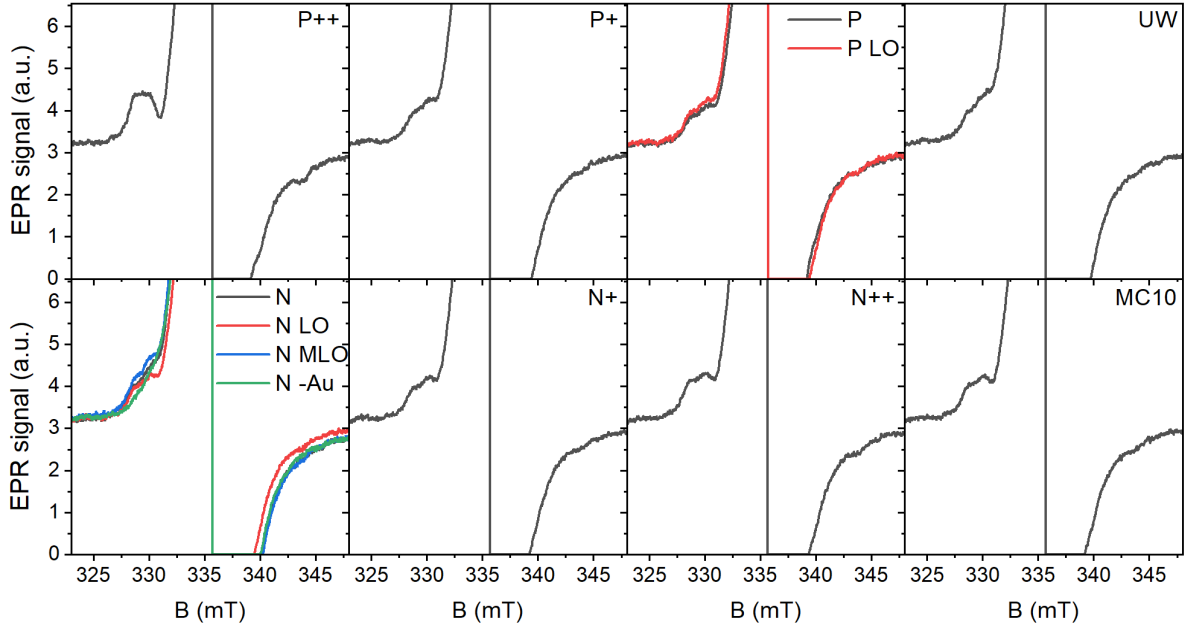

Figure S8: Electron paramagnetic resonance spectra of all the samples depicting peaks for hyperfine interaction of electron spin with the central $^{29}$Si nuclei. The superhyperfine interaction with the backbond $^{29}$Si nuclei is invisible due to large width of $P_b^{iso}$ spectra.

center[13]. The Gaussian and Lorentzian peak-to-peak linewidths for the $P_b$ are $\Delta B_{pp}^G = 0.045\,\text{mT}$ and $\Delta B_{pp}^L = 0.16\,\text{mT}$, respectively [14, 15] (Table S2). These linewidths closely match the values evaluated by Stesmans et al.[14, 15] during their study of dipolar interaction between (111) $P_b$ and its influence on the low-temperature EPR spectra. Indeed, they found $\Delta B_{pp}^L \approx 0.16\,\text{mT}$ for $[P_b] \approx 7 \cdot 10^{12}\,\text{cm}^{-2}$. With the weight decrease of the $P_b$ centers in LL-MACE samples, the $P_b$ fitting becomes less straightforward and the fitting parameters start to deviate from the ones for the planar $P_b$ centers. This is expected due to high structural irregularity of the samples' porous surfaces.

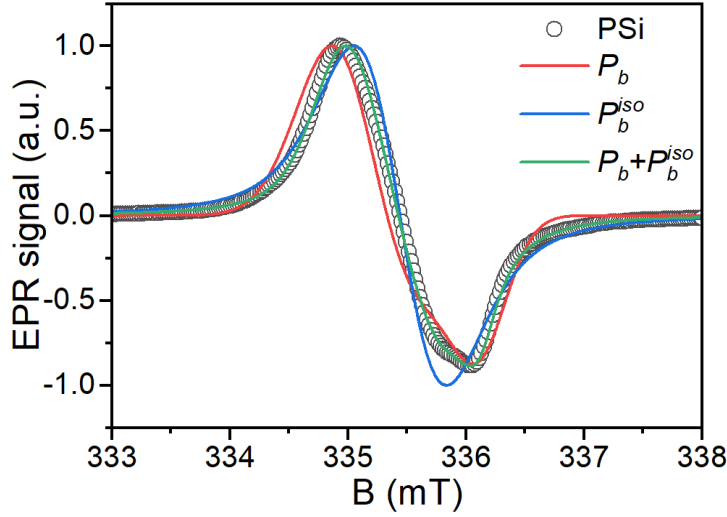

Figure S9: Fitting of the experimental EPR spectra of TOPSi sample (open dark circles) with powder pattern of (111) $P_b$ centers (red line), $P_b^{iso}$ centers (blue line), and combination of the two centers (green line). Obtained weights for $P_b$ and $P_b^{iso}$ were 0.37 and 0.63, respectively. The $g$ strain for $P_b$ defect was $\Delta g_\perp = 0.00296$ and $g_\parallel = 0.0005$.

Compared to $P_b$ centers, $P_b^{iso}$ was described by phenomenological Gaussian $\Delta B_{pp}^G$ and Lorentzian $\Delta B_{pp}^L$ peak-to-peak linewidths. They were found to be of the same order of magnitude in the range of (0.1–0.5) mT, which constituted the total linewidth of $\approx 0.6$ mT and corresponded to the average linewidth of randomly oriented $P_b$ centers found on different crystalline planes[13, 16]. The Gaussian part in the linewidth was assumed to come from the $g$-factor strain that was not included as an additional fitting parameter for $P_b^{iso}$, while the Lorentzian part showed even larger values than the ones that take into account dipolar interaction induced broadening [14, 15] (Table S2). We were not able to find a feasible explanation for such a large broadening from the porous Si literature. One possible explanation is the clustering of $P_b$ centers due to the irregularity of the porous surface in a similar way it was demonstrated by LOD-EPR for partially amorphous Si sample[17]. Overall, it is reasonable to assume the presence of dipolar interaction in our samples with similar $[P_b]$ or higher concentrations compared to Stesmans et al.[14, 15] (Table S1).

The results of EasySpin simulation of the EPR spectra were then used to calculate the EPR spectra at the DNP conditions. For this, the best fit models for each sample were fed to EasySpin to simulate powder pattern structure representing the high-field frequency-swept experimental conditions. All the high field spectra looked similar and, therefore, only the ones for P sample are presented (Fig. S10). Similar to the X-band EPR, the high-field spectra show the strong central peak and the two weak satellite peaks, which correspond to the $P_b$ centers located on the central $^{28}$Si and $^{29}$Si atoms, respectively. We highlight the slight shift towards higher frequency of the strongest EPR peak compared to zero DNP frequency (Fig. S3.2) possibly due to slightly lower experimental magnetic field strength than 6.7 T used for the EPR simulation.

As a final remark, conduction band electrons with $g = 1.9995$ have been observed in heavily doped $n$-type porous Si and $p$-type porous Si under illumination at 4.2 K [18, 19]. It is, however, not possible to identify conduction band electrons in our samples. Although the fitting of EPR does give the $g$-factor close to 1.9995 (Fig. S11), the peak width is at least three times larger than 0.1 mT measured by Young et al.[18, 19] Thus, it is concluded that neither conduction band electrons nor the phosphorus donor electrons could be identified.

Table S2: EasySpin simulation results of the experimental EPR data. The fitting was performed according to the mixture of anisotropic $P_b^{(111)}$ and isotropic $P_b^{iso}$ centers. Anisotropic centers were fitted with $g_\parallel = 2.00185$, $A_\parallel = 230 \pm 25$ MHz and $g_\perp = 2.0081$, $A_\perp = 420 \pm 15$ MHz with $g$-strain[8] and Lorentzian line broadening[15] to include $g$-factor stain and $P_b$ dipolar interaction, respectively. $P_b^{iso}$ centers were fitted with Voigtian lineshape to include homogeneous and inhomogeneous line broadening effects due to strain and dipolar interaction.

| Sample | $P_b^{(111)}$ | | | | $P_b^{iso}$ | | | |
|---|---|---|---|---|---|---|---|---|
| | $g_\perp \cdot 10^{-3}$ strain | $g_\parallel \cdot 10^{-3}$ strain | $\Delta B_{pp}^L$, mT | weight, % | $g$-factor | $\Delta B_{pp}^G$, mT | $\Delta B_{pp}^L$, mT | weight, % |
| PSi   | 3.0 | 0.1 | 0.0012 | 13 | 2.0055 | 0.69 | 0.28 | 87 |
| P++   | 5.4 | 1.3 | 0.0041 | 11 | 2.0054 | 0.46 | 0.37 | 89 |
| P+    | 3.7 | 1.3 | 0.0001 | 9  | 2.0054 | 0.13 | 0.49 | 91 |
| P     | 5.9 | 1.7 | 0.0009 | 11 | 2.0055 | 0.24 | 0.45 | 89 |
| P 2LO | 3.1 | 0.1 | 0.0014 | 11 | 2.0054 | 0.00 | 0.49 | 89 |
| UW    | 5.9 | 1.6 | 0.0014 | 8  | 2.0055 | 0.21 | 0.45 | 92 |
| N     | 5.6 | 1.4 | 0.0006 | 8  | 2.0054 | 0.23 | 0.45 | 92 |
| N 2LO | 5.2 | 1.8 | 0.0017 | 9  | 2.0055 | 0.17 | 0.47 | 91 |
| N 1LO | 4.3 | 1.6 | 0.0018 | 7  | 2.0054 | 0.12 | 0.48 | 93 |
| N -Au | 5.5 | 1.9 | 0.0027 | 4  | 2.0054 | 0.13 | 0.47 | 96 |
| N+    | 3.6 | 0.1 | 0.0003 | 13 | 2.0054 | 0.19 | 0.47 | 87 |
| N++   | 4.7 | 1.6 | 0.0009 | 12 | 2.0055 | 0.26 | 0.45 | 88 |
| MC10  | 5.2 | 0.6 | 0.0022 | 15 | 2.0056 | 0.47 | 0.37 | 85 |

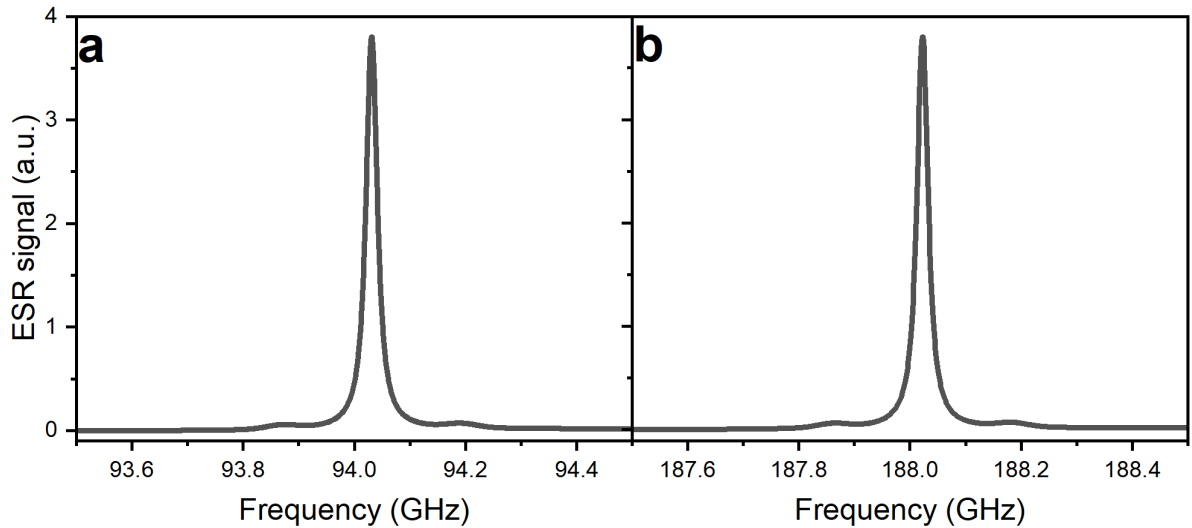

Figure S10: Simulation of EPR spectra for P sample at 3.3451 T (a) and at 6.6919 T (b). Simulation has been done using EasySpin after the best (111) $P_b$ and $P_b^{iso}$ system was obtained from fitting of the experimental EPR data (Fig. S6). The magnetic field values were calculated from the spectrometer frequency.

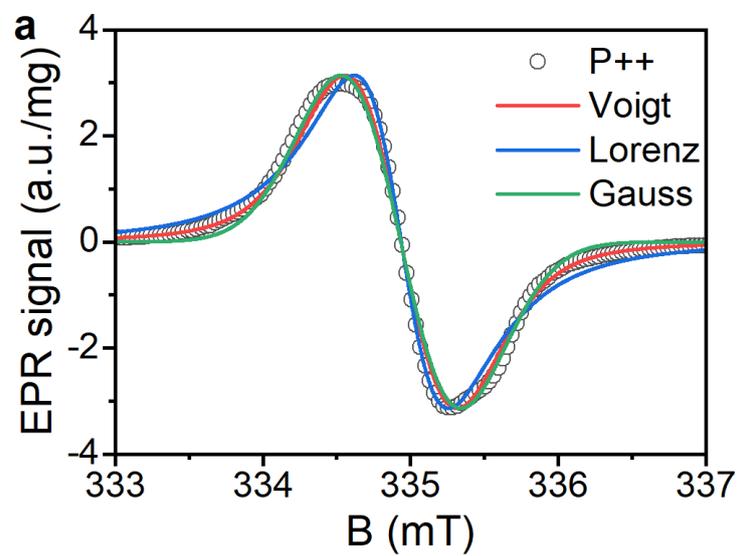

Figure S11: Fitting of the EPR spectrum of P++ PSi NPs with two pseudo-Voigt lines (red line). The first and second Voigt lines give $g$-factors of 2.0054 and 1.9991, respectively. The EPR spectrum was obtained by integrating the corresponding EPR spectrum. The FWHM for $g = 1.9991$ peak is 0.4 mT.

## S2.5 $P_b$ centers in silicon

$P_b$ centers have been widely investigated by EPR both on atomically flat specific crystalline planes and in porous Si because of their importance in metal-oxide-semiconductor devices[20, 21] and to elucidate their influence on photoluminescent properties of porous Si[7, 22–24]. Different types of $P_b$ centers have been identified with some ambiguity in their naming. Following Brower[9], a $P_b$ center is a localized dangling bond of a Si atom backbonded to three Si atoms at the Si/SiO$_2$ interface on the (111) crystalline plane (Si$_3 \equiv$ Si·). We denote this center as $P_b^{(111)}$ for clarity. $P_{b0}$ and $P_{b1}$ centers are two distinct centers on the oxidized (100) plane with axial and rhombic symmetries[25], respectively. Due to structural similarity of $P_b^{(111)}$ and $P_{b0}$ centers, quite commonly these centers are interchangeably denoted as $P_b$ or $P_{b0}$ in the published literature *e.g.*, in Refs. [13, 20].

The trigonal symmetry of $P_b^{(111)}$ centers dictates the trigonal symmetry of the $g$-factor and HF tensor resulting in $g_\parallel = 2.00185$, $A_\parallel = 230 \pm 25$ MHz and $g_\perp = 2.0081$, $A_\perp = 420 \pm 15$ MHz as determined by angular resolved EPR[13]. The line shape has been found to vary depending on the orientation of the [111] crystalline direction with respect to the external magnetic field. The EPR line has Lorentzian shape with a peak-to-peak line width $\Delta B_{pp}^L = 0.22 \pm 0.015$ mT for $\mathbf{B} \parallel [111]$ and Gaussian shape with $\Delta B_{pp}^G = 0.82 \pm 0.05$ mT for $\mathbf{B} \perp [111]$ indicating the presence of $g_\perp$ strain with $\Delta g \approx 0.0045$. Hyperfine (HF) interaction with the nearest (backbonded) neighbor $^{29}$Si nuclei has also been resolved with a HF constant $A_{2n} = 41.5 \pm 0.5$ MHz. Note, that in the literature $A_{2n}$ is denoted as superhyperfine interaction in some cases[26].

The analysis of the HF tensor in terms of one-electron molecular orbitals[9] gave a 12% $s$-like and 88% $p$-like wave function character with around 80% of the total spin-density localized on the central Si· atom. The spin density distribution together with the large HF interaction results in a large Fermi-contact interaction compared to the dipolar part of the HF interaction: The dipolar HF interaction is up to $\sim 65$ MHz and $\sim 1.5$ MHz for the central and the nearest neighbor nuclei, respectively. Further from the nearest neighbors, the HF interaction is supposed to be governed by the dipolar part, which decreases rapidly with distance. For a $^{29}$Si at a distance of two lattice constants away from the $P_b$ center, the estimated dipolar HF interaction is $\sim 10$ kHz.

In oxidized porous Si films, X-band EPR (9 GHz) performed at room temperature[11, 20, 22, 24, 27, 28] and (4–20) K [16, 18, 29] found two general classes of $P_b$ centers depending on the oxidation conditions. The first class has been observed in both (100)- and (111) crystallographic planes of porous Si oxidized under controlled oxygen, hydrogen and moisture content. It is reminiscent of $P_b^{(111)}$, $P_{b0}$, $P_{b1}$ centers found on the corresponding oxidized crystalline planes[9, 13–15, 20, 21, 30]. Among these, the dominant center is the $P_b^{(111)}$ due to the simultaneous presence of four possible interfaces (111), (1$\bar{1}$1), (1$\bar{1}\bar{1}$), ($\bar{1}\bar{1}$1)[11]. This center exhibits similar axial symmetry, $g$-factors and HF constants, $s$ and $p$ spin densities as the $P_b^{(111)}$ center on the corresponding crystalline plane[11, 27]. Highlighted difficulties to detect (100) $P_{b0}$ and $P_{b1}$ centers[11, 31, 32] have been attributed to the dominance of $P_b^{(111)}$ center and to the reconstruction of (100) $P_{b0}$ centers in porous Si[24] (reconstruction is not efficient on a planar (100) Si surface[22]). Therefore, the measured EPR spectra in controllably oxidized porous Si closely follows the features of crystalline Si samples including angular dependence of $g$-factors and line widths[11]. Furthermore, EPR spectra from oxidized porous Si are comparable with the spectra obtained at K- (24 GHz) and Q-band (35 GHz) at room and (1.4–20) K[9, 13–15] on the planar Si surfaces if spectra were acquired under non-saturating conditions. At liquid He temperatures, low MW powers are required due to strong saturability and long $T_{1e}$ times up to approximately 80 ms[33].

The second class of $P_b$ centers in porous Si develops under uncontrolled native[22] or thermal oxidation in air[18, 23, 28], and during thermal annealing[7, 27, 28]. This $P_b^{iso}$ center is characterized by isotropic $g = 2.0055$. Despite the $g$-factor is isotropic, the linewidth can retain anisotropy which follows the trigonal structure similar to the $P_b^{(111)}$ center with the smallest value of 0.6 mT for $B \parallel [100]$ and the largest value of for 1.2 mT $B \parallel [111]$[16, 22]. Compared to $P_b^{iso}$, the $P_b$ centers formed on a corresponding Si plane thus have much narrower line widths. Electron spin relaxation times of $P_b^{iso}$ have received less attention and, therefore, are compared to commercial samples previously investigated for DNP[34], although they might have substantial amount of paramagnetic amorphous Si centers[35]: The measured $T_{1e}^{slow}$ and $T_{2e}$ show rather similar values to $P_b^{(111)}$ equal to (10–70) ms and (0.1–2) µs at 10 K, respectively.

The measured EPR spectra of our samples are represented by the relatively broad lines with noticeable asymmetry (Fig. 1f and S6, Suppl. Inf.). The total surface densities of the $P_b$ centers are in the range

of $(1.8–6.8) \cdot 10^{12}$ cm$^{-2}$ (Fig. 1g) leading to a $1.9 - 3.7$ nm average distance between them. According to EasySpin[10] fitting (Section S2.4, Suppl. Inf.), $P_b$ centers in the oxidized LL-MACE samples are represented by high number of $P_b^{\text{iso}}$ and few $P_b^{(111)}$ centers. For planar $P_b^{(111)}$ centers, the fitted line broadening values indicate the presence of $g$-strain typical for such centers ($\Delta g \approx 0.0047$)[13] . This strain contributes to the Gaussian peak-to-peak line width $\Delta B_{pp}^G \approx 0.8$ mT, similar to planar $P_b^{(111)}$ strain. The Lorentzian part below $3\,\mu$T is much lower than obtained by Stesmans and Gorp[14, 15] ($\Delta B_{pp}^L \approx 0.16$ mT, Table S2, Suppl. Inf.), indicating the possible isolation of the $P_b^{(111)}$ centers.

$P_b^{\text{iso}}$ centers are fitted with a phenomenological Voigtian lineshape, which gave $\Delta B_{pp}^G = (0.12–0.47)$ mT and $\Delta B_{pp}^L = (0.37–0.49)$ mT. The smaller $\Delta B_{pp}^G$ is possibly due to less strain for the $P_b^{\text{iso}}$ than for the $P_b^{(111)}$, while the high value $\Delta B_{pp}^L$ could indicate stronger dipolar coupling compared to $P_b^{(111)}$. $\Delta B_{pp}^L(P_b^{\text{iso}})$ corresponds to $T_2(P_b^{\text{iso}}) \approx 30$ ns. It is possible that the large $\Delta B_{pp}^L$ for the $P_b^{\text{iso}}$ may indicate the clustering of $P_b^{\text{iso}}$ centers with orders of magnitude faster electron spin-lattice relaxation rate than for the standalone centers. The investigation of clusters requires further (pulsed) EPR studies to detect the spin-lattice relaxation.

HF interaction with the central $^{29}$Si atom was also clearly identified in our X-band EPR measurements (Fig. S8, Suppl. Inf.). The HF constants are in the range of $(11.6–15.4)$ mT or $(325–431)$ MHz and correspond to the typical values of the (111) $P_b$ center with $A_\parallel = 210$ MHz and $A_\perp = 417$ MHz[13]. HF coupling with the nearest neighbor (backbonded) $^{29}$Si nuclei was not observed due to the large broadening of $P_b^{\text{iso}}$ but could be assumed to be present with $A_{2n} \approx 42$ MHz[9]. On the other hand, it could be possible that the HF interaction with backbonded $^{29}$Si is diminished in our samples due to the backbond oxidation[1] of the central $^{28}$Si atom. In the case of backbond oxidation, HF interaction with a distant $^{29}$Si nuclei can be assumed to be of a purely dipolar nature and scale as $r^{-3}$ with $r$ the distance from a $P_b$ center.

The average distance between the $P_b$ centers assuming their uniform distribution is $d_{\text{ee}} = (1.9–3.7)$ nm deduced from their amount per surface area. This distance gives the estimated dipolar coupling $D_{\text{ee}}$ on the order of $D_{\text{ee}} = (1.0–7.4)$ MHz, which is about $(1 - 10)$ times lower than the homogeneous line broadening calculated from $\Delta B_{pp}^L$ for the $P_b^{\text{iso}}$ centers (Tbl. S2, Suppl. Inf.). Such a discrepancy between $D_{\text{ee}}$ and $\Delta B_{pp}^L$ may indicate clustering of the $P_b$ centers on the ridges and edges of the irregular pore walls and pore openings. Nevertheless, $\Delta B_{pp}^L$ clearly correlates with the $D_{\text{ee}}$ and surface density of $P_b$ centers with the correlation coefficient of $\sim 0.63$ demonstrating a consistent increase of dipolar interaction with the decrease of their mutual distance. For the $P_b^{(111)}$ centers, $\Delta B_{pp}^L \approx (0.001–0.004)$ mT, which is more than two orders of magnitude smaller than $\Delta B_{pp}^L$ for $P_b^{\text{iso}}$ centers, might indicate a relative isolation of the $P_b^{(111)}$ centers from other $P_b$ centers.

## S3 Dynamic nuclear polarization

### S3.1 Thermal polarization buildup

The polarization enhancements and absolute polarizations were calculated by integrating the pseudo-Voigt fits of FFT-processed FID data. The integrated values were then divided by the thermal polarization signal processed the same way and taken after 72 h of polarization inside a polarizer with microwave radiation switched off (Fig. S12).

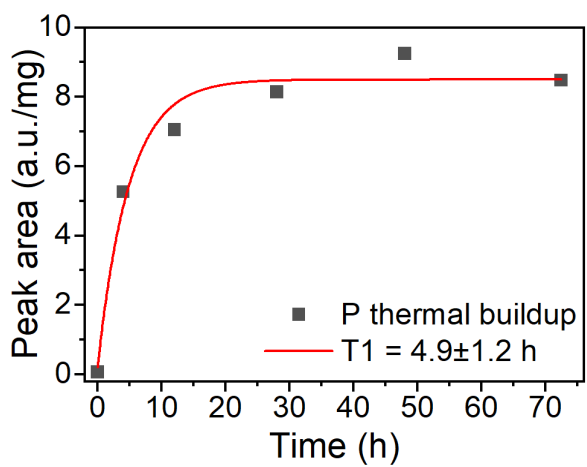

Figure S12: Thermal polarization buildup for the P sample at 6.7 T and 1.4 K.

## S3.2 DNP profiles

The normalized DNP profiles (sweep spectra) at 6.7 T and 1.4 K for thermally oxidized PSi NPs are depicted in Fig. S13. There are minor differences between the Si types in the asymmetry of positive and negative peak values. This asymmetry was attributed to the slight difference of surface area induced by increased remote etching for highly doped Si[6], and the corresponding possible change in the structure of $P_b$ centers.

In all the spectra, however, the absolute value of the negative peak is smaller than the positive of the peak. The main reason was the non-uniform output power dependence of the microwave generator, which decreased for higher frequencies. When the microwave generator was upgraded, the typical shape of the sweep curve for DNP with $P_b$ centers was observed (Fig. S14, Fig. S15 N 1LO and N -Au samples). Nevertheless, most of the data was obtained with the old microwave generator, and, therefore, the positive peak was selected to study buildup, in agreement with our data at 3.4 T. Almost complete absence of the negative peak for N++ PSi NPs can at least partially be attributed to its generally low polarization combined with the decrease of MW power.

Similar to Fig. 6, for 3.35 T (1.6 K), the extrema of the DNP profiles with MW modulation correspond to the $m_I = \pm 1/2$ hyperfine lines of the simulated EPR spectra (Fig. S16.

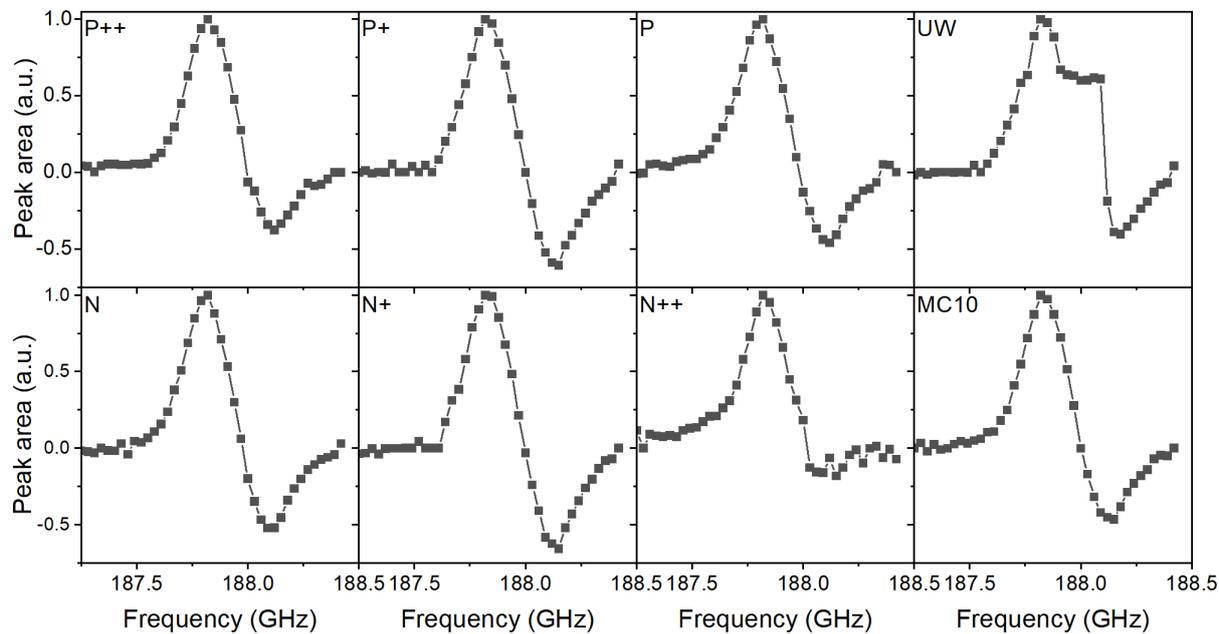

Figure S13: Microwave sweep spectra for thermally oxidized samples of different doping types at 6.7 T and 1.4 K. Each point of a spectrum includes microwave modulation with a frequency of 3 kHz and bandwidth of 150 MHz[34, 36]. Significant decrease of amplitude of the negative peak can be due to decrease of the microwave power with the increase of frequency.

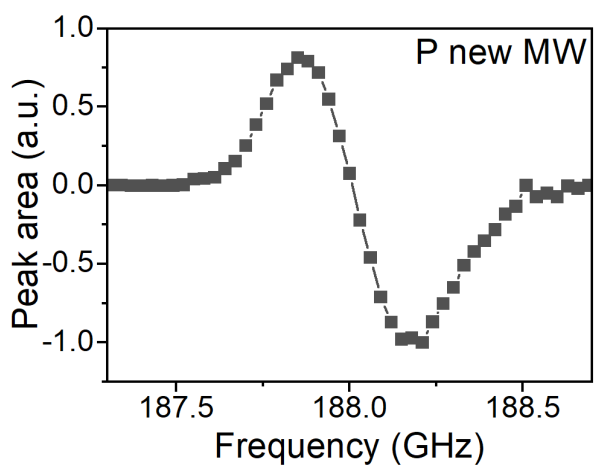

Figure S14: Microwave sweep spectrum of P PSi NPs at 6.7 T and 1.4 K after replacement of microwave generator. The spectrum shows typical asymmetry for DNP of Si using $P_b$ centers[34, 36, 37].

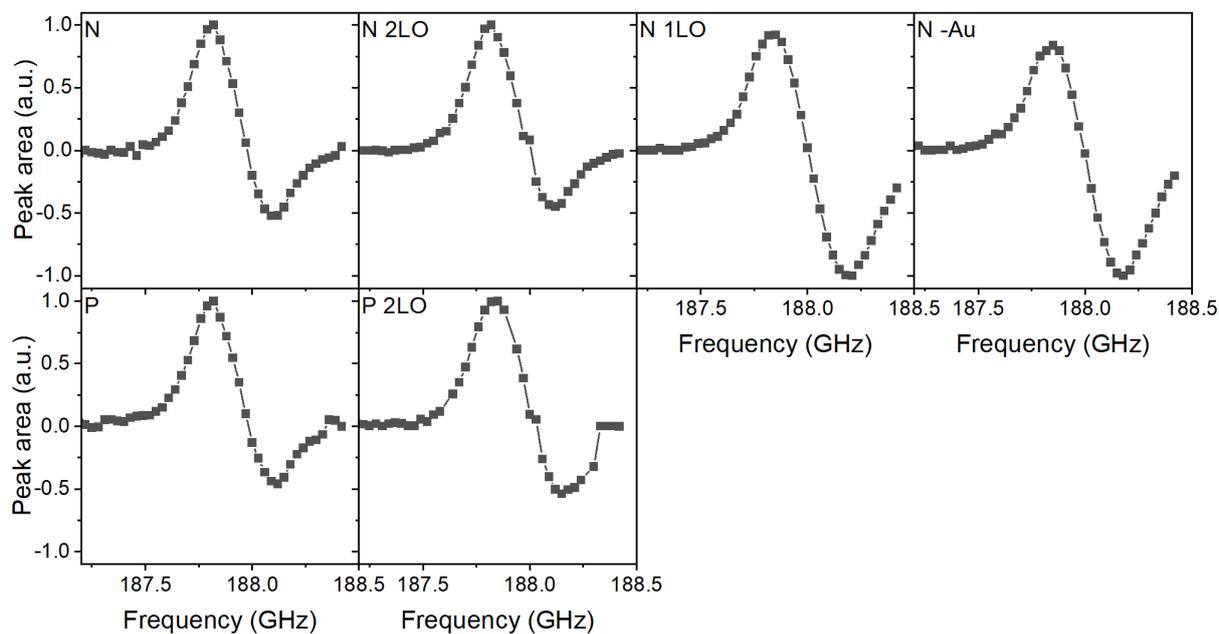

Figure S15: Microwave sweep spectra for differently oxidized N and P PSi NPs at 6.7 T and 1.4 K. Each point of a spectrum includes microwave modulation with a frequency of 3 kHz and bandwidth of 300 MHz[34, 36].

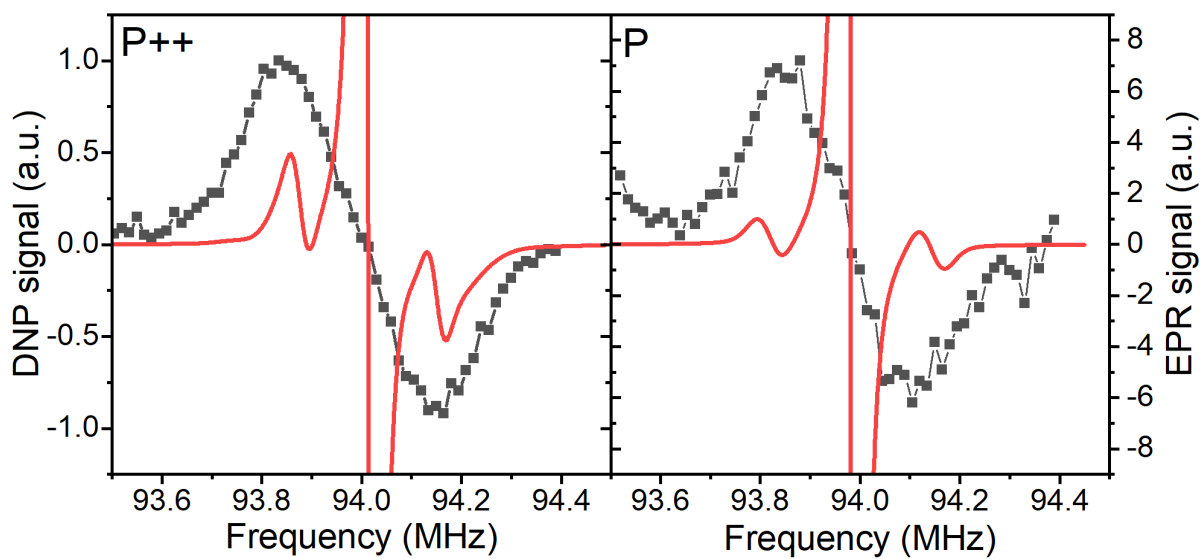

Figure S16: The sweep spectra overlapped with the simulated EPR spectra for P++ and P PSi NPs at 3.35 T and 1.6 K. The sweep was recorded with 100 MHz frequency modulation, 1 kHz sweep rate and 80 mW power.

## S3.3 Dynamic nuclear polarization buildup and decay data

Fig. S17 compares different pulse delays (1 and 30 min) and demonstrates that the perturbations be the monitoring RF pulses can be accurately corrected for [38]. Figs. S18 and S19 compare the buildups for the different samples at 3.34 T (3.4 K) and 7 T (3.4 K), respectively. The effects of different oxidation on the build-ups are depicted in Figs. S20 and S21.

Since most of the experiments presented in the current work were recorded with the full MW power available at a given setup, it needs to be investigated if this has a strong influence on the DNP performance. At 7 T (3.4 K), reducing the MW power by a factor of ten (20 instead of 200 mW output power) leads to a minor increase of the steady-state polarization (enhancement) at the expense of a longer buildup time (Fig. S22). This is consistent with the discussion of relaxation enhancement by MW irradiation[39] in Sec. S3.4.

Figs. S19 and S24 compare the hyperpolarization decays at 7 T (3.4 K) and 6.7 T (1.4 K) with a summary of the fitted decay times given in Tab. S3. The decay data is later used to calculated the decay rate constants for the rate-equation model (Secion S3.4). The temperature decrease from 3.4 K to 1.4 K causes the increase in thermal electron polarization (88 % and 99.7 %, respectively) and leads to a drastic reduction in paramagnetic relaxation of nuclei. Relaxation is reduced since virtually all electrons are polarized such that triple spin flips can no longer relax the nuclear polarization[40].

Fig. S25 compares the room temperature decays (at 7 T) of the three most promising samples (P, UW, N). The samples were polarized for around 20 h at 3.34 T (3.4 K) and transferred to the nearby 7 T magnet. Since both magnets were unshielded, we avoided using a strong permanent magnet carrier device for the shuttling which might result in different relaxation behaviors during the transfer for different samples. Furthermore, as we learned later, the 7 T set-up at the time of these measurements had problems with a poor electrical contact, eventually causing increased noise floors in certain measurements.

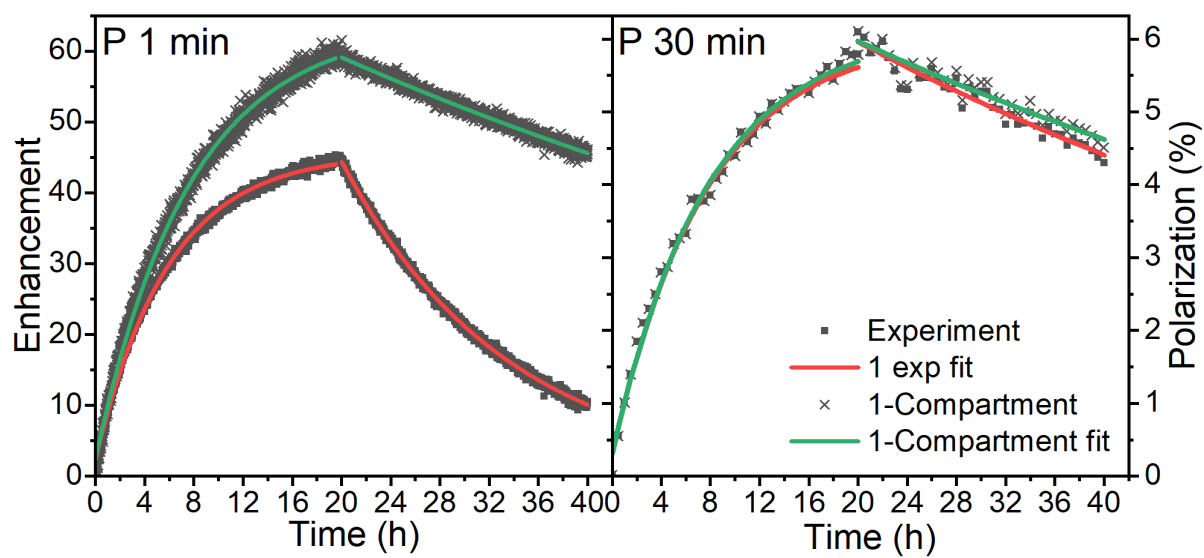

Figure S17: Evaluation of the RF pulse correction and one-compartment model using different sampling rate for the P PSi NPs sample: NMR measurement each 1 min (left) and each 30 min (right) with flip angle $\sim 3°$. The RF pulse correction accurately predicts the polarization buildup and decay for the high sampling rate compared to the low sampling rate provided that NMR flip angle was correctly estimated.

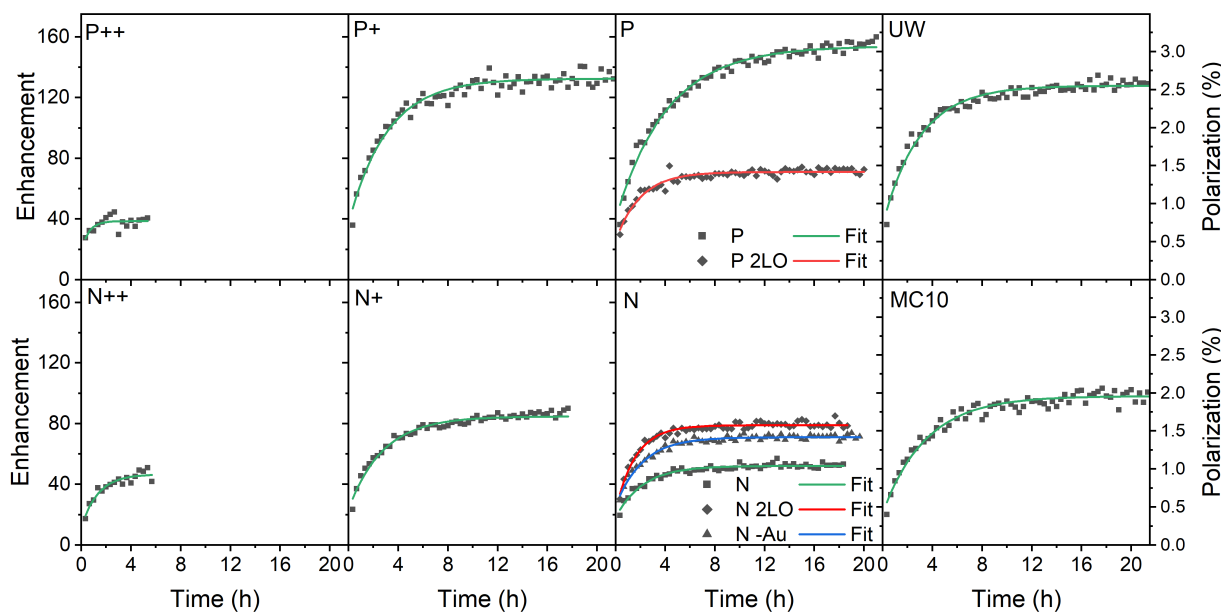

Figure S18: Dynamic nuclear polarization of thermally oxidized PSi NPs with different dopants (dark squares) and single exponential fit with RF pulse correction according to the one-compartment model[38] (green lines). Magnetic field is 3.34 T, temperature is 3.4 K, microwave frequency is 93.83 GHz with around 200 MHz modulation, microwave power is 200 mW.

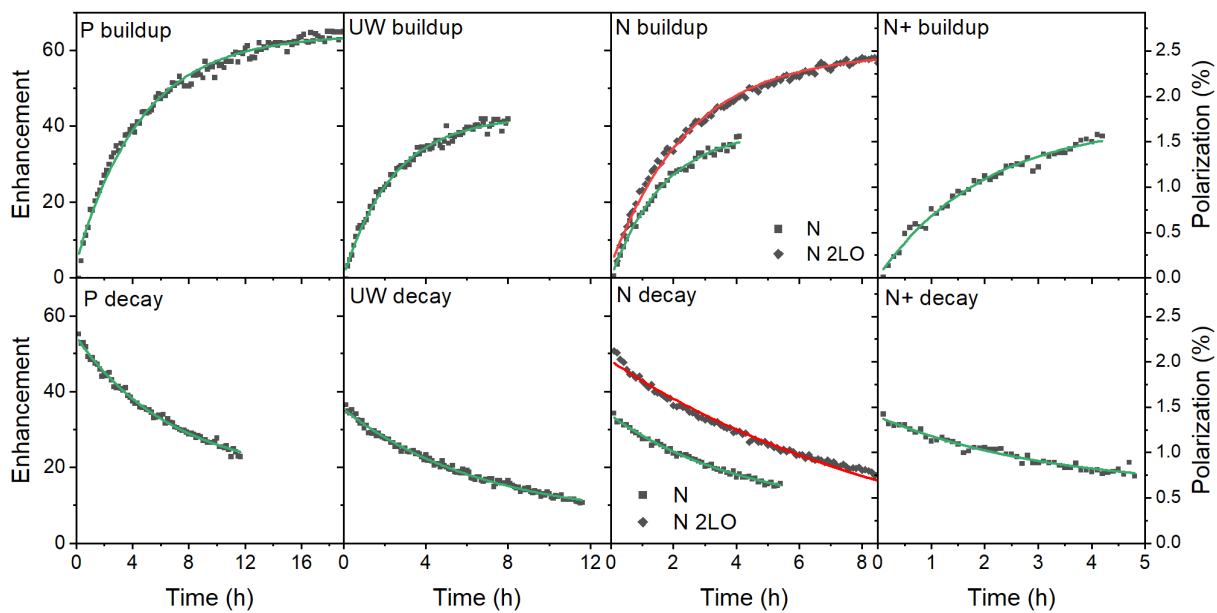

Figure S19: Dynamic nuclear polarization buildup and polarization decay of thermally oxidized PSi NPs with different dopants (dark squares) and single exponential fit with RF pulse correction according to the one-compartment model[38] (green lines). Magnetic field is 7 T, temperature is 3.4 K. The buildup microwave frequency is 197.025 GHz with 300 MHz modulation, microwave power is 200 mW.

17

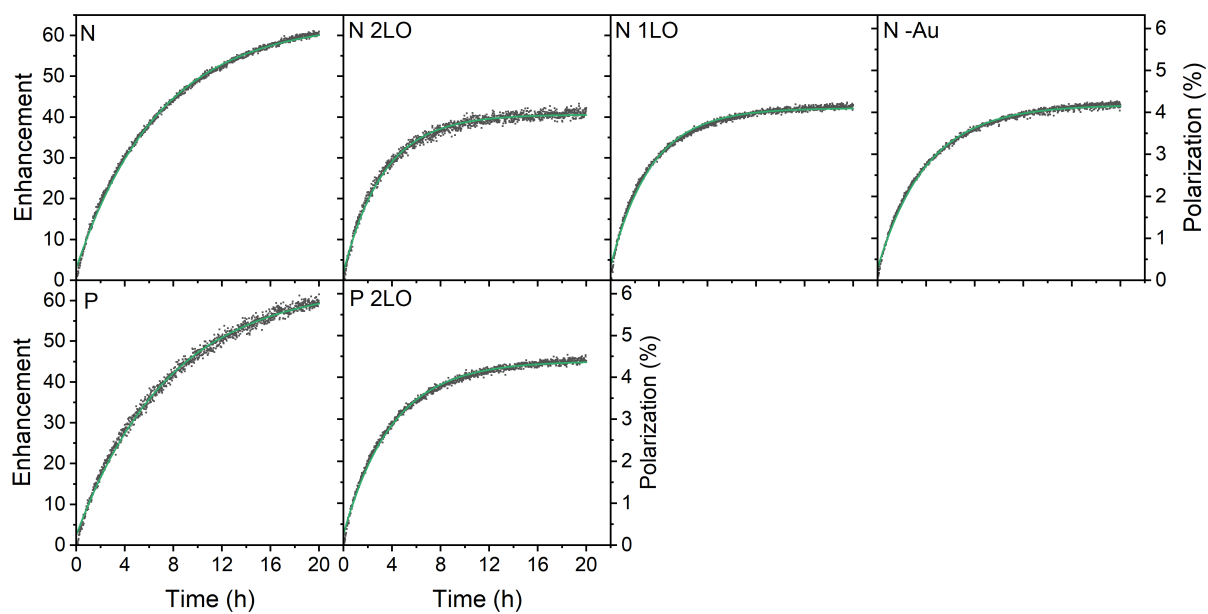

Figure S20: Dynamic nuclear polarization of differently oxidized N and P PSi NPs (dark squares) and single exponential fit with RF pulse correction according to the one-compartment model[38] (green lines). Magnetic field is 6.7 T, temperature is 1.4 K, microwave frequency is 187.82 GHz with 200 MHz modulation, microwave power is 30 mW.

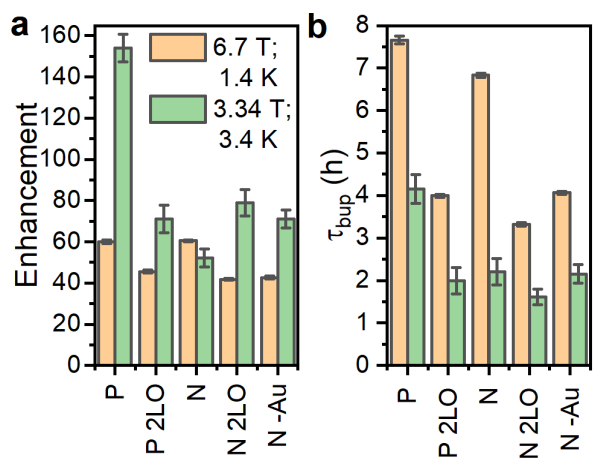

Figure S21: Oxidation induced change of the enhancement (a) and buildup time (b) for the P and N samples at 6.7 T (1.4 K) (orange bars) and 3.34 T (3.4 K) (green bars). The 2LO demotes the two-step liquid oxidation applied after the thermal oxidation either to P or to N sample (Section S1.1). The N -Au sample is the N sample with dissolved Au NPs after LL-MACE, for which the dissolution medium performed the surface oxidation (no thermal oxidation applied).

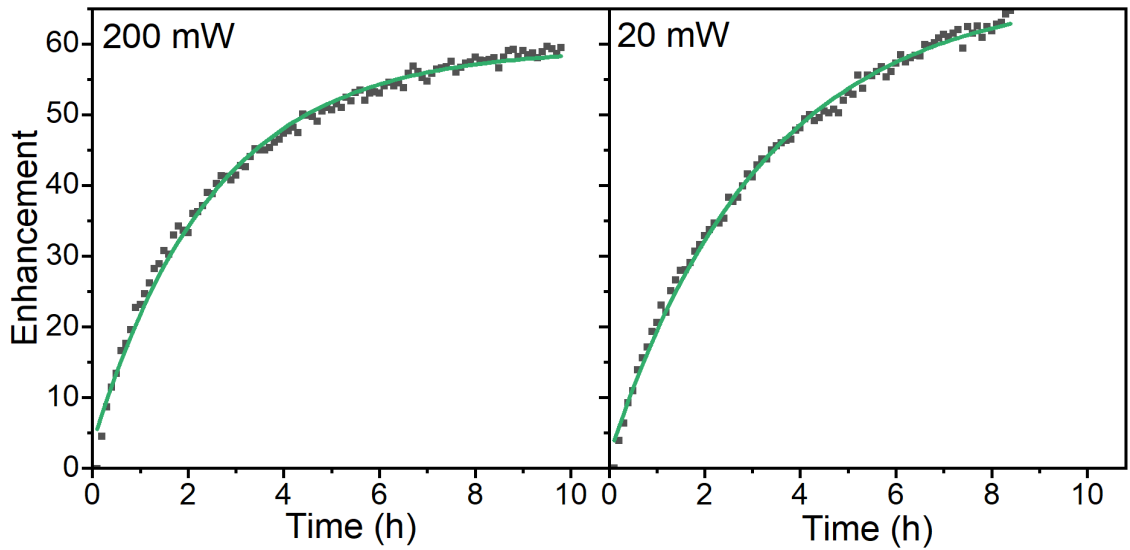

Figure S22: Microwave power dependence of the $^{29}$Si polarization buildup at 7 T (3.4 K). The microwave powers were 200 mW (left) and 20 mW (right). The N 2LO sample was used. The DNP buildup times calculated using the one-compartment model were $2.6 \pm 0.1$ h and $3.4 \pm 0.1$ h for the 200 mW and 20 mW power, respectively. $k_W = 10.7 \cdot 10^{-3}$ h$^{-1}$ and $8.9 \cdot 10^{-3}$ h$^{-1}$; $k_R^{bup} = 0.44$ h$^{-1}$ and $0.36$ h$^{-1}$ for 200 mW and 20 mW, respectively.

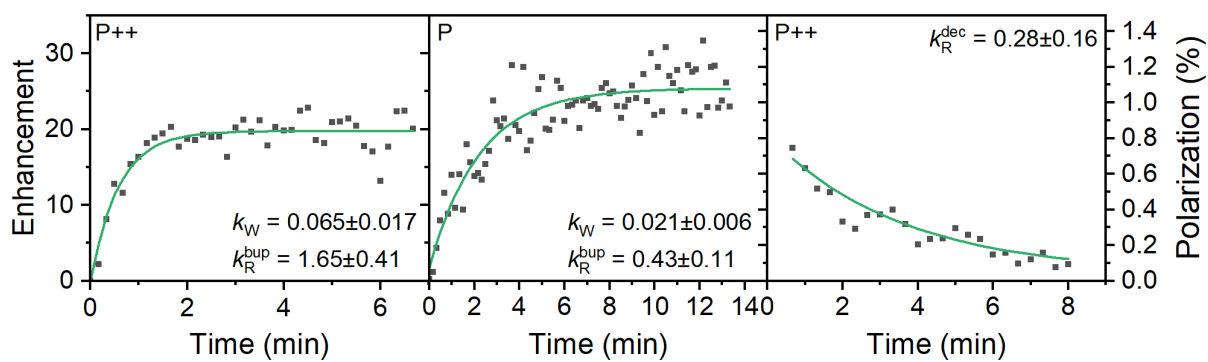

Figure S23: The polarization buildups for P++ and P PSi NPs and the polarization for P++ sample. The buildup was performed with 100 MHz frequency modulation, 1 kHz sweep rate and 80 mW power. The calculated rate constants from the one-compartment model are depicted in the graphs.

Table S3: Relaxation time of the selected PSi NPs at various DNP conditions and switched off microwave radiation.

| Abbreviation | $\tau_{dec}$, h | |
| --- | --- | --- |
| | 6.7 T (1.4 K) | 7 T (3.4 K) |
| P++ | $17.5 \pm 0.8$ | - |
| P+ | $21.8 \pm 0.8$ | - |
| P | $76.9 \pm 17.9$ | $7.4 \pm 0.2$ |
| P 2LO | $13.6 \pm 3.3$ | - |
| UW | $20.7 \pm 2.4$ | $6.8 \pm 0.1$ |
| N | $79.5 \pm 48.2$ | $4.2 \pm 0.2$ |
| N 2LO | $12.3 \pm 0.4$ | - |
| N -Au | - | $3.7 \pm 0.4$ |
| N+ | $32.9 \pm 2.6$ | - |
| N++ | $20.0 \pm 7.5$ | - |
| MC10 | $24.8 \pm 3.1$ | - |

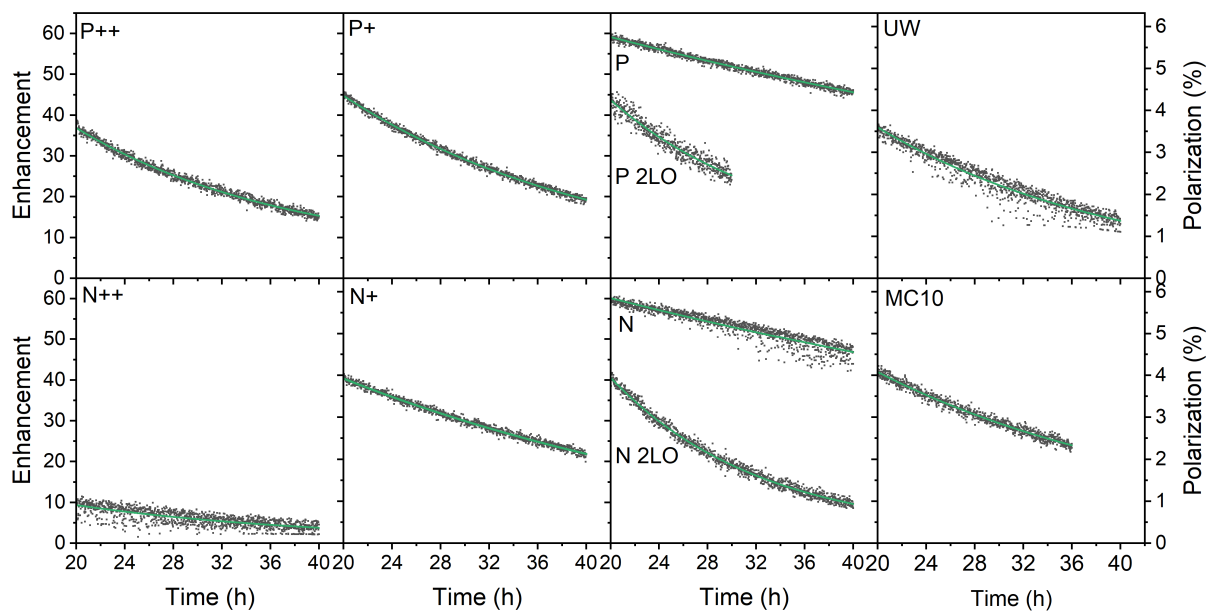

Figure S24: Relaxation of the $^{29}$Si polarization for different PSi NPs (dark squares) at 6.7 T and 1.4 K. Single exponential fits (green lines) with RF pulse correction are according to the one-compartment model.

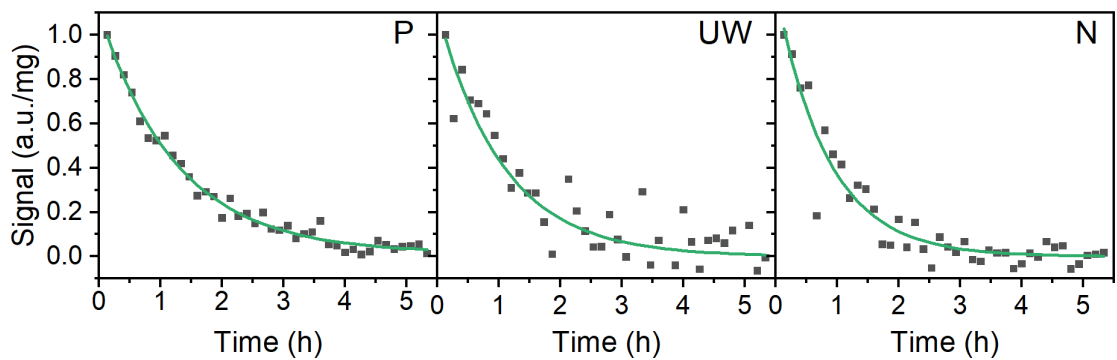

Figure S25: Relaxation of the $^{29}$Si polarization for different PSi NPs (dark squares) at 7 T and room temperature (300 K). The decay times are listed in Table 3 in the main text.

## S3.4 Quantitative analysis of DNP injection and relaxation

The complex interplay between experimental conditions and structural properties of PSi NPs makes it hard to identify the important parameters that affect the hyperpolarization enhancement and buildup time. On the macroscopic level, it is possible to disentangle enhancements and buildup times using a rate-equation model provided the DNP process can be described with a single time constant[38]. Applying the model, for example, the limited benefit of lower temperatures (1.4 K instead of 3.4 K) can be traced back to a relaxation enhancement by MW irradiation[39].

In the rate-equation model the buildup time $\tau_{\text{bup}}$ consists of the two competing processes: the nuclear polarization injection rate constant, $k_{\text{W}}$, and the nuclear relaxation rate constant $k_{\text{R}}^{\text{bup}}$. Together with the thermal electron polarization $P_{0\text{e}}$, we can describe the buildup as[38]:

$$\frac{\mathrm{d}P_{\text{n}}(t)}{\mathrm{d}t} = (P_{0\text{e}} - P_{\text{n}})k_{\text{W}} - k_{\text{R}}^{\text{bup}} P_{\text{n}}, \tag{S1a}$$

$$\tau_{\text{bup}}^{-1} = k_{\text{W}} + k_{\text{R}}^{\text{bup}}, \tag{S1b}$$

$$P_{1\text{n}} = P_{0\text{e}} \frac{k_{\text{W}}}{k_{\text{W}} + k_{\text{R}}^{\text{bup}}}, \tag{S1c}$$

where $P_{1\text{n}}$ is the steady-state nuclear polarization reached by the end of the DNP process. Eqs. S1b, S1c can be rewritten to[38] $k_{\text{W}} = \tau_{\text{bup}}^{-1} P_{1\text{n}}/P_{0\text{e}}$ and $k_{\text{R}}^{\text{bup}} = \tau_{\text{bup}}^{-1}(1 - P_{1\text{n}}/P_{0\text{e}})$. Since the steady-state nuclear polarizations are low with respect to the electron thermal polarization ($P_{1\text{n}} \ll P_{0\text{e}}$), we can simplify Eqs. S1 to

$$\tau_{\text{bup}}^{-1} \stackrel{P_{1\text{n}} \ll P_{0\text{e}}}{\approx} k_{\text{R}}^{\text{bup}}, \tag{S2a}$$

$$P_{1\text{n}} = P_{0\text{e}} k_{\text{W}} \tau_{\text{bup}} \stackrel{P_{1\text{n}} \ll P_{0\text{e}}}{\approx} P_{0\text{e}} \frac{k_{\text{W}}}{k_{\text{R}}^{\text{bup}}}. \tag{S2b}$$

For the decays, the model is much simpler than for the buildups since due to the absence of a polarization injection, i.e. the decay time and relaxation rate constant during decay are thus simply related as $k_{\text{R}}^{\text{dec}} = \tau_{\text{dec}}^{-1}$. The decay rate constant $k_{\text{R}}^{\text{dec}}$ is then directly obtained from the single exponential fit of the polarization decay data (Fig. S24, Suppl. Inf.). For the buildup, rate constants are calculated from the maximum thermal electron polarization $P_{0\text{e}}$ at the DNP conditions using the fitted values of $\tau_{\text{bup}}$ and steady-state DNP polarization $P_{0\text{n}}$. In the following, we discuss the effects of experimental conditions and sample doping along the calculated rate constants.

We start by comparing the influence of experimental conditions on the buildup rate constants ($k_{\text{W}}$ and $k_{\text{R}}^{\text{bup}}$) for a limited set of samples with high enhancements (Fig. S26a and S23, Suppl. Inf.). We find the highest injection rates $k_{\text{W}}$ to be at 3.4 K and 3.34 T except for the N sample, which has the highest $k_{\text{W}}$ at 7 T. For $B_0 \approx 7$ T, $k_{\text{W}}$ are similar at 1.4 K and 3.4 K despite different MW powers available (see Methods). The single experiment with 20 mW microwave power at 7 T (3.4 K) gives similar results compared to the full MW power of 200 mW (Fig. S22, Suppl. Inf.). Contrary to $B_0 \approx 7$ T, decreasing the temperature at 3.35 T from 3.4 to 1.4 K severely reduces the enhancements despite similar $k_{\text{W}}$ compared to $\sim 7$ T. To explain such a large difference in enhancements we turn from $k_{\text{W}}$ to $k_{\text{R}}^{\text{bup}}$.

The relaxation rate constants during buildup $k_{\text{R}}^{\text{bup}}$ significantly differ between the experimental conditions and strongly affect the achieved enhancements and nuclear polarizations. $k_{\text{R}}^{\text{bup}}$ rates are typically several times lower at 1.4 K (6.7 T) than at 3.4 K (Fig. S26a) consistent with the decrease of electron spin lattice relaxation time[41]. Contrary to 6.7 T, at 3.35 T and 1.4 K $k_{\text{R}}^{\text{bup}}$ is the highest, which detrimentally affects the enhancements (Fig. S23, Suppl. Inf.). At 3.4 K, $k_{\text{R}}^{\text{bup}}$ is similar at 3.34 and 7 T.

The relaxation rate constants during the decay $k_{\text{R}}^{\text{dec}}$ conformed to the following trend: slowest at 1.4 K (6.7 T) (Fig. S26b), intermediate at 7 T (3.4 K) and fastest at 3.34 T (3.4 K) (Figs. S28, S27 and Tbl. S3). This trend is consistent with the gradual increase of electron spin polarization from $\sim 60\%$ at 3.34 T, 3.4 K up to $\sim 99.7\%$ at 6.7 T, 1.4 K. The similar relaxation reduction at temperatures below $\sim 2$ K at $\sim 7$ T has been observed in our setups for $^1$H glassy matrices[39] as well as for diamonds before (data not shown). These results are attributed to high electron polarization which strongly reduces thermal paramagnetic relaxation approximated by $1 - P_{0\text{e}}^2$[42].

The discrepancy between relaxation rate constants during buildup ($k_{\text{R}}^{\text{bup}}$) and decay ($k_{\text{R}}^{\text{dec}}$) is the most pronounced at 1.4 K. During the buildup, relaxation rates are five to ten times faster at 6.7 T (Fig. S26b)

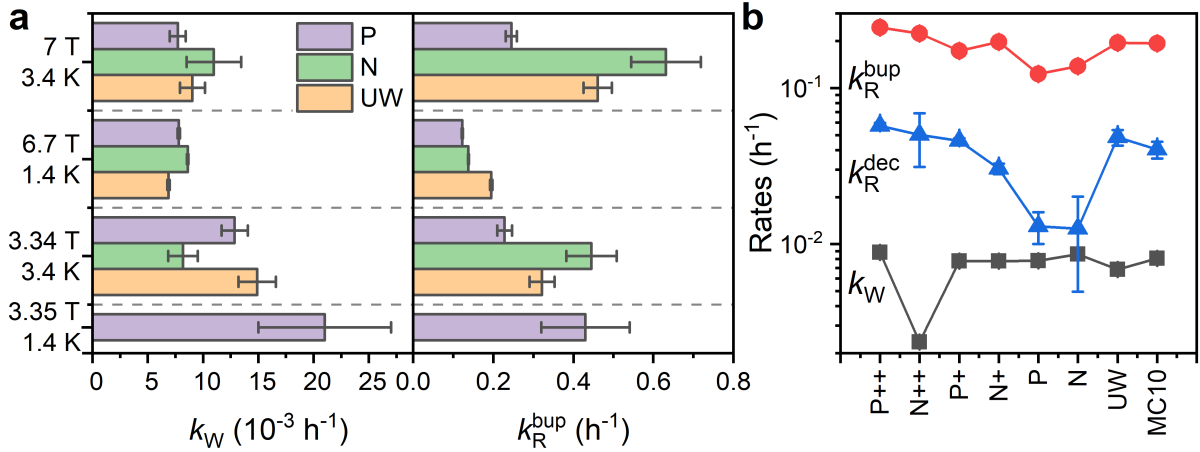

Figure S26: One-compartment model parameters calculated from the mono-exponential fitting of the experimental hyperpolarization buildups and decays. (a) Comparison of $k_W$ and $k_R^{bup}$ between three experimental conditions for the selected P, UW and N PSi NPs. Note the $10^{-3}$ scale for $k_W$. (b) Polarization buildup (dark squares and red circles) and decay (blue triangles) rates for the PSi NPs with different doping and oxidation. The data was acquired at 6.7 T, 1.4 K. Lines are a guide for the eye.

and more than an order of magnitude faster at 3.35 T (Fig. S23, Suppl. Inf.) compared to the MW-off decay. At 3.4 K and 7 T the $k_R^{dec}/k_R^{bup}$ ratio is in the range of $2.5 \pm 0.5$ while at 3.34 T the ratio is close to 1 (Sec. S3.3, Suppl. Inf.). The observed strong increase of the $k_R^{bup}$ relaxation rate either with the decrease of temperature at 3.35 T or with the decrease of magnetic field at 1.4 K is consistent with the results found in Ref. [39]. There, the relaxation enhancement during buildup is ascribed to the increase of the triple-spin flip rate induced by MW irradiation. At lower temperatures, on the one hand, paramagnetic relaxation of nuclei is reduced as a result of higher electron polarization ($1 - P_{0e}^2$). On the other hand, longer electronic relaxation times increase the saturation of the EPR line causing an increased relaxation by MW irradiation. With the increase of magnetic fields, the anisotropic EPR line is broadened, which reduces its spectral density and, consequently, the number of electron spin pairs that could cause an efficient three-spin nuclear paramagnetic relaxation. Thus, the higher relaxation enhancement at lower temperatures and lower fields increases $k_R^{bup}$ and results in the lowest enhancements observed at 3.35 T (1.4 K).

Converting the analyzed rate constants back to the measured DNP parameters, it is now possible to trace the influence of experimental conditions on the buildup time $\tau_{bup}$ and enhancements (or nuclear polarization $P_{1n}$). The injection rates $k_W$ vary relatively weakly between the utilized temperatures, magnetic fields and MW powers. In contrast, the relaxation rates $k_R^{bup}$ are strongly affected by the experimental conditions which in turn strongly influence $\tau_{bup}$ and $P_{1n}$. The large $k_R^{bup}$ prevent excessively long buildup at the expense of rather low achievable polarization levels (Eq. S1c). Conversely, the lowest observed $k_R^{bup}$ and $k_R^{dec}$ at 6.7 T (1.4 K) result in the highest polarization and longest buildup time. At 3.4 K, the relatively field independent $k_W$ and $k_R^{bup}$ result in similar steady-state polarizations and buildup times. However, the higher thermal nuclear polarization at 7 T means that the enhancements are approximately halved for the same gained nuclear polarization. Cooling to 1.4 K at 3.35 T results in a very large relaxation rate during the buildup reflected by a fast buildup time and low steady-state polarization. The field independent injection rate facilitates further study since typical DNP models for the electron-nuclear HF-mediated polarization transfer predict the decrease of triple spin flip transition rate with increasing magnetic field strength[40, 43].

Next, we compare the rate equation parameters at 6.7 T (1.4 K) measurements across the different samples (Fig. S26b). The DNP injection $k_W$ is nearly identical for all samples except for the N++. The relaxation during the buildup ($k_R^{bup}$) shows a weak dependence on the doping level with lower doping levels having lower relaxation rate constants. The UW sample has higher relaxation and lower injection compared to P and N samples, resulting in its comparatively lower enhancement. This is even more pronounced for the relaxation during decay ($k_R^{dec}$) for which the UW sample has a nearly and order of magnitude faster relaxation than P and N with the latter two samples standing out among all samples with their smallest

$k_R^{dec}$. For all the samples, $k_R^{bup}/k_R^{dec}$ is between 5 and 10, suggesting a strong relaxation enhancement by MW irradiation[39]. In contrast, increasing the temperature to 3.4 K (7 T) which corresponds to a reduction in thermal electron polarization from around 99.7 to 88%, reduces the relaxation enhancement to around two- or three-fold (Fig. S27b). Reducing the field to 3.34 K at 3.4 K (around 59% thermal electron polarization) leads to the absence of a relaxation enhancement (Fig. S28).

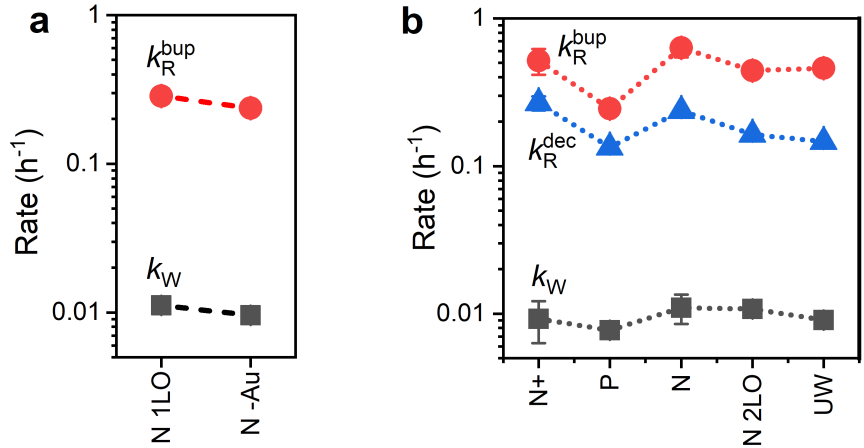

Figure S27: The polarization injection rates $k_W$ and the decay rates during the buildup $k_R^{bup}$ and decay $k_R^{dec}$. (a) Rates for the N 1LO and N -Au samples at 6.7 T (1.4 K). (b) Rates for the selected samples at 7 T (3.4 K). The rates were extracted from the one-compartment model[38]. Lines are guide to the eye.

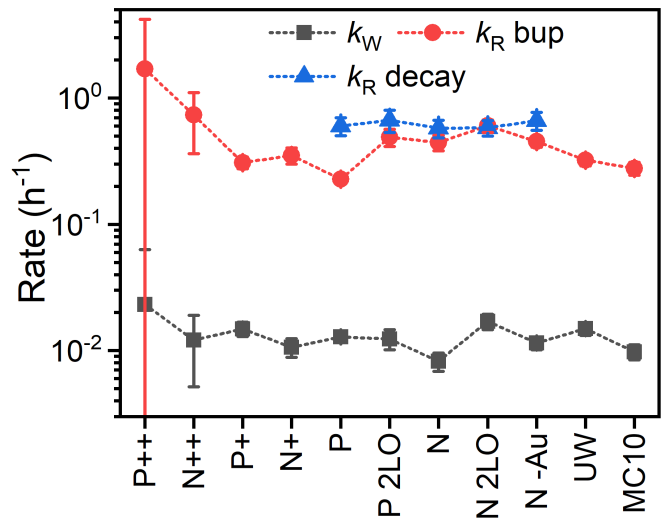

Figure S28: One-compartment model parameters calculated from the polarization build (dark squares and red circles) for the PSi NPs with different doping and oxidation. Decays were recorded only for around 5 hours only, resulting in underestimated decay relaxation rates. Magnetic field strength is 3.34 T, temperature is 3.4 K. Lines are guide to the eye.

At 6.7 T (1.4 K), the additional liquid oxidation and the oxidation induced by the Au removal increased the relaxation rates for the P and N samples (Fig. 4) with little difference between the two oxidation methods at 6.7 T (1.4 K) for the N sample (Fig. S27a). This increased relaxation turned into lower enhancements and faster buildup times compared to the thermally oxidized PSi NPs of the same doping density. The influence of additional liquid oxidation remains unclear provided that these oxidations yield little effect on the structure and number of the $P_b$ centers (Fig. 1 and Sec. S2.2, Suppl. Inf.). In contrast, at 3.34 T (3.4 K), the additional liquid oxidation and oxidation from the gold removal increased the achievable enhancements for N samples without causing a prolonged buildup. We note that the removal of the

gold nanoparticle catalyst has little effect on the DNP suggesting that the nuclear polarization inside the particle is well protected from surface modifications including metallic NPs.

To quantify the analysis of all samples at 6.7 T (1.4 K), we performed a correlation analysis (see Methods) between different sample properties and the experimentally measured DNP parameters (buildup time, polarization and rate constants). Below, we summarize the most significant correlations found:

- The specific surface area and, correspondingly, the density of $P_b$ centers is controlled by the doping density through its known influence on the outcome of LL-MACE (correlation coefficient equals to 0.9)[5, 6]. However, the role of doping is more complicated because high doping would lead to a large number of quenched nuclear spins which Larmor frequencies that are shifted by HF interaction with the spatially extended wave functions of shallow dopants (phosphorous and boron). This might be the reason of exceptionally small $k_W$ in the N++ sample. With the increase of temperature, the presence of delocalized thermally excited charge carriers would lead to fast room temperature relaxation for high doping density[44]. On the other hand, depletion space-charge layers formed due to Fermi level pinning by charged $P_b$ centers may almost completely deplete the pore walls from the charge carriers for the moderate and low doping densities[6] *i.e.*, for all the PSi NPs except N++ and P++. Indeed, no correlation is found between the doping densities and the enhancements or the rate-equation model rates.

- $k_W$ shows a correlation of 0.59 and -0.32 with the steady-state polarization in $P_0$ and the buildup time $\tau_{\text{bup}}$. For $k_R^{\text{bup}}$ these correlations are -0.48 and 0.97. Together, this is in good agreement with the expectation of Eq. S2 in which $\tau_{\text{bup}}$ is inversely proportional to $k_R^{\text{bup}}$ while $P_0$ depends on the ratio $k_W/k_R^{\text{bup}}$.

- The injection rate $k_W$ seems to depend mainly on the density of $P_b$ centers with a positive correlation coefficient of 0.74.

- The $k_R^{\text{bup}}$ and $k_R^{\text{dec}}$ relaxation rates show almost no correlation with the density of $P_b$ centers (absolute value below 0.1) and only weak correlation with the $\Delta B_{pp}^L$ (absolute values up to 0.18).

- All the rates depend moderately on the values of HF constant $A_{\text{iso}}$ and its strain as well on the Lorenzian line width of the $P_b^{(111)}$ centers.

These results might suggest that the DNP injection $k_W$ and relaxation $k_R^{\text{bup}}$ can be controlled separately in the NP synthesis as $k_W$ is sensitive to the overall $P_b$ density while $k_R^{\text{bup}}$ and $k_R^{\text{dec}}$ depend on the local properties of the $P_b$ centers.

Overall, DNP injection appears rather uniform across the samples while the relaxation rates during the buildup govern the buildup time and polarization enhancement. Furthermore, the DNP injection varies much less with experimental conditions than the relaxation rate. However, suppressing the relaxation with lower temperatures (1.4 K instead of 3.4 K) shows at best only a modest improvement as relaxation enhancement by MW irradiation[39] becomes more pronounced. Combining lower temperatures with higher fields partially suppresses the relaxation enhancement by MW irradiation[39] and results in the highest nuclear polarizations at 6.7 T and 1.4 K.

## S4  Density functional theory (DFT) simulations

Spin polarized density functional theory (DFT) simulations to calculate the HF and SHF interaction from first principles were performed with the *CP-PAW* code (http://www2.pt.tu-clausthal.de/paw/), employing the projector augmented wave (PAW) approach[45] and Perdew-Burke-Ernzerhof (PBE) exchange functional[46]. The plane-wave cutoffs were set to 40 Ry for the wave functions and to 80 Ry for the charge density. The silicon lattice constant was set to 0.5431 nm. The simulation box consisted of five conventional eight-atomic unit cells in each spatial direction, resulting in 1000 lattice sites. A single silicon atom in the centre of unit cell was replaced by either a boron or a phosphorous atom. The isotropic Fermi-contact HF interaction of the P dopant was calculated to 91.1 MHz - in agreement with the experimental value of 117.5 MHz[47] considering the finite unit cell of the simulation and PBE functional[48]. For the B dopant, the calculated isotropic Fermi-contact HF interaction is 1.4 MHz for an applied strain of 4 kbar as employed in previous DNP experiments[49]. The largest computed Si SHF

for the P dopant is 7.3 MHz, which is close to the measured 6 MHz[47]. The values for the P dopant are more than an order of magnitude larger than the computed 0.5 MHz in the B case.


\fi

\end{document}